\documentclass[twocolumn,showpacs,preprintnumbers]{aa}
\usepackage{natbib,twoopt}
\usepackage{graphicx,units,subfigure,color,amsmath,amssymb,epsfig}
\usepackage[varg]{txfonts}
\usepackage[usenames,dvipsnames,svgnames,table]{xcolor}

\usepackage[breaklinks=true]{hyperref} 
\bibpunct{(}{)}{;}{a}{}{,} 
\makeatletter
\newcommandtwoopt{\citeads}[3][][]{\href{http://adsabs.harvard.edu/abs/#3}%
{\def\hyper@linkstart##1##2{}%
\let\hyper@linkend\@empty\citealp[#1][#2]{#3}}}
\newcommandtwoopt{\citepads}[3][][]{\href{http://adsabs.harvard.edu/abs/#3}%
{\def\hyper@linkstart##1##2{}%
\let\hyper@linkend\@empty\citep[#1][#2]{#3}}}
\newcommandtwoopt{\citetads}[3][][]{\href{http://adsabs.harvard.edu/abs/#3}%
{\def\hyper@linkstart##1##2{}%
\let\hyper@linkend\@empty\citet[#1][#2]{#3}}}
\newcommandtwoopt{\citealtads}[3][][]{\href{http://adsabs.harvard.edu/abs/#3}%
{\def\hyper@linkstart##1##2{}%
\let\hyper@linkend\@empty\citealt[#1][#2]{#3}}}
\newcommandtwoopt{\citeyearads}[3][][]%
{\href{http://adsabs.harvard.edu/abs/#3}
{\def\hyper@linkstart##1##2{}%
\let\hyper@linkend\@empty\citeyear[#1][#2]{#3}}}
\makeatother

\definecolor{light-gray}{gray}{0.95}
\definecolor{dark-gray}{gray}{0.4}
\newcommand{\tab}[1]{\textcolor{TealBlue}{\textsc{\footnotesize \textbf{#1}}}}
\newcommand{\tabtext}[1]{\textcolor{TealBlue}{\textbf{\tiny {#1}}}}
\newcommand{\boxtext}[1]{\textit{\small{#1}}}
\newcommand{\button}[1]{\textcolor{dark-gray}{\textsf{#1}}}
\newcommand{\clicktext}[1]{[\textcolor{TealBlue}{\footnotesize \textsf{#1}}]}
\newcommand{\link}[1]{\textcolor{TealBlue}{\textsf{#1}}}

\begin{document}
\input epsf
\title{CRDB: a database of charged cosmic rays}

\author{
       D. Maurin\inst{1}
	\and F. Melot\inst{1}
  \and R. Taillet\inst{2}
} 

\offprints{D. Maurin, {\tt dmaurin@lpsc.in2p3.fr}}

\institute{
  LPSC, Universit\'e Grenoble-Alpes, CNRS/IN2P3,
      53 avenue des Martyrs, 38026 Grenoble, France
  \and
  LAPTh, UMR 5108 Universit\'e de Savoie - CNRS, BP 110, 74941, Annecy-le-Vieux, France
}

\date{Received / Accepted}

\abstract
{}
{This paper gives a description of a new on-line database
(\url{http://lpsc.in2p3.fr/crdb}) and associated on-line tools (data selection, data
export, plots, etc.) for charged cosmic-ray measurements. The experimental setups (type, flight dates,
techniques) from which the data originate are included in the database, along with the references to
all relevant publications.}
{The database relies on the \textsf{MySQL}5 engine. The web pages and queries are based on
\textsf{PHP}, \textsf{AJAX} and the \textsf{jquery}, \textsf{jquery.cluetip}, \textsf{jquery-ui}, and
\textsf{table-sorter} third-party libraries.}
{In this first release, we restrict ourselves to Galactic cosmic rays with $Z\leq 30$ and a kinetic
energy per nucleon up to a few tens of TeV/n. This corresponds to more than 200 different
sub-experiments (i.e., different experiments, or data from the same experiment flying at different
times) in as many publications.}
{We set up a cosmic-ray database and provide tools to sort and visualise the data. New data can be
submitted, providing the community with a collaborative tool to archive past and future cosmic-ray
measurements. Any help/ideas to further expand and/or complement the database is welcome (please
contact \href{mailto:crdatabase@lpsc.in2p3.fr}{crdatabase@lpsc.in2p3.fr}).}

\keywords{Astroparticle physics - Galaxy: cosmic rays - databases: miscellaneous}

\maketitle


\section{Introduction}

Since the discovery of cosmic rays (CR) a century ago, instrumental capabilities have steadily
improved. A large variety of types of experiments (balloon- or satellite-borne, flown on a shuttle,
installed on the international space station, or ground-based experiments) and techniques have been
used (nuclear emulsions, drift chambers, Cerenkov counters, spectrometers...) to refine our knowledge
of the CR composition and spectrum.

The presence of heavy $Z<30$ \citepads{1948PhRv...74.1818F,1948PhRv...74..213F}, and extremely heavy
$Z\ge30$ elements \citepads{1967RSPSA.301...39F,1969PhRvL..23..338B} in the cosmic radiation were among
the main discoveries related to Galactic CR nuclei, which culminated with the discovery of a few $Z>90$
events \citepads{1970RSPSA.318....1F,1971PhRvL..26..463O,1971PhRvD...3..815P}. Isotopes and in particular
the radioactive CR clocks were identified, with more or less difficulties due to their decreasing
abundance and mass separation with increasing atomic number: $^{10}$Be \citepads{1973Ap&SS..24...17W},
$^{36}$Cl \citepads{1981ApJ...246.1014Y}, $^{26}$Al \citepads{1982ApJ...252..386W}, and $^{54}$Mn
\citepads{1979ICRC....1..430W}. CR leptons were identified in the early 60's, with the first measurement
of electrons---also called negatrons at that time--- \citepads{1961PhRvL...6..125E,1961PhRvL...6..193M}, and
of the positron fraction 
\citepads{1964PhRvL..12....3D,1965JGR....70.2713H,1965ICRC....1..331A,1965ICRC....1..335D}. Anti-protons
are $\sim 10^{-4}$ times less abundant than protons and were only observed in the late 70's
\citepads{1979ICRC....1..330B,1979PhRvL..43.1196G}. Anti-deuterons, which are expected to be yet another
factor $\sim10^{-4}$ below \citepads{1997PhLB..409..313C}, are still to be detected: the best limit is
given by the {\sf BESS} balloon \citepads{2005PhRvL..95h1101F}---for limits on anti-helium, see
\citetads{2012PhRvL.108m1301A}---, and is still three orders of magnitude above what is required to reach
the expected astrophysical production. This level could be within reach of the AMS-02 detector on the
International Space Station \citepads{2008arXiv0801.3243A,2008ICRC....4..765C} and/or the {\sf GAPS}
balloon-borne experiment \citepads{2012NIMPA.682...90A}. Note that other milestones in CR studies are the
discovery of the $\gamma$-ray diffuse emissions reported first by \citetads{1972ApJ...177..341K}---and
studied by the contemporary {\sf Fermi-LAT} instrument \citepads{2012ApJ...750....3A}---, and the first
evidence of high-energy CRs from extensive air showers \citepads{1939RvMP...11..288A}---currently studied at
the {\sf Pierre Auger Observatory} \citepads[e.g.,][]{2010PhLB..685..239A}.

In the last twenty years, a lot of efforts have been devoted in measuring the CR composition at
higher energy ($\gtrsim $ TeV). Very accurate isotopic data were also provided in the low energy range $\sim$
100--500 MeV/n ({\sf Voyager 1\&2}, {\sf Ulysses}, and {\sf ACE}), over an extended period
of time. The recently installed {\sf AMS-02} experiment on the International Space Station (May 2011)
has started to provide impressive measurements in the GeV/n--TeV/n range \citepads{2013PhRvL.110n1102A}.
Interestingly, most of the even oldest CR measurements are not outdated yet. Indeed, many instruments
are designed to focus on specific CR species: neither all instruments have the isotopic resolution
capabilities, nor all species have been measured repeatedly. Some old experiments are also useful when
one wishes to inspect a possible charge-sign dependence (22 year cycle) of the Solar modulation effect
as a function of the Sun polarity, as first proposed by \citetads{1996ApJ...464..507C} and further
studied in \citepads{2002ApJ...568..216C}. For all these reasons, we believe it is worth providing an
archival database of CR measurements to the community.

CR data are the backbone of Galactic CR propagation studies (e.g.,
\citealtads{2001ApJ...547..264J}; \citealtads{2001ApJ...555..585M};
\citealtads{2007ARNPS..57..285S}; \citealtads{2008JCAP...10..018E}). In
the last twenty years, anti-protons and positron fraction measurements have also become a strong probe
for dark matter indirect searches (e.g., \citealtads{2011ARA&A..49..155P} and \citealtads{2012CRPhy..13..740L}). 
A database would
therefore be useful to any researcher in these fields, but also to CR experimentalists who wish to
compare their data to previously published ones. Another independent effort to provide a CR database
was presented in \citetads{2009arXiv0907.0565S}. We present here a contextualised and more complete
version of the data\footnote{The data were gathered independently of the data presented in the
\citetads{2009arXiv0907.0565S} database \url{https://sourceforge.net/projects/cosmicraydataba}},
along with many user-friendly interfaces and tools to use them.

The paper is organised as follows: Sect.~\ref{sec:db_content} describes the database content;
Sect.~\ref{sec:website} describes the website and available tools. We conclude and comment on possible
improvements of the database in Sect.~\ref{sec:concl}. Appendix~\ref{app:rules} provides the rules
to combine CR quantities from a given experiment (e.g., to form quantity A/B from A and
B fluxes), and App.~\ref{app:bibtex} gives a summary list of all the experiments/references contained in
this first release.


\section{Content of the database}
\label{sec:db_content}

In this section, we first describe the information gathered in the database, and the data themselves.
We then present how this information is organised in a \textsf{MySQL} framework\footnote{The descriptions
correspond to {\tt CRDB V2.1}, which was the version at the time of re-submission of the paper.}.


\subsection{Definitions}
\label{sec:definition}

CR data are connected to experiments, analyses, and publications. The first step for creating the
database is to define what an experiment is. Then, the need to define a sub-experiment arises because
i) an experiment may consist of several detectors, or ii) an instrument may have flown several times,
or over distinct periods. Data from a sub-experiment often involve several CR species, the analyses of
which are published in one or several papers. For the sake of clarity, the following
keywords/definitions are used in the database:

 \begin{description}

   \item[\bf Experiment] Name associated with the instrument ({\em CREAM}, {\em AMS}). To identify
   unnamed balloons, we use the syntax {\em Balloon (YYYY)}, and a further distinction is made if a
   balloon was flown several times: a comma-separated list of years {\em Balloon (1966,1967)} is used
   if the data were analysed and published for each flight; a plus-separated list {\em Balloon
   (1967+1968)} is used if the data resulted from the combined analysis of the flights\footnote{The
   naming convention chosen for the experiment and the sub-experiment ensures the many unnamed balloons
   flown before the 90's to be uniquely defined.}.

   \item[\bf Sub-experiment] Sub-detector name or experiment name concatenated with the flight number
   and data taking period {\em (YYYY/MM)}, with start and stop dates separated by a hyphen for
   durations over a month\footnote{Some of the start and stop dates for balloon flights are not given
   in the publication and were taken from the {\sc StratoCat} database of stratospheric balloons
   launched worldwide since 1947 (\url{http://stratocat.com.ar/globos/indexe.html}). For long-lived
   instruments, we do not include in the database the excluded time periods (within the start and stop
   dates) based on the analysis quality criteria (solar flares, high solar activity, instrument
   stability\dots) because they are never given in the publication.}: {\em Balloon (1972/07)}, {\em
   CREAM-I (2004/12-2005/01)}, {\em CREAM-II (2005/12-2006/01)}, {\em Ulysses-HET
   (1990/10-1997/12)}.

   \item[\bf Cosmic-ray quantity] Combination (sum, ratio, etc.) of measured CR species\footnote{A CR
   must be a stable species with respect to the confinement time in the Galaxy, i.e. with an effective
   lifetime $\gtrsim$~kyr (note that the electronic capture decay mode is suppressed because CR nuclei
   are fully stripped of $e^-$ above $\sim 0.1$ GeV/n).}. It can be an elemental (e.g., C), isotopic
   (e.g., $^1$H), or leptonic (e.g., $e^-$) flux, or any ratio of these quantities such as B/C,
   $^{10}$Be/Be, $e^+/(e^-+e^+)$, etc. The keyword {\em SubFe} is used for the group $Z=21-23$, but no
   other charge group is defined for now.

   \item[\bf Energy axis] Detectors often measure the CR total energy $E_{\rm tot}$ or rigidity ${\cal
   R}= pc/Ze$ ($p$ is the momentum, $Z$ the charge, $c$ the speed of light, and $e$ the electron
   charge). Data are also very often presented as a function of the kinetic energy $E_{\rm k}=
   E_{\rm tot} - m$ ($m$ is the CR mass) or the kinetic energy per nucleon $E_{\rm k/n}= E_{\rm
   k}/A$ ($A$ is the atomic number). In the database, we allow four representations of the energy unit
   and axis: {\em [GeV]} for $E_{\rm tot}$, {\em [GV]} for ${\cal R}$, {\em [GeV]} for $E_{\rm k}$, and
   {\em [GeV/n]} for $E_{\rm k/n}$.

   \item[\bf Publication] Refereed or non-refereed reference (i.e., journal or conference proceedings)
   providing CR quantity data from (sub-)experiments. A publication is usually attached to a single
   (sub-)experiment and it contains different CR measurements, but there are a few exceptions. Over
   time, some of these publications may be superseded by newer analyses: a specific entry of the
   database allows to keep track of deprecated analyses and references.

   \item[\bf Data] CR quantity measurement and uncertainties at one or several energy bins
   (see Sect.~\ref{sec:data} for a complete description).

 \end{description}

The fact that combinations of CR quantities are themselves CR quantities introduce a subtleties in the
choice of how to handle the database. One could be tempted to fill the database with all useful
combinations of data (e.g., the often used B/C ratio) from published quantities (e.g., B and C fluxes).
However, the number of combinations that can be formed is large (for $Z<30$, as many elements and about
a hundred isotopes can be combined), and the procedure to combine the errors on the measurements is not
always sound. For these reasons, we decided to fill the database with the published quantities only. We
leave the task of extracting the most complete dataset (for a given CR quantity) to the \tabtext{Data
Extraction} interface (Sect.~\ref{sec:data_extraction}), which combines the data found directly in the
database, and those obtained by looking for all combinations of data leading to this quantity (see
Sect.~\ref{sec:data_extraction}, and also App.~\ref{app:rules} for a discussion of the priority rules
and criteria to decide how and when to form new quantities and evaluate their uncertainties).

\subsection{Data description and units}
\label{sec:data}

The structure of a CR data entry (energy, energy range, measurement and uncertainties) for any measured
quantity is as follows:

 \begin{description}

   \item[$\mathbf{\langle E\rangle}$] `Central' energy given in the publication (unit is {\em [GeV]} if
   the energy axis is $E_{\rm tot}$ or $E_{\rm k}$, {\em [GV]} for ${\cal R}$, and {\em [GeV/n]} for
   $E_{\rm k/n}$). If only the bin range (see below) is given in the publication, the geometric mean
   $\langle E\rangle=\sqrt{E_{\rm min}E_{\rm max}}$ is used.

   \item[\bf Bin range] Energy range (same unit as $\langle E\rangle$). If only $\langle E\rangle$ is
   given in the publication, $E_{\rm min}=E_{\rm max}=\langle E\rangle$.

   \item[\bf Value] Measured CR quantity in unit of [$(\langle E \rangle\,{\rm m}^2\,{\rm s}\,{\rm
   sr})^{-1}$] if this is a flux, or unit-less if this is a ratio. The data correspond to
   top-of-atmosphere (TOA) quantities, i.e. modulated by the Sun's activity.

   \item[\bf Stat Err] Statistical error (same unit as {\em Value}).

   \item[\bf Syst Err] Systematic error (same unit as {\em Value}); set to $0$ if not given in
   publication.

\end{description}

In the database, these values must be filled for each data point from the published data. Whenever
available, we used the values given in the publication tables. However, most publications provide none,
and the data had to be retrieved from the plots (using the {\sc DataThief III}
software\footnote{\url{http://datathief.org}}).

\subsection{Solar modulation description}
\label{sec:solmod}

A flight period for a given instrument is uniquely associated to a Solar activity period. Hence,
for each sub-experiment of the database, a unique modulation level can be attached. In practice,
its determination depends primarily on the choice of the (unknown) Solar modulation model. It also
depends on the assumption made for the unknown interstellar (IS) flux, and on the set of data used
for its calculation. At least three different model-dependent approaches exist: very often, the
modulation potential given in the experimental papers is determined from an assumed {\em ad-hoc} local
interstellar spectrum (LIS), which is modulated to match the data. Another widely used approach is to
take a demodulated spectrum from another publication and modulate it to match the data. Another strategy
is to rely on the worldwide network of neutron monitor data that gives an indirect estimation of the
Solar modulation level \citepads{2002SoPh..207..389U,2005JGRA..11012108U,2011JGRA..11602104U}.

\subsubsection{Sets of modulation levels in the database}
As there is no consensus yet regarding how to proceed best, we provide in the database several sets
of values (of the modulation parameters) and the underlying assumptions made for their determination.
The two sets introduced in {\tt CRDB~V2.1} are:
  \begin{enumerate}
   \item {\em From publication:} in this set, the solar modulation values are directly taken from
   the publications. This set is not {\em homogeneous} since each value relies on IS fluxes and
   Solar modulation models (see below) which differ from one publication to another. Note that the
   underlying hypotheses are not always provided, and that only a small fraction of the publications
   offers a modulation value at all.
   \item {\em NM [Uso11]:} the modulation levels in this set ($\phi$ in the Force-Field approximation))
   are based on the NM data analysis of \citetads{2011JGRA..11602104U}. Monthly average values in the
   period July 1936 -- December 2009 are interpolated for the flight dates of each sub-experiment.
   The ensemble of values forms an {\em homogeneous} set of $\phi$ for all sub-experiments, because
   each value is based on the same modelling.
  
  \end{enumerate}
It is very important to pursue the efforts towards the determination of homogeneous sets of modulation
parameter values: first, because these values are routinely used {\em as is} by cosmic-ray physicists for 
CR propagation studies (and also in the context of indirect dark matter detection); second, because
assessing which is the correct modulation model requires to disprove simple models, and systematic
studies of CR data on several decades are mandatory to achieve this goal. Indeed, as underlined in
\citetads{2014arXiv1403.1612M}, there is a degeneracy between the IS flux and the solar modulation level,
and assuming the wrong IS flux lead to a shift of all determined $\phi$ values. Combining the results
of direct fits to CR data and the use of NM data should provide more robust (and homogeneous) determinations
of $\phi$ time series \citetads{2014arXiv1403.1612M}. As progress is made along this line (Ghelfi
et al., in prep.), new sets of values will complement the two above.

\subsubsection{Ingredients: modulation model, IS flux, and data}
The modulation level for any set and any sub-experiments is associated to assumptions
regarding the ingredients of its calculation. To avoid too much complexity, a few general
categories were created for each ingredient:
\begin{itemize}
   \item {\bf Modulation models and values\footnote{For the time being, only one parameter value
    per Solar modulation model is allowed. If no value is provided in the publication, or if a
    3D model is used, the value is set to 0.}}
      \begin{enumerate}[(i)]
         \item {\em N/A:} no modulation model in the publication;
         \item {\em Diffusion/convection:} model used in early CR data publications, with the free parameter $\eta$
              \citepads{1958PhRv..110.1445P,1958PhRv..109.1874P,1972ApJ...171..363L};
         \item {\em Force-Field approximation:} widely used model, with one
              free parameter $\phi_{\rm FF}$ \citepads{1967ApJ...149L.115G,1968ApJ...154.1011G,1987A&A...184..119P,2004JGRA..109.1101C}
         \item {\em Spherically Symmetric solution:} model with a single effective\footnote{The parameter $\phi$ is related to the Solar wind
         velocity $V_w$, the spatial-dependent diffusion coefficient $\kappa(r)$, and the Solar cavity boundary $R_0$,
         by $\phi=(1/3)\int_1^{R_0} V_w/\kappa(r)\, dr$. Note however that this formula depends
         on assumptions made on the wind spatial dependence and the diffusion coefficient rigidity dependence.}
              free parameter $\phi_{\rm 1D}$ \citepads{1969JGR....74.4973F,1971JGR....76..221F,1993ApJ...413..268B};
         \item {\em Sign-charge dependent:} 3D models have plenty of parameters to
         describe the tilt angle, the current sheet, anisotropic diffusion, etc. \citepads{1979ApJ...234..384J,1985ApJ...294..425P,2013LRSP...10....3P}.
         For simplicity, we use the same category for all of them.
     \end{enumerate}
    
   \item {\bf IS fluxes}
      \begin{enumerate}[(i)]
         \item {\em N/A:} if unknown or if $\phi$ not calculated;
         \item {\em LIS flux hypothesis:} if taken from another study;
         \item {\em LIS flux fitted:} if taken from a fit on CR data;
         \item {\em Leaky-Box calculation:} if taken from a Leaky-Box model calculation \citepads[e.g.,][]{1987ApJS...64..269G};
         \item {\em GALPROP calculation:} if taken from a {\tt GALPROP}\footnote{\url{http://galprop.stanford.edu}} run \citepads{1998ApJ...509..212S};
         \item {\em Monte Carlo Diffusion:} if taken from a Monte Carlo propagation run \citepads[e.g.,][]{1997AdSpR..19..817W}.
     \end{enumerate}

   \item {\bf Data}
      \begin{enumerate}[(i)]
         \item {\em N/A:} if unknown or if $\phi$ not calculated;
         \item {\em CR data from the publication:} if $\phi$ is determined from the published set of CR data;
         \item {\em CR data from another publication:} if $\phi$ is based on another published set of CR data;
         \item {\em Neutron Monitor (NM) data:} if $\phi$ from NM data.
     \end{enumerate}
\end{itemize}
Examples of how these informations are organised and displayed in {\tt CRDB} are given in the next sections.
\begin{figure*}[!t]
\begin{center}
\includegraphics[width=0.9\textwidth]{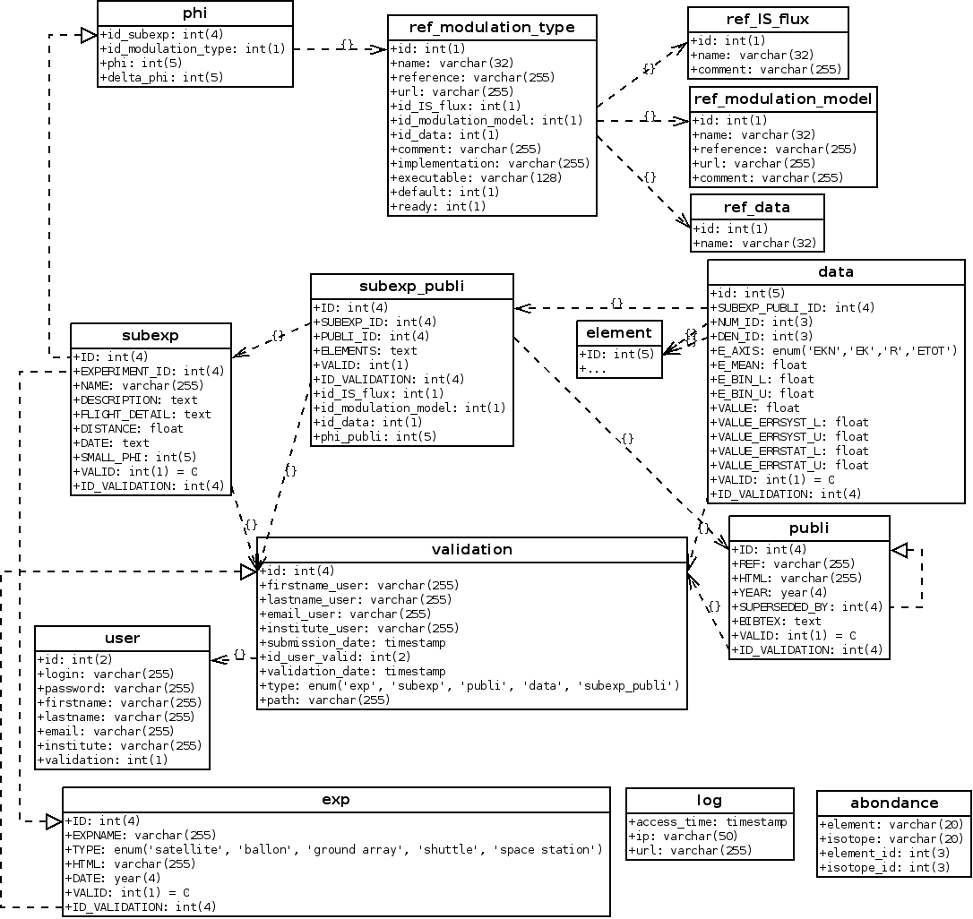}
\caption{Tables and keys of the database \textsf{MySQL} structure (see text for details).}
\label{fig:mysql}
\end{center}
\vspace{-3mm}
\end{figure*}

\subsection{Database structure description}
\label{sec:structure}

The database engine is {\sf MySQL5}, hosted at the Laboratoire de Physique and Cosmologie (LPSC) on a
backed up server. The structure and keys are shown in Fig.~\ref{fig:mysql} (keys in {\bf exp}, {\bf
subexp}, and {\bf publi} were discussed in Sect.~\ref{sec:definition}, those in {\bf data} in
Sect.~\ref{sec:data}, and those in {\bf ref\_modulation\_type} in Sect.~\ref{sec:solmod}).
Each entry in a table is associated with a unique identifier. These identifiers
are used to link elements from one table to another (for example, several sub-experiments can be linked
to a single experiment). For completeness, all tables of Fig.~\ref{fig:mysql} are briefly described
below:

\begin{figure*}[!t]
\begin{center}
\includegraphics[width=0.75\textwidth]{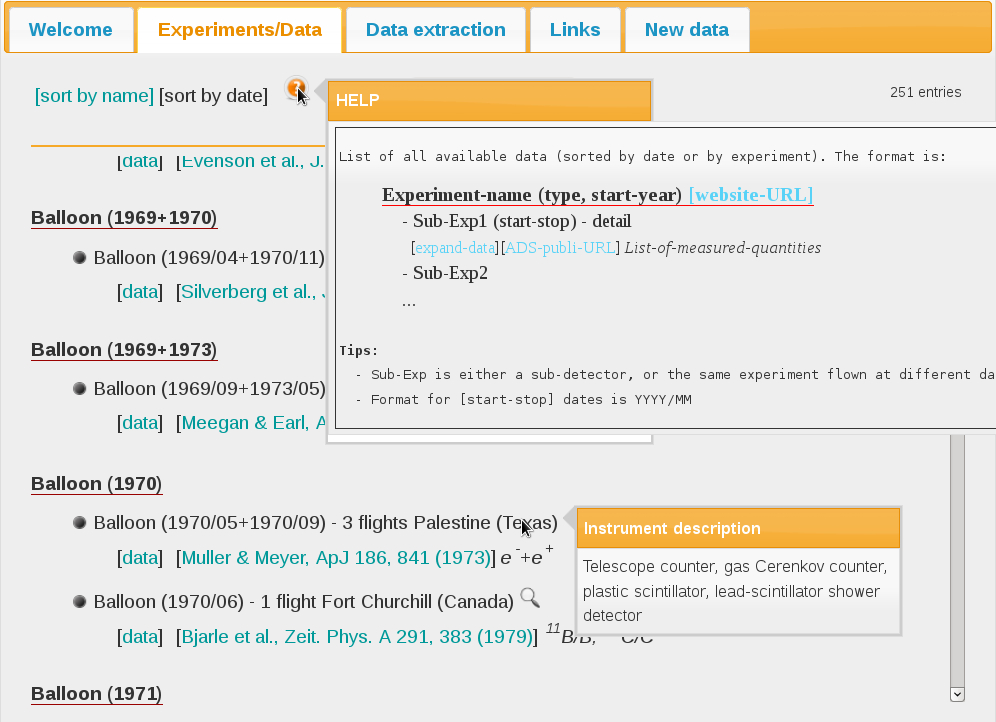}
\caption{Snapshot of the \tabtext{Experiments/Data} tab content (Sect.~\ref{sec:exp_data_tab}). The
\boxtext{HELP} box is activated clicking on the {\em question mark} icon, and a picture of the
instrumental setup (not shown) pops-up when clicking on the {\em magnifying glass} icon. The
\boxtext{Instrument description} box appears for a mouse-over action on the sub-experiment name. A
click on \clicktext{data} pops-up a new window with the data entries and a summary of all the
(sub-)experiment/publication informations (not shown).}
\label{fig:exp_data}
\end{center}
\vspace{-3mm}
\end{figure*}

\begin{description}

   \item {\bf exp} Name, type, web site (if available), and flight date.

   \item {\bf subexp} Link to the experiment it belongs to, name, description of the apparatus, flight
   details (launch place and the number of flights for balloons), flight dates, distance to the Sun
   [AU], and Solar modulation level [MV].

   \item {\bf publi} Bibliographic reference, web link, publication year, 
   \textsc{Bib}\TeX\footnote{\url{http://www.bibtex.org}} entry (taken from the Astrophysics
   Data System ADS\footnote{\url{http://cdsads.u-strasbg.fr}}), and link to other publications
   (if more recent analyses exist).

   \item {\bf subexp\_publi} Bridge table linking entries from {\bf publi} to one or several entries of
   {\bf subexp}.

   \item {\bf phi} Solar modulation level and error for the sub-experiment.

   \item {\bf ref\_modulation\_type} Bridge table linking entries from {\bf phi} to entries
   in the three tables below.

   \item {\bf ref\_IS\_flux} List/description of IS flux categories.

   \item {\bf ref\_modulation\_model} List/description of Solar modulation models.

   \item {\bf ref\_data} List/description of data used to calculate $\phi$.

   \item {\bf element} Name, mass, atomic number, charge, etc. for CR quantities (isotopes, elements,
   $\bar{p}$, $e^{-}$, and $e^{+}$). 

   \item {\bf data} Type (flux or ratio of {\bf element}), energy axis, energy, bin range, value,
   statistic and systematic errors.

   \item {\bf abondance} Associate each element to its most abundant isotope to allow
   (approximate) conversions between different energy axes (since {\tt V1.2}).

   \item {\bf user} Contact details of administrators (persons authorised to change and validate
   submitted data).

   \item {\bf log} Statistics of the number of the users (since {\tt V1.2}).

   \item {\bf validation} Contact details of persons submitting new data (see
   Sect.~\ref{sec:add_data_tab} for the \tabtext{New Data} interface), validation date, and identity of
   the person (user) who validated the data.

\end{description}

\section{Website, interfaces, and example plots}
\label{sec:website}

The CR database website \url{http://lpsc.in2p3.fr/crdb} is hosted by the LPSC
laboratory website, and is based on a {\sf LAMP} solution\footnote{The acronym {\sf LAMP} refers to a
stack of free open source softwares: {\sf Linux} operating system, {\sf Apache HTTP} server, {\sf MySQL}
database software, and {\sf PHP}.}. Authentication uses the {\tt https} protocol to ensure a good level
of confidentiality (only administrators own credentials to access protected areas). All web pages are
written using the {\sf PHP} (Hypertext PreProcessor) language, with a global structure made in {\sf
AJAX} (Asynchronous {\sf JavaScript} and {\sf XML}). The third-party libraries {\sf jquery},
{\sf jquery-ui}, {\sf jquery.cluetip}, and {\sf table-sorter} are also used.

The website is based on tabs, in which the user is guided by \boxtext{HELP} boxes (identified by
{\em question mark} icons). We give below a brief description of the implemented tabs:

 \begin{description}

   \item \tab{Welcome} Quick description and organisation of the database, log of the latest changes,
   and link to download the database content formatted for the 
   {\sf USINE}\footnote{\url{http://lpsc.in2p3.fr/usine}} propagation code.

   \item \tab{Experiments/Data} List of available data sorted by experiment names or dates. A list of
   experiment acronyms is given.

   \item \tab{Data extraction} Main interface to retrieve data in {\sf ASCII} files, {\sf
   ROOT}\footnote{\url{http://root.cern.ch}} macros and plots, and \textsc{Bib}\TeX\ references for the
   selection.

   \item \tab{Admin} Shown for authenticated users only: internal checks of the database content,
   validation of submitted data.

   \item \tab{Links} Standard useful (here GCR-related) web links. 

   \item \tab{New Data} Interface to submit new data which will appear in the database after validation by
   authorised users.

 \end{description}

\begin{figure*}[!t]
\begin{center}
\includegraphics[width=0.85\textwidth]{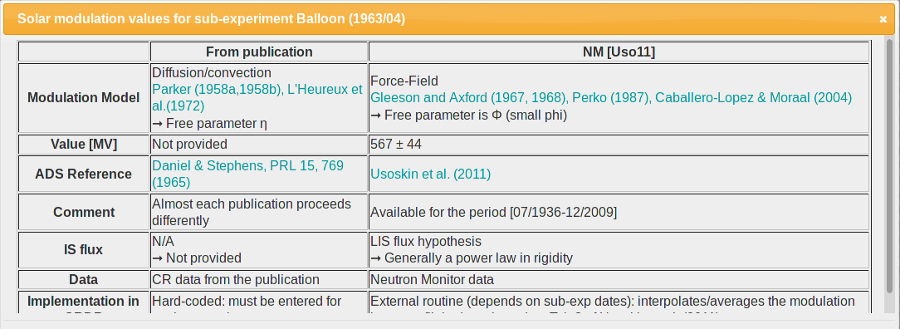}
\caption{Snapshot of the Solar modulation values for a given sub-experiment 
(after clicking on \clicktext{data}, see Fig.~\ref{fig:exp_data}, then 
\clicktext{Parameters}).}
\label{fig:snapshot_modul}
\end{center}
\vspace{-3mm}
\end{figure*}

As underlined previously, native data (i.e. data directly from publications) are listed and accessed
from \tabtext{Experiments/Data} (Sect.~\ref{sec:exp_data_tab}). In the \tabtext{Data extraction} tab
(Sect.~\ref{sec:data_extraction}), native data and matching combinations of native data are combined to
provide the most complete list of data found for user-selected quantities/criteria. Adding new data
is possible from the \tabtext{New data} tab (Sect.~\ref{sec:add_data_tab}).

\subsection{Data access from \tab{Experiments/Data} tab}
\label{sec:exp_data_tab}

\begin{figure*}[!t]
\begin{center}
\includegraphics[width=0.8\textwidth]{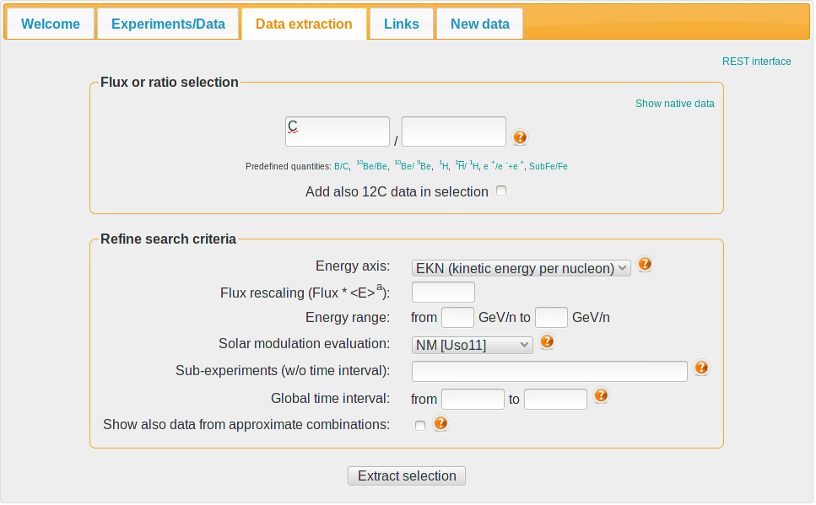}
\caption{Snapshot of the \tabtext{Data extraction} tab interface. In the upper panel, a CR quantity is
chosen by means of selection boxes (auto completion enabled). In the lower panel, more search criteria
are possible (energy range, list of experiments or sub-experiments, time period, etc.).
The selector {\em Solar modulation evaluation} allows the user to choose
which set of $\phi$ values (see Sect.~\ref{sec:solmod}) is displayed. A tick box
allows to add in the search the data points obtained from combinations of `native' data (see
App.~\ref{app:rules}). Clicking on the \button{Extract selection} button pops up
the results, as shown in the example Fig.~\ref{fig:extraction_results}.}
\label{fig:extraction}
\end{center} 
\vspace{-3mm}
\end{figure*}

Figure~\ref{fig:exp_data} shows a snapshot of the \tab{Experiments/Data} tab (and some enabled actions within this tab). The
list of published data is ordered by experiment name or date. For each experiment, the list
of sub-experiments is shown and sorted by start time\footnote{We refer the reader to
Sect.~\ref{sec:definition} for the definition of what is meant by experiment, sub-experiment,
publication, etc., and to Sect.~\ref{sec:structure} for the structure of the data in the \textsf{MySQL}
frame.}. The publication references related to this sub-experiment are then listed along with the
quantities measured (older analyses/publications of the same data are indicated). The most useful
actions/pop-up informations available for the user are:

 \begin{itemize}

   \item experiment description (name, type, official web page);

   \item sub-experiment description (name, data periods\footnote{Full details of the flight dates are
   given clicking on \clicktext{data}. The start and stop time format is {\em YYYY/MM/DD-hh:mm:ss}. A
   new line is used for each flight.}, instrument description [{\em mouse-over} name], experimental
   setup picture [{\em magnifying glass} icon]);

   \item data for each publication (from a sub-experiment). A click on \clicktext{data} (see
   Fig.~\ref{fig:exp_data}) pops-up a window (not shown in the examples): its upper half summarises all
   the information on the sub-experiment (contained in the database) and gives the ADS link of the
   reference; the lower half shows the data (see Sect.~\ref{sec:data} for their format): ratios are
   sorted first and fluxes second (one can jump directly to the data of interest clicking on one of the
   {\em Quantities} suggested in the upper half panel). In the upper half panel,
   clicking on \clicktext{Solar modulation: Parameters} pops up a new window with the sub-experiment
   modulation values, as shown in Fig.~\ref{fig:snapshot_modul}.
\clicktext{Solar modulation: Parameters}).

 \end{itemize}

By default, CR data for nuclei and anti-nuclei are given as a function of kinetic energy per nucleon,
whereas leptons are given as a function of the kinetic energy. Whenever the energy axis is rigidity, the
flag \textcolor{red}{\small[Rigidity]} is added after the data.

\begin{figure*}[!t]
\begin{center}
\includegraphics[width=0.7\textwidth]{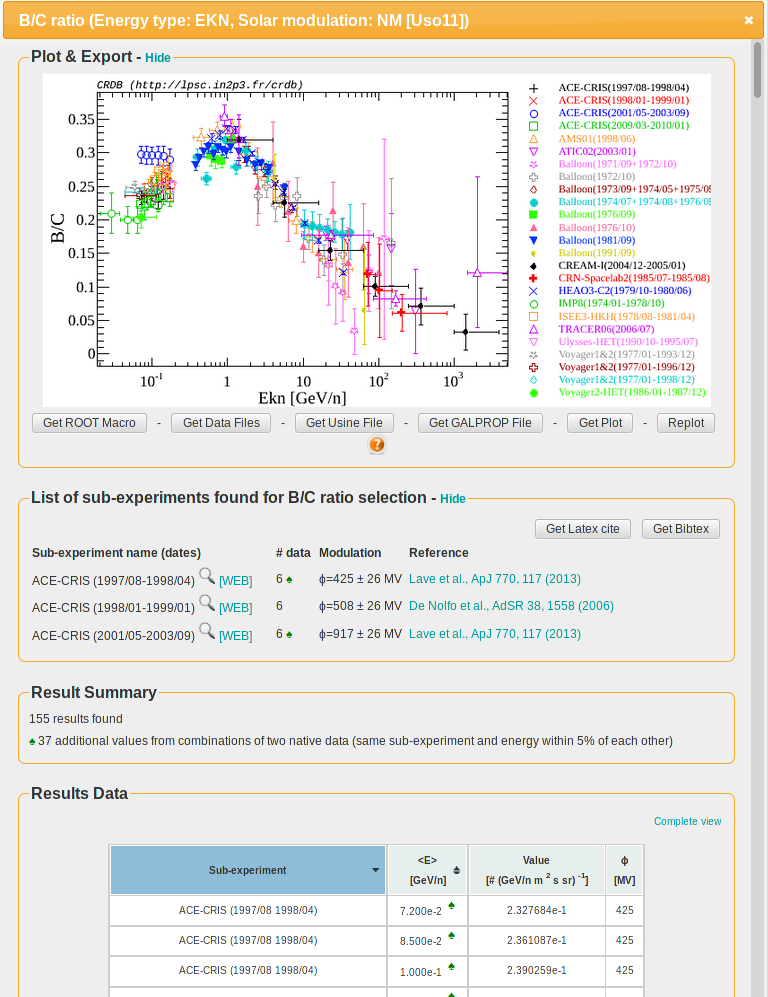}
\caption{Snapshot of the result of the \tabtext{Data extraction} operation. This pop-up window appears
after the selection step shown in Fig.~\ref{fig:extraction} is completed. Buttons, links, and tables
give access to raw data and plots, see text for details.}
\label{fig:extraction_results}
\end{center} 
\vspace{-3mm}
\end{figure*}

\subsection{Selection and tools from \tab{Data extraction} tab} \label{sec:data_extraction}

Figure~\ref{fig:extraction} shows a snapshot of the selection interface within the tab. A mandatory step
is the quantity selection ({\sf Flux or ratio selection}), for which a few predefined choices are
proposed. For a ratio, both the numerator and denominator selection boxes must be filled (auto
completion is enabled). The other optional selection criteria ({\sf Refine search criteria}) are:

  \begin{itemize}

    \item \button{Energy axis}: to be selected among {\sf EKN, EK, R} or {\sf Etot}\footnote{As already
    said, most data are published in EKN for nuclei and anti-nuclei, and EK for leptons (very few data
    are published on several energy axes). For elements, a conversion to move from one energy axis
    to another is implemented since {\tt V1.2}. The result is approximate, since
    a unique  $A$, $Z$, and $m$ must be assumed for all isotopes of this elements (we use the most
    abundant isotope, which is a very good approximation for some elements).};

    \item \button{Flux rescaling}: multiplies the flux values and errors by $\langle E\rangle^{a}$
    (useful for presentation purpose);

    \item \button{Energy range}: restricts the energy range allowed;
    
    \item \button{(Sub-)Experiment names}: list of comma-separated names (partial names allowed, e.g.,
    {\em CREAM,BESS});

    \item \button{Time interval}: selects only experiments falling into the selected period (format is
    {\em YYYY/MM}).

    \item \button{Show also data from combinations}: tick box to add in the search the data points
    obtained from combinations of `native' data (see App.~\ref{app:rules}).

  \end{itemize}

Hitting the \button{Extract Selection} button pops-up a new window with the data extracted from the user
selection. This is shown in Fig.~\ref{fig:extraction_results}, organised in three panels (click on
\clicktext{hide}/\clicktext{show} to collapse/expand each panel):

  \begin{enumerate}

    \item {\sf Plots and exports for the selection}: 
       \begin{itemize}
         \item Export selected data in various formats: \button{Get Data Files},
             \button{USINE, GALPROP File} provide respectively
             a tar-ball of {\sf ASCII} files containing the
             data (one file per sub-experiment), and a {\sf USINE}- or {\sf GALPROP}-compliant file
             (i.e., input for these propagation codes);
         \item Exports plots: \button{Get ROOT Macro} and \button{Plot} provide respectively
           a {\sf ROOT} executable {\sf C++} file to re-generate and/or modify
           the plot\footnote{Based on the {\sf ROOT} library {\url{http://root.cern.ch}}. To execute,
           type {\tt root database\_plot.C} (the data are hard-coded). The errors displayed correspond
           to the quadratic sum of statistical and systematic uncertainties.}, and a high-resolution
           image {\tt database\_plot.png};
         \item Replot (since {\tt V1.2}): new panel to change the plot (size, y-axis range, select only
             a subset of sub-experiments, etc.).
     \end{itemize}

    \item {\sf List of experiments found for the selection}: summary of the data sorted by
     (sub-)experiment (name, publication, number of data, etc.). The \button{Get Bibtex} and
     \button{Latex cite} buttons provide respectively a \textsc{Bib}\TeX\ file ({\sf bibtex.bib}) to be
     included in the references, and the text to cite this selection in the \LaTeX\
     document\footnote{These files are useful to quickly prepare scientific manuscripts based on
     \LaTeX\ and \textsc{Bib}\TeX. App.~\ref{app:bibtex} and the references of this paper were prepared
     with the full list of references retrieved from the \button{Get Bibtex} and \button{Latex cite}
     buttons in the \tabtext{Welcome} tab.}. As for the \tabtext{Experiments/Data} tab
     (Sect.~\ref{sec:exp_data_tab}), links to the experiment website and the ADS publication are
     provided.

    \item {\sf Data for the selection}: data in a table (see Sect.~\ref{sec:data} for the content
    description) sorted by experiment name or energy. An asterisk denotes the data obtained by
    combinations of native data of the database (see App.~\ref{app:rules}).

    \end{enumerate}

We remind that, with the default search criteria (i.e., none), all analyses of a CR quantity by the same
instrument show up in the result as long as they correspond to different data taking (or analysed)
periods (i.e., different sub-experiments). Most of the time, these data are independent, but in a few
cases, the analysed periods overlap. It happens, for instance, for the {\em Voyager} data (launched in
1977 and still taking data). In that case, it is up to the user to decide which data sets are relevant
for her/his analysis, and exclude it using the \button{(Sub-)Experiment names} or \button{Time interval}
selection box (see Fig.~\ref{fig:extraction}).

\subsection{New data from the \tab{Add Data} tab}
\label{sec:add_data_tab}

This tab allows anyone to interactively enter new data. This is an essential part of the database as it
provides the community with the possibility to contribute to the completion of the database (either by
adding data from new instruments, or adding missing data from older experiments).

\begin{figure}[!t]
\begin{center}
\includegraphics[width=0.75\columnwidth]{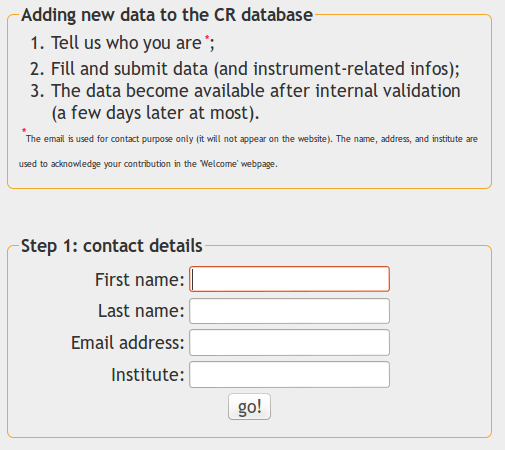}\\
\vspace{0.1cm}
\includegraphics[width=.85\columnwidth]{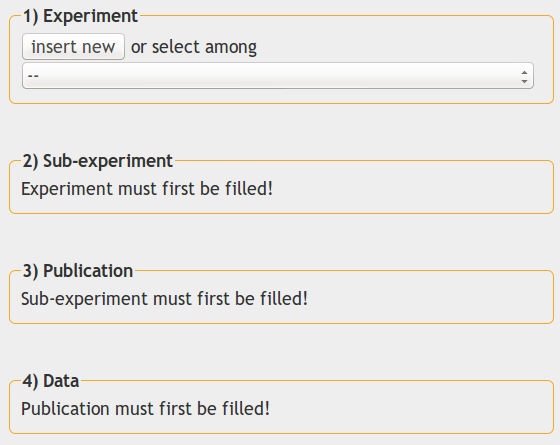}\\
\vspace{0.1cm}
\includegraphics[width=1.\columnwidth]{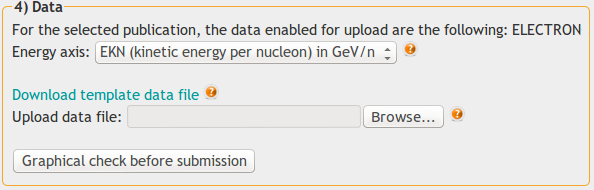}
\caption{Snapshots of the user interfaces in the \tabtext{Add data} tab. The first stage for adding new
data is the submitter identification (upper panel). The ordered 4-steps submission process comes next
(middle panel). The user has to either select among existing entries, or insert a new one (this pops-up
a window with fields to fill), except for the fourth step that concerns the CR data. The latter must be
filled one CR quantity at a time (and must belong to the list of quantities declared at the publication
step): the bottom panel shows the state of the panel once this step is reached, with the energy axis to
select, the data file to upload, and the \button{Graphical check before submission}. Note that at all
steps, {\em HELP} buttons exist for most of the fields to fill.}
\label{fig:add_data}
\end{center} 
\end{figure}

Submitting new data consists in two parts: as shown in Fig.~\ref{fig:add_data} (top panel), the first
part concerns the submitter identification (contact details). The second part (same figure, middle and
bottom panel) is data submission. Four steps must be passed in order. For the first three steps
(experiment, sub-experiment, publication), the submitter is left with the choice of selecting her/his
entry among those already in the database, or to add a new entry (\button{Insert new} button): the
latter action pops-up a new window in which the submitter is guided---\boxtext{HELP} boxes are
provided for each item---to fill the necessary informations, which match the keys of the database structure described
in Sect.~\ref{sec:structure} (see also Fig.~\ref{fig:mysql}). Each time a new entry is submitted, the
submitted element becomes available for further submissions, though it does not appear yet in the
database (i.e., in \tabtext{Experiments/Data} and \tabtext{Extract data} tabs). 

Once the three previous steps are completed, the last action is the submission of the CR data (see
bottom panel of Fig.~\ref{fig:add_data}. A template file for the required format (and units) of the data
is provided (see Sect.~\ref{sec:data}). Only one CR quantity at a time (with as many energy bins as
desired) can be submitted. Before the final submission of the data, the \button{Data graphical check}
button pops-up a summary of the uploaded file, along with a plot of the submitted data for a last visual
inspection. At this stage, the submission process can still be cancelled if any mistake is spotted. 

For each submitted entry (experiment, sub-experiment, publication, data), an email is sent to the
administrators of the database content\footnote{For now, the only authorised persons are the
developers of the database. Any person wishing to get involved in further developments of the database
is welcome to contact us.}. Validation tools from the \tabtext{Admin} tab are then used (format check,
completeness, etc.) to authorise the addition in the database (pending validation, the data are inserted
in the database with a `not validated' flag, and do not appear on the web site).


\section{Conclusions and future improvements}
\label{sec:concl}

We have developed a database of charged CRs (\url{http://lpsc.in2p3.fr/crdb}) that
includes $e^-$, $e^+$, $\bar{p}$, and nuclides data up to $Z=30$ for energies below a few TeV/n. Each CR
data is linked to a description of the instrument that measured it (flight dates, picture of the
experiment setup, techniques used, etc.) and to the ADS reference in which it was published: this first
release contains more than 200 experiments and 200 publications. The data can be extracted according to
a selection on the CR quantity, the energy range, the experiments and the epoch of measurement:
{\sf ASCII} files, {\sc ROOT} macros, plots, and \textsc{Bib}\TeX\ for the corresponding publications
are then readily available. Since {\tt V1.3}, we have also implemented (on request) a {\tt REST}
interface\footnote{An example of call using a command line in an xterm is given by
{\tiny \em wget 'http://lpsc.in2p3.fr/crdb/rest.php?num=C\&energy\_type=EKN' -\,-output-document=file}.
For a list of parameters, click on \tabtext{REST interface} (Fig.~\ref{fig:extraction_results}).}.

The possibility to add new data by means of a user-friendly interface enables the database to be a
collaborative tool for the CR community, provided enough people take an interest, use it, and help
expand it. New data will be added as new results become available (we encourage experimentalists to
submit their data once they become public). The database could also be extended to a larger energy
domain (data from ground-based detectors $\gtrsim$~TeV/n data) or to heavier species ($Z>30$). Time
series, e.g., low-energy proton, helium and electron data with a monthly or finer time resolution from
long-lived high-precision instruments (e.g., {\sf AMS-02}) could also be very interesting inputs for
Solar modulation studies.

We welcome any help to further develop the database. Comments, questions, suggestions, and
corrections\footnote{Despite our best efforts, many data published in CR conferences (e.g., ICRC) are
probably missing, and many typos and errors probably remain to be corrected.} can be addressed to
\href{mailto:crdatabase@lpsc.in2p3.fr}{\tt crdatabase@lpsc.in2p3.fr}. 

\begin{acknowledgements}
We thank our colleagues B. Coste, F. Donato, and A. Putze who contributed to the collection of some of
the data gathered in the first release of this database. We thank A. Putze for useful comments on the
paper. This work is part of the {\sf USINE} project (CR propagation code). It has been financially
supported by the {\sf PNHE}.
\end{acknowledgements}


\appendix


\section{Forming new combination of data from `native' data in the database}
\label{app:rules}

CR experiments provide measurements over several energy bins. The elementary brick of information in
GCR physics is the isotopic flux. However, isotopic identification is rarely achieved even in
contemporary experiments, and published results are often ratios of elements (e.g., B/C) and elemental
fluxes (e.g., B, C). 

Flux measurements suffer from many systematics (energy-dependent acceptance, dead time, rigidity
cut-off in the Geomagnetic field, etc.) that mostly factor out from the ratio of species close in
charge. As a result, uncertainties of published ratios are smaller than if calculated by quadratically
combining the errors on the fluxes. Obviously, published ratios should always be preferred if provided
in the publication. But as underlined in Sect.~\ref{sec:definition}, only the most-used ratios are
generally published (e.g., B/C).

In this Appendix, we detail how `native' data are searched for and combined to lead to the user-selected
quantity (App.~\ref{app:priority_rule}), be it for fluxes or ratios. We then discuss the necessary rules
to calculate the uncertainties from the `native' data uncertainties (App.~\ref{app:error_rule}). Note
that data can only be combined if coming from the same sub-experiment (otherwise, they would correspond to different
Solar modulation periods and suffer from different systematics). Obviously, the benefit of adding extra
data (for a CR quantity) to a CR analysis must also always be weighted against using only the `native'
and more accurate data found in the same energy range. A check box in the selection interface (shown in
Fig.~\ref{fig:extraction}) enables or disables the search for these
data. We did not systematically check
how many extra data points are obtained in this way, but the algorithm finds some. These
data are singled out by an asterisk in the result panel (shown in Fig.~\ref{fig:extraction_results}).

\subsection{Priority rules to form new quantities}
\label{app:priority_rule}

The database extraction tool (\tabtext{Extract data} tab) is designed to return the most complete set
of data for a given CR quantity $Q^{\rm searched}$. It needs a set of rules to browse through the
database content and retrieve the queried quantity. The search is prioritised as follows (we use
the superscript {\em native} to indicate quantity in the database):

\begin{enumerate}

   \item look directly for the queried quantity $Q^{\rm native}$;
   \item if $Q$ is a flux, search for all $i$ for which the pair $(Q/Q_i)^{\rm native}$, $(Q_i)^{\rm
   native}$ exist. The new data obtained are $Q^{\rm combo} = \cup_{i} \left\{ (Q/Q_i)^{\rm
   native}/(Q_i)^{\rm native}\right\}$;

   \item if CR quantity is a ratio $Q=Q_{\rm num}/Q_{\rm denom}$, search for 
   
      \begin{itemize}
         \item $Q^{\rm combo} = Q_{\rm num}^{\rm native}/Q_{\rm denom}^{\rm native}$;
         \item $Q^{\rm combo} = \cup_{i} \left\{ (Q_{\rm num}/Qi)^{\rm native}/(Q_{\rm denom}/Q_i)^{\rm native}\right\}$; 
         \item $Q^{\rm combo} = \cup_{i} \left\{ (Q_{\rm num}/Qi)^{\rm native}\times(Q_i/Q_{\rm denom})^{\rm native}\right\}$.
      \end{itemize}

\end{enumerate} 
To be complete, more complex combinations should be implemented (addition and subtraction of isotopic
fraction of an element, addition of group of charges, etc.). These combinations require a more complex
query that remains to be implemented (it is under development).

Another subtlety is that valid combinations are restricted to data points of similar energy ranges. As
underlined in Sect.~\ref{sec:data}, sometimes only the `central' energy bin of the data point is
available. Moreover, values extracted from published plots (using {\sf DataThief}) cannot match exactly
one another. For these reasons, we relax the constraint of a perfect energy match, and use instead:

\begin{itemize}

  \item {\sf data point has only a `central' value}: we demand this value to be i) within the energy
range, or ii) within 20\% of the `central' value of the data point it is combined to;

  \item {\sf data point has an energy range}: we demand an overlap of the energy ranges of the data
points to be combined.

\end{itemize}

\subsection{Rules to calculate uncertainties}
\label{app:error_rule}

For all experiments, the default rule is that relative errors are quadratically summed whenever
combinations of CR data are formed. In this first release, the only exception concerns {\sf HEAO-3} data
\citepads{1990A&A...233...96E}, where the authors provide tables for the oxygen flux and for the relative
fluxes (to O). The rules to calculate uncertainties, depends on the charge proximity of the combined
elements. The original text, spread over the original publication, is reproduced below:

{\em \small
\begin{description}

  \item[-]``In Table~5 [...] the standard error is given with each flux value. This error includes the
  statistical error and, in the case of the last two energy channels, the systematic error."

  \item[-]``The spectrum of any element between Z=4 and Z=28 can be obtained by multiplying the oxygen
  flux values of Table~5 by the corresponding abundance ratios given in Table~2. The relative errors on
  these two numbers shall be quadratically summed to get the error on the flux derived by this way."

  \item[-]``For widely different charges $Z_1$ and $Z_2$, a reasonable estimate of the error on the
  abundance ratio is the quadratic sum of the errors on both type of nuclei $Z_1$ and $Z_2$ (for
  example the total error on Fe/O ratio is $\sim 5.2\%$). For nearly adjacent charges, an upper limit
  of the systematic error is given by the difference between the systematic errors on each element (for
  example, the error on B/C is $\sim 0.4\%$)."

 \end{description}
}


\section{List of experiments and publications}
\label{app:bibtex}

We provide below a sorted list of experiments (unnamed balloons appear first), sub-experiments, and
associated publications present in the database first release\footnote{The list and bibtex references
were obtained from the \link{Get all bibtex entries} and \link{Latex cite} links in the \tabtext{Welcome}
tab.}.

{\scriptsize 
\begin{itemize}
   \item Balloon (1963/04): \citetads{1965PhRvL..15..769D}
   \item Balloon (1964,1965,1966)
      \begin{itemize}
         \item Balloon (1964/07): \citetads{1967ApJ...148..399L}
         \item Balloon (1965/06+1965/07): \citetads{1967ApJ...148..399L}
         \item Balloon (1966/06): \citetads{1968CaJPS..46..892L}
      \end{itemize}
   \item Balloon (1965/07+1965/08+1966/06): \citetads{1967ApJ...150..371H,1969ApJ...158..771F,1968ApJ...152..783F}
   \item Balloon (1965,1966,1967,1968) 
      \begin{itemize}
         \item Balloon (1965/07): \citetads{1965ICRC....1..327B}
         \item Balloon (1966/08): \citetads{1970ICRC....1..209B}
         \item Balloon (1967/08): \citetads{1970ICRC....1..209B}
         \item Balloon (1968/07): \citetads{1970ICRC....1..209B}
      \end{itemize}
   \item Balloon (1965,1966,1968,1969,1971+1972)
      \begin{itemize}
         \item Balloon (1965/07): \citetads{1973ICRC....2..760W}
         \item Balloon (1965/07+1966/07): \citetads{1968CaJPS..46.1014B,1967JGR....72.2783W}
         \item Balloon (1966/07): \citetads{1973ICRC....2..760W}
         \item Balloon (1968/07): \citetads{1973ICRC....2..760W}
         \item Balloon (1969/07): \citetads{1973ICRC....2..760W}
         \item Balloon (1971/07+1972/07): \citetads{1973ICRC....2..760W}
      \end{itemize}
   \item Balloon (1966/03): \citetads{1968PhRvL..20..764A}
   \item Balloon (1966/09): \citetads{1968CaJPh..46..530D}
   \item Balloon (1966,1967)
      \begin{itemize}
         \item Balloon (1966/07): \citetads{1972JGR....77.1087E}
         \item Balloon (1966/09): \citetads{1972JGR....77.1087E}
         \item Balloon (1967/07): \citetads{1972JGR....77.1087E}
      \end{itemize}
   \item Balloon (1967/06+1967/07): \citetads{1968PhRvL..20.1053I}
   \item Balloon (1967/07): \citetads{1969NCimL...1...53A}
   \item Balloon (1967/11+1968/06): \citetads{1971Ap&SS..14..301F}
   \item Balloon (1968/05): \citetads{1973ICRC....1..355A}
   \item Balloon (1968/06): \citetads{1971A&A....11...53S}
   \item Balloon (1968/07): \citetads{1969PhRvL..22..412B}
   \item Balloon (1968+1969\dots1999+2001): \citetads{1980ApJ...238..394N,1985ICRC....9..539N,1999ICRC....3...61K,2001AdSpR..26.1827N,2012ApJ...760..146K}
   \item Balloon (1968-1975,1977,1979,1982,1984,1987,1990,1992,1994)
      \begin{itemize}
         \item Balloon (1968/06+1968/07): \citetads{1975JGR....80.1701F,1971ApL.....9..165H,1972JGR....77.3295S}
         \item Balloon (1969/06+1969/07): \citetads{1975JGR....80.1701F,1971ApL.....9..165H,1972JGR....77.3295S}
         \item Balloon (1970/06+1970/07): \citetads{1975JGR....80.1701F,1971ApL.....9..165H,1972JGR....77.3295S}
         \item Balloon (1971/06+1971/07): \citetads{1975JGR....80.1701F,1971ApL.....9..165H}
         \item Balloon (1972/07): \citetads{1975JGR....80.1701F,1971ApL.....9..165H}
         \item Balloon (1973/07): \citetads{1975ICRC....3.1000C}
         \item Balloon (1974/07): \citetads{1977ICRC...11..203C}
         \item Balloon (1975/07): \citetads{1977ICRC...11..203C}
         \item Balloon (1977/07): \citetads{1979ICRC....1..462E}
         \item Balloon (1979/08): \citetads{1984JGR....89.2647E}
         \item Balloon (1982/08): \citetads{1984JGR....89.2647E}
         \item Balloon (1982/10): \citetads{1986JGR....91.2858G}
         \item Balloon (1984/09): \citetads{1986JGR....91.2858G}
         \item Balloon (1987/08): \citetads{1995JGR...100.7873E}
         \item Balloon (1990/08): \citetads{1995JGR...100.7873E}
         \item Balloon (1992/08): \citetads{1995JGR...100.7873E}
         \item Balloon (1994/08): \citetads{1995JGR...100.7873E}
      \end{itemize}
   \item Balloon (1969/04+1970/11): \citetads{1973JGR....78.7165S}
   \item Balloon (1969/09+1973/05): \citetads{1975ApJ...197..219M}
   \item Balloon (1970/05+1970/09): \citetads{1973ApJ...186..841M}
   \item Balloon (1970/06): \citetads{1979ZPhyA.291..383B}
   \item Balloon (1971/05): \citetads{1973ICRC....1..126A}
   \item Balloon (1971/09+1972/10): \citetads{1974ApJ...191..331J}
   \item Balloon (1971,1972)
      \begin{itemize}
         \item Balloon (1971/07): \citetads{1973ICRC....2..760W}
         \item Balloon (1971/07+1972/07): \citetads{1973ICRC....2..760W}
         \item Balloon (1972/07): \citetads{1973ICRC....2..760W}
      \end{itemize}
   \item Balloon (1972/07): \citetads{1975ICRC....1..312W,1987ApJ...312..178W}
   \item Balloon (1972/07): \citetads{1975ApJ...198..493D}
   \item Balloon (1972/10): \citetads{1977ApJ...213..588F,1978ApJ...226.1147O}
   \item Balloon (1972/11+1973/05): \citetads{1974PhRvL..33...34B,1975ApJ...199..669B}
   \item Balloon (1973/06): \citetads{1973ICRC....5.3073I}
   \item Balloon (1973/08): \citetads{1976ApJ...205..938F,1977ApJ...212..262H,1978ApJ...221.1110L}
   \item Balloon (1973/09+1974/05): \citetads{1978ApJ...224..691D}
   \item Balloon (1973/09+1974/05+1975/09+1975/10): \citetads{1987ApJ...322..981D}
   \item Balloon (1974/07+1974/08): \citetads{1976ApJ...204..927H}
   \item Balloon (1974/07+1974/08+1976/09): \citetads{1978ApJ...223..676L}
   \item Balloon (1975/09+1975/10): \citetads{1981ApJ...248..847M}
   \item Balloon (1975/10): \citetads{1979ApJ...227..676P,1977PhRvL..38.1368H}
   \item Balloon (1975/12): \citetads{1979ICRC....1..330B}
   \item Balloon (1976/05): \citetads{1984ApJ...287..622G,1987A&A...188..145G}
   \item Balloon (1976/10): \citetads{1980ApJ...239..712S}
   \item Balloon (1977/05): \citetads{1978ApJ...226..355B}
   \item Balloon (1977/07): \citetads{1983ApJ...275..391W}
   \item Balloon (1977/09): \citetads{1979ICRC....1..389W,1982ApJ...252..386W}
   \item Balloon (1980/10): \citetads{1984ApJ...278..881T}
   \item Balloon (1981/04): \citetads{1985ApJ...291..207J}
   \item Balloon (1981/09): \citetads{1985ICRC....2...16W,1985ICRC....2...88W}
   \item Balloon (1984/04): \citetads{1987ApJ...312..183M}
   \item Balloon (1987/05+1988/05+1989/05+1991/05): \citetads{1993PhRvD..48.1949I}
   \item Balloon (1989/05+1991/05): \citetads{1997APh.....6..155K}
   \item Balloon (1989/09): \citetads{1995PhRvD..52.6219H}
   \item Balloon (1990/07): \citetads{1995ICRC....2..598B}
   \item Balloon (1991/09): \citetads{1994ApJ...429..736B}
   \item ACE
      \begin{itemize}
         \item ACE-CRIS (1997/08-1998/04): \citetads{2009ApJ...698.1666G}
         \item ACE-CRIS (1997/08-1998/12): \citetads{1999ApJ...523L..61W}
         \item ACE-CRIS (1997/08-1999/07): \citetads{2001ApJ...563..768Y}
         \item ACE-CRIS (1998/01-1999/01): \citetads{2006AdSpR..38.1558D}
         \item ACE-CRIS (2001/05-2003/09): \citetads{2009ApJ...698.1666G}
         \item ACE-SIS (1997/08-1999/07): \citetads{2001ApJ...563..768Y}
      \end{itemize}
   \item AESOP
      \begin{itemize}
         \item AESOP94 (1994/08): \citetads{1996ApJ...464..507C}
         \item AESOP97+98 (1997/09+1998/08): \citetads{2000JGR...10523099C}
         \item AESOP99 (1999/08): \citetads{2002ApJ...568..216C}
         \item AESOP00 (2000/08): \citetads{2002ApJ...568..216C}
         \item AESOP02 (2002/08): \citetads{2004JGRA..109.7107C}
         \item AESOP06 (2006/08): \citetads{2009JGRA..11410108C}
      \end{itemize}
   \item ALICE
      \begin{itemize}
         \item ALICE (1987/08): \citetads{1996A&A...314..785H}
         \item ALICE (1987/08+1987/08): \citetads{1992APh.....1...33E}
     \end{itemize}
   \item AMS-01
      \begin{itemize}
         \item AMS01 (1998/06): \citetads{2000PhLB..490...27A,2000PhLB..494..193A,2002PhR...366..331A,2003JHEP...11..048X,2010ApJ...724..329A,2011ApJ...736..105A}
         \item AMS01-BremsstrahlungPhotons (1998/06): \citetads{2007PhLB..646..145A}
         \item AMS01-singleTrack (1998/06): \citetads{2000PhLB..484...10A,2002PhR...366..331A}
      \end{itemize}
   \item ATIC
      \begin{itemize}
         \item ATIC01\&02 (2001/01+2003/01): \citetads{2008Natur.456..362C}
         \item ATIC02 (2003/01): \citetads{2008ICRC....2....3P,2009BRASP..73..564P}
      \end{itemize}
   \item BESS
      \begin{itemize}
         \item BESS93 (1993/07): \citetads{1997ApJ...474..479M,2002ApJ...564..244W}
         \item BESS94 (1994/07): \citetads{2003ICRC....4.1805M}
         \item BESS95 (1995/07): \citetads{1998PhRvL..81.4052M,2003ICRC....4.1805M}
         \item BESS97 (1997/07): \citetads{2000PhRvL..84.1078O,2003ICRC....4.1805M,2007APh....28..154S}
         \item BESS98 (1998/07): \citetads{2000ApJ...545.1135S,2001APh....16..121M,2003ICRC....4.1805M,2007APh....28..154S}
         \item BESS99 (1999/08): \citetads{2002PhRvL..88e1101A,2007APh....28..154S}
         \item BESS00 (2000/08): \citetads{2002PhRvL..88e1101A,2007APh....28..154S}
         \item BESS-TeV (2002/08): \citetads{2004PhLB..594...35H,2005ICRC....3...13H,2007APh....28..154S}
         \item BESS-PolarI (2004/12): \citetads{2008PhLB..670..103B}
         \item BESS-PolarII (2007/12-2008/01): \citetads{2012PhRvL.108e1102A}
      \end{itemize}
   \item BETS
      \begin{itemize}
         \item BETS97\&98 (1997/06+1998/05): \citetads{2001ApJ...559..973T}
         \item BETS04 (2004/01): \citetads{2008AdSpR..42.1670Y}
      \end{itemize}
   \item CAPRICE
      \begin{itemize}
         \item CAPRICE94 (1994/08): \citetads{1997ApJ...487..415B,1999ApJ...518..457B,2000ApJ...532..653B}
         \item CAPRICE98 (1998/05): \citetads{2001AdSpR..27..669B,2001ApJ...561..787B,2003ICRC....4.1809M,2003APh....19..583B,2004ApJ...615..259P}
      \end{itemize}
   \item CREAM
      \begin{itemize}
         \item CREAM-I (2004/12-2005/01): \citetads{2008APh....30..133A,2011ApJ...728..122Y}
         \item CREAM-II (2005/12-2006/01): \citetads{2009ApJ...707..593A,2010ApJ...715.1400A}
      \end{itemize}
   \item CRISIS (1977/05): \citetads{1980ApJ...240L..53F,1981ApJ...246.1014Y}
   \item CRN-Spacelab2 (1985/07-1985/08): \citetads{1990ApJ...349..625S,1991ApJ...374..356M}
   \item CRRES (1990/07-1992/10): \citetads{1996ApJ...466..457D}
   \item Fermi
      \begin{itemize}
         \item Fermi-LAT (2008/06-2011/04): \citetads{2012PhRvL.108a1103A}
         \item Fermi-LAT (2008/06-2009/06): \citetads{2010PhRvD..82i2004A}
      \end{itemize}
   \item H.E.S.S. (2004/10-2007/08): \citetads{2008PhRvL.101z1104A,2009A&A...508..561A}
   \item HEAO3-C2 (1979/10-1980/06): \citetads{1988A&A...193...69F,1990A&A...233...96E}
   \item HEAT
      \begin{itemize}   
         \item HEAT94 (1994/05): \citetads{1995PhRvL..75..390B,1997ApJ...482L.191B,1998ApJ...498..779B}
         \item HEAT95 (1995/08): \citetads{1997ApJ...482L.191B,2001ApJ...559..296D}
         \item HEAT94\&95 (1994/05+1995/08): \citetads{1997ApJ...482L.191B,2001ApJ...559..296D}
         \item HEAT-pbar (2000/06): \citetads{2001PhRvL..87A1101B,2004PhRvL..93x1102B}
      \end{itemize}
   \item HEIST (1988/08): \citetads{1992ApJ...391L..89G}
   \item IMAX92 (1992/07): \citetads{1996PhRvL..76.3057M,1998ApJ...496..490R,2000ApJ...533..281M,2000AIPC..528..425D}
   \item IMP
      \begin{itemize}      
         \item IMP1 (1963/11-1964/05): \citetads{1964PhRvL..13..786C}
         \item IMP3 (1965/06-1965/12): \citetads{1971ApJ...166..221H}
         \item IMP3 (1965/07-1966/03): \citetads{1968ApJ...151..737F}
         \item IMP4 (1967/06-1967/10): \citetads{1971ApJ...166..221H}
         \item IMP5 (1969/06-1969/09): \citetads{1971ApJ...166..221H}
         \item IMP7 (1972/09-1972/12): \citetads{1975ApJ...202..815T}
         \item IMP7 (1972/10-1976/10): \citetads{1979ApJ...232L..95G}
         \item IMP7 (1973/05-1973/08): \citetads{1975ApJ...202..265G}
         \item IMP7\&8 (1972/09-1975/09): \citetads{1977ApJ...217..859G}
         \item IMP7\&8 (1973/02-1977/09): \citetads{1981ApJ...244..695G}
         \item IMP7\&8 (1974/01-1980/05): \citetads{1981ICRC....2...72G}
         \item IMP8 (1974/01-1977/11): \citetads{1985ApJ...294..455B}
         \item IMP8 (1974/01-1978/10): \citetads{1987ApJS...64..269G}
      \end{itemize}
   \item ISEE
      \begin{itemize}      
         \item ISEE3-HIST (1978/08-1978/12): \citetads{1980ApJ...236L.121M,1980ApJ...235L..95M,1981ApJ...251L..27M,1986ApJ...311..979M}
         \item ISEE3-HKH (1978/08-1979/08): \citetads{1980ApJ...239L.139W}
         \item ISEE3-HKH (1978/08-1980/05): \citetads{1981ApJ...247L.119W,1981PhRvL..46..682W}
         \item ISEE3-HKH (1978/08-1981/04): \citetads{1983ICRC....9..147W,1988ApJ...328..940K,1993ApJ...405..567L}
         \item ISEE3-MEH (1978/08-1979/02): \citetads{1979ICRC....1..462E}
         \item ISEE3-MEH (1978/08-1984/04): \citetads{1986ApJ...303..816K}
         \item ISEE3-MEH (1979/02-1980/03): \citetads{1981ICRC...10...77E}
         \item ISEE3-MEH (1980/03-1981/03): \citetads{1981ICRC...10...77E}
      \end{itemize}
   \item ISOMAX (1998/08): \citetads{2004ApJ...611..892H}
   \item JACEE (1979+1980+1982\dots1990+1994+1995): \citetads{1998ApJ...502..278A}
   \item MASS
      \begin{itemize}      
         \item MASS89 (1989/09): \citetads{1991ApJ...380..230W,1994ApJ...436..769G}
         \item MASS91 (1991/09): \citetads{1996ApJ...467L..33H,1999PhRvD..60e2002B,1999ICRC....3...77B,2002A&A...392..287G}
      \end{itemize}
   \item MUBEE (1975/09+1978/08+1986/07+1987/07): \citetads{1993ICRC....2...13Z}
   \item OGO5
      \begin{itemize}      
         \item OGO5 (1968/04-1968/05): \citetads{1974JGR....79.1533B}
         \item OGO5 (1968/04-1969/04): \citetads{1972ApJ...171..363L}
         \item OGO5 (1968/06-1968/10): \citetads{1975JGR....80.1701F}
         \item OGO5 (1969/06-1969/07): \citetads{1975JGR....80.1701F,1974JGR....79.1533B}
         \item OGO5 (1970/06-1970/07): \citetads{1975JGR....80.1701F,1974JGR....79.1533B}
         \item OGO5 (1971/05-1971/08): \citetads{1975JGR....80.1701F}
         \item OGO5 (1971/07-1971/08): \citetads{1974JGR....79.1533B}
         \item OGO5 (1972/06): \citetads{1975JGR....80.1701F}
         \item OGO5 (1972/06-1972/07): \citetads{1974JGR....79.1533B}
      \end{itemize}
   \item PAMELA
      \begin{itemize}      
         \item PAMELA (2006/07-2008/02): \citetads{2009Natur.458..607A}
         \item PAMELA (2006/07-2008/06): \citetads{2009PhRvL.102e1101A}
         \item PAMELA (2006/07-2008/12): \citetads{2011ASTRA...7..465C,2010PhRvL.105l1101A,2011Sci...332...69A}
         \item PAMELA (2006/07-2010/01): \citetads{2011PhRvL.106t1101A}
      \end{itemize}
   \item Pioneer
      \begin{itemize}      
         \item Pioneer8\&9 (1967/12-1969/04): \citetads{1973JGR....78.1487W}
         \item Pioneer10-HET (1972/03-1973/03): \citetads{1975ApJ...202..815T}
         \item Pioneer10-HET (1985/04-1988/11): \citetads{1994ApJ...435..464W}
      \end{itemize}
   \item RICH-II (1997/10): \citetads{2003APh....18..487D}
   \item RUNJOB (1995+1996+1997+1998+1999): \citetads{2005ApJ...628L..41D}
   \item SMILI
      \begin{itemize}      
         \item SMILI-I (1989/09): \citetads{1993ApJ...413..268B}
         \item SMILI-II (1991/07): \citetads{2000ApJ...534..757A}
      \end{itemize}
   \item SOKOL (1984/03-1984/04): \citetads{1993ICRC....2...17I}
   \item TRACER
      \begin{itemize}      
         \item TRACER03 (2003/12): \citetads{2008ApJ...678..262A}
         \item TRACER06 (2006/07): \citetads{2011ApJ...742...14O,2012ApJ...752...69O}
      \end{itemize}
   \item Trek-MIR (1991/06-1992/12): \citetads{1996ApJ...468..679W}
   \item TS93 (1993/09): \citetads{1996ApJ...457L.103G}
   \item Ulysses
      \begin{itemize}      
         \item Ulysses-KET (1990/10-1991/02): \citetads{1996A&A...307..981R}
         \item Ulysses-HET (1990/10-1995/07): \citetads{1996ApJ...465..982D,1996A&A...316..555D,1997ApJ...475L..61C,1997ApJ...481..241D,1997ApJ...482..792T}
         \item Ulysses-HET (1990/10-1996/12): \citetads{1998ApJ...497L..85S}
         \item Ulysses-HET (1990/10-1997/12): \citetads{1998ApJ...501L..59C,1998ApJ...509L..97C,1999ICRC....3...33C}
         \item Ulysses-KET (1992/07-1992/10): \citetads{1996A&A...307..981R}
         \item Ulysses-KET (1993/07-1993/10): \citetads{1996A&A...307..981R}
         \item Ulysses-KET (1994/07-1994/10): \citetads{1996A&A...307..981R}
      \end{itemize}
   \item Voyager
      \begin{itemize}      
         \item Voyager1-HET (1977/10-1977/11): \citetads{1983ApJ...275..391W}
         \item Voyager1-HET-Aend (1977/10-1977/11): \citetads{1983ApJ...275..391W}
         \item Voyager1-HET (1994/01-1994/09): \citetads{1995ApJ...451L..33S}
         \item Voyager1-HET (2010/01-2010/03): \citetads{2010ApJ...725..121C}
         \item Voyager2-HET (1986/01-1987/12): \citetads{1991A&A...247..163F}
         \item Voyager2-HET (1987/01-1987/12): \citetads{1994ApJ...432..656S}
         \item Voyager2-HET (2010/01-2010/03): \citetads{2010ApJ...725..121C}
         \item Voyager1\&2 (1977/01-1991/12): \citetads{1994ApJ...423..426L}
         \item Voyager1\&2 (1977/01-1993/12): \citetads{1994ApJ...430L..69L,1996ApJ...457..435W}
         \item Voyager1\&2 (1977/01-1996/12): \citetads{1997ApJ...488..454L,1997ICRC....3..389L,1997ApJ...476..766W}
         \item Voyager1\&2 (1977/01-1998/12): \citetads{1999ICRC....3...41L}
         \item Voyager1\&2 (1986/01-1989/12): \citetads{1994ApJ...426..366L}
         \item Voyager1\&2-HET (1987/01-1987/12): \citetads{2009JGRA..11402103W}
      \end{itemize}
\end{itemize}
}

\bibliographystyle{aa}
\bibliography{database}

\begin{thebibliography}{273}
\expandafter\ifx\csname natexlab\endcsname\relax\def\natexlab#1{#1}\fi

\bibitem[{{Abe} {et~al.}(2012{\natexlab{a}}){Abe}, {Fuke}, {Haino}, {Hams},
  {Hasegawa}, {Horikoshi}, {Itazaki}, {Kim}, {Kumazawa}, {Kusumoto}, {Lee},
  {Makida}, {Matsuda}, {Matsukawa}, {Matsumoto}, {Mitchell}, {Myers},
  {Nishimura}, {Nozaki}, {Orito}, {Ormes}, {Sakai}, {Sasaki}, {Seo}, {Shikaze},
  {Shinoda}, {Streitmatter}, {Suzuki}, {Takasugi}, {Takeuchi}, {Tanaka},
  {Thakur}, {Yamagami}, {Yamamoto}, {Yoshida}, \&
  {Yoshimura}}]{2012PhRvL.108m1301A}
{Abe}, K., {Fuke}, H., {Haino}, S., {et~al.} 2012{\natexlab{a}}, Physical
  Review Letters, 108, 131301

\bibitem[{{Abe} {et~al.}(2012{\natexlab{b}}){Abe}, {Fuke}, {Haino}, {Hams},
  {Hasegawa}, {Horikoshi}, {Kim}, {Kusumoto}, {Lee}, {Makida}, {Matsuda},
  {Matsukawa}, {Mitchell}, {Nishimura}, {Nozaki}, {Orito}, {Ormes}, {Sakai},
  {Sasaki}, {Seo}, {Shinoda}, {Streitmatter}, {Suzuki}, {Tanaka}, {Thakur},
  {Yamagami}, {Yamamoto}, {Yoshida}, \& {Yoshimura}}]{2012PhRvL.108e1102A}
{Abe}, K., {Fuke}, H., {Haino}, S., {et~al.} 2012{\natexlab{b}}, Physical
  Review Letters, 108, 051102

\bibitem[{{Abraham} {et~al.}(2010){Abraham}, {Abreu}, {Aglietta}, {Ahn},
  {Allard}, {Allen}, {Alvarez-Mu{\~n}iz}, {Ambrosio}, {Anchordoqui},
  {Andringa}, \& et~al.}]{2010PhLB..685..239A}
{Abraham}, J., {Abreu}, P., {Aglietta}, M., {et~al.} 2010, Physics Letters B,
  685, 239

\bibitem[{{Ackermann} {et~al.}(2012{\natexlab{a}}){Ackermann}, {Ajello},
  {Allafort}, {Atwood}, {Baldini}, {Barbiellini}, {Bastieri}, {Bechtol},
  {Bellazzini}, {Berenji}, {Blandford}, {Bloom}, {Bonamente}, {Borgland},
  {Bouvier}, {Bregeon}, {Brigida}, {Bruel}, {Buehler}, {Buson}, {Caliandro},
  {Cameron}, {Caraveo}, {Casandjian}, {Cecchi}, {Charles}, {Chekhtman},
  {Cheung}, {Chiang}, {Ciprini}, {Claus}, {Cohen-Tanugi}, {Conrad}, {Cutini},
  {de Angelis}, {de Palma}, {Dermer}, {Digel}, {Do Couto E Silva}, {Drell},
  {Drlica-Wagner}, {Favuzzi}, {Fegan}, {Ferrara}, {Focke}, {Fortin},
  {Fukazawa}, {Funk}, {Fusco}, {Gargano}, {Gasparrini}, {Germani}, {Giglietto},
  {Giommi}, {Giordano}, {Giroletti}, {Glanzman}, {Godfrey}, {Grenier}, {Grove},
  {Guiriec}, {Gustafsson}, {Hadasch}, {Harding}, {Hayashida}, {Hughes},
  {J{\'o}hannesson}, {Johnson}, {Kamae}, {Katagiri}, {Kataoka},
  {Kn{\"o}dlseder}, {Kuss}, {Lande}, {Latronico}, {Lemoine-Goumard}, {Llena
  Garde}, {Longo}, {Loparco}, {Lovellette}, {Lubrano}, {Madejski}, {Mazziotta},
  {McEnery}, {Michelson}, {Mitthumsiri}, {Mizuno}, {Moiseev}, {Monte},
  {Monzani}, {Morselli}, {Moskalenko}, {Murgia}, {Nakamori}, {Nolan}, {Norris},
  {Nuss}, {Ohno}, {Ohsugi}, {Okumura}, {Omodei}, {Orlando}, {Ormes}, {Ozaki},
  {Paneque}, {Parent}, {Pesce-Rollins}, {Pierbattista}, {Piron}, {Pivato},
  {Porter}, {Rain{\`o}}, {Rando}, {Razzano}, {Razzaque}, {Reimer}, {Reimer},
  {Reposeur}, {Ritz}, {Romani}, {Roth}, {Sadrozinski}, {Sbarra}, {Schalk},
  {Sgr{\`o}}, {Siskind}, {Spandre}, {Spinelli}, {Strong}, {Takahashi},
  {Takahashi}, {Tanaka}, {Thayer}, {Thayer}, {Tibaldo}, {Tinivella}, {Torres},
  {Tosti}, {Troja}, {Uchiyama}, {Usher}, {Vandenbroucke}, {Vasileiou},
  {Vianello}, {Vitale}, {Waite}, {Winer}, {Wood}, {Wood}, {Yang}, \&
  {Zimmer}}]{2012PhRvL.108a1103A}
{Ackermann}, M., {Ajello}, M., {Allafort}, A., {et~al.} 2012{\natexlab{a}},
  Physical Review Letters, 108, 011103

\bibitem[{{Ackermann} {et~al.}(2010){Ackermann}, {Ajello}, {Atwood}, {Baldini},
  {Ballet}, {Barbiellini}, {Bastieri}, {Baughman}, {Bechtol}, {Bellardi},
  {Bellazzini}, {Belli}, {Berenji}, {Blandford}, {Bloom}, {Bogart},
  {Bonamente}, {Borgland}, {Brandt}, {Bregeon}, {Brez}, {Brigida}, {Bruel},
  {Buehler}, {Burnett}, {Busetto}, {Buson}, {Caliandro}, {Cameron}, {Caraveo},
  {Carlson}, {Carrigan}, {Casandjian}, {Ceccanti}, {Cecchi}, {{\c C}elik},
  {Charles}, {Chekhtman}, {Cheung}, {Chiang}, {Cillis}, {Ciprini}, {Claus},
  {Cohen-Tanugi}, {Conrad}, {Corbet}, {Deklotz}, {Dermer}, {de Angelis}, {de
  Palma}, {Digel}, {di Bernardo}, {Do Couto E Silva}, {Drell}, {Drlica-Wagner},
  {Dubois}, {Fabiani}, {Favuzzi}, {Fegan}, {Fortin}, {Fukazawa}, {Funk},
  {Fusco}, {Gaggero}, {Gargano}, {Gasparrini}, {Gehrels}, {Germani},
  {Giglietto}, {Giommi}, {Giordano}, {Giroletti}, {Glanzman}, {Godfrey},
  {Grasso}, {Grenier}, {Grondin}, {Grove}, {Guiriec}, {Gustafsson}, {Hadasch},
  {Harding}, {Hayashida}, {Hays}, {Horan}, {Hughes}, {J{\'o}hannesson},
  {Johnson}, {Johnson}, {Johnson}, {Kamae}, {Katagiri}, {Kataoka}, {Kerr},
  {Kn{\"o}dlseder}, {Kuss}, {Lande}, {Latronico}, {Lemoine-Goumard}, {Llena
  Garde}, {Longo}, {Loparco}, {Lott}, {Lovellette}, {Lubrano}, {Makeev},
  {Mazziotta}, {McEnery}, {Mehault}, {Michelson}, {Minuti}, {Mitthumsiri},
  {Mizuno}, {Moiseev}, {Monte}, {Monzani}, {Moretti}, {Morselli}, {Moskalenko},
  {Murgia}, {Nakamori}, {Naumann-Godo}, {Nolan}, {Norris}, {Nuss}, {Ohsugi},
  {Okumura}, {Omodei}, {Orlando}, {Ormes}, {Ozaki}, {Paneque}, {Panetta},
  {Parent}, {Pelassa}, {Pepe}, {Pesce-Rollins}, {Petrosian}, {Pinchera},
  {Piron}, {Porter}, {Profumo}, {Rain{\`o}}, {Rando}, {Rapposelli}, {Razzano},
  {Reimer}, {Reimer}, {Reposeur}, {Ripken}, {Ritz}, {Rochester}, {Romani},
  {Roth}, {Sadrozinski}, {Saggini}, {Sanchez}, {Sander}, {Sgr{\`o}}, {Siskind},
  {Smith}, {Spandre}, {Spinelli}, {Stawarz}, {Stephens}, {Strickman}, {Strong},
  {Suson}, {Tajima}, {Takahashi}, {Takahashi}, {Tanaka}, {Thayer}, {Thayer},
  {Thompson}, {Tibaldo}, {Tibolla}, {Torres}, {Tosti}, {Tramacere}, {Turri},
  {Uchiyama}, {Usher}, {Vandenbroucke}, {Vasileiou}, {Vilchez}, {Vitale},
  {Waite}, {Wallace}, {Wang}, {Winer}, {Wood}, {Yang}, {Ylinen}, \&
  {Ziegler}}]{2010PhRvD..82i2004A}
{Ackermann}, M., {Ajello}, M., {Atwood}, W.~B., {et~al.} 2010, \prd, 82, 092004

\bibitem[{{Ackermann} {et~al.}(2012{\natexlab{b}}){Ackermann}, {Ajello},
  {Atwood}, {Baldini}, {Ballet}, {Barbiellini}, {Bastieri}, {Bechtol},
  {Bellazzini}, {Berenji}, {Blandford}, {Bloom}, {Bonamente}, {Borgland},
  {Brandt}, {Bregeon}, {Brigida}, {Bruel}, {Buehler}, {Buson}, {Caliandro},
  {Cameron}, {Caraveo}, {Cavazzuti}, {Cecchi}, {Charles}, {Chekhtman},
  {Chiang}, {Ciprini}, {Claus}, {Cohen-Tanugi}, {Conrad}, {Cutini}, {de
  Angelis}, {de Palma}, {Dermer}, {Digel}, {Silva}, {Drell}, {Drlica-Wagner},
  {Falletti}, {Favuzzi}, {Fegan}, {Ferrara}, {Focke}, {Fortin}, {Fukazawa},
  {Funk}, {Fusco}, {Gaggero}, {Gargano}, {Germani}, {Giglietto}, {Giordano},
  {Giroletti}, {Glanzman}, {Godfrey}, {Grove}, {Guiriec}, {Gustafsson},
  {Hadasch}, {Hanabata}, {Harding}, {Hayashida}, {Hays}, {Horan}, {Hou},
  {Hughes}, {J{\'o}hannesson}, {Johnson}, {Johnson}, {Kamae}, {Katagiri},
  {Kataoka}, {Kn{\"o}dlseder}, {Kuss}, {Lande}, {Latronico}, {Lee},
  {Lemoine-Goumard}, {Longo}, {Loparco}, {Lott}, {Lovellette}, {Lubrano},
  {Mazziotta}, {McEnery}, {Michelson}, {Mitthumsiri}, {Mizuno}, {Monte},
  {Monzani}, {Morselli}, {Moskalenko}, {Murgia}, {Naumann-Godo}, {Norris},
  {Nuss}, {Ohsugi}, {Okumura}, {Omodei}, {Orlando}, {Ormes}, {Paneque},
  {Panetta}, {Parent}, {Pesce-Rollins}, {Pierbattista}, {Piron}, {Pivato},
  {Porter}, {Rain{\`o}}, {Rando}, {Razzano}, {Razzaque}, {Reimer}, {Reimer},
  {Sadrozinski}, {Sgr{\`o}}, {Siskind}, {Spandre}, {Spinelli}, {Strong},
  {Suson}, {Takahashi}, {Tanaka}, {Thayer}, {Thayer}, {Thompson}, {Tibaldo},
  {Tinivella}, {Torres}, {Tosti}, {Troja}, {Usher}, {Vandenbroucke},
  {Vasileiou}, {Vianello}, {Vitale}, {Waite}, {Wang}, {Winer}, {Wood}, {Wood},
  {Yang}, {Ziegler}, \& {Zimmer}}]{2012ApJ...750....3A}
{Ackermann}, M., {Ajello}, M., {Atwood}, W.~B., {et~al.} 2012{\natexlab{b}},
  \apj, 750, 3

\bibitem[{{Adriani} {et~al.}(2010){Adriani}, {Barbarino}, {Bazilevskaya},
  {Bellotti}, {Boezio}, {Bogomolov}, {Bonechi}, {Bongi}, {Bonvicini},
  {Borisov}, {Bottai}, {Bruno}, {Cafagna}, {Campana}, {Carbone}, {Carlson},
  {Casolino}, {Castellini}, {Consiglio}, {de Pascale}, {de Santis}, {de
  Simone}, {di Felice}, {Galper}, {Gillard}, {Grishantseva}, {Hofverberg},
  {Jerse}, {Karelin}, {Koldashov}, {Krutkov}, {Kvashnin}, {Leonov}, {Malvezzi},
  {Marcelli}, {Mayorov}, {Menn}, {Mikhailov}, {Mocchiutti}, {Monaco}, {Mori},
  {Nikonov}, {Osteria}, {Papini}, {Pearce}, {Picozza}, {Pizzolotto}, {Ricci},
  {Ricciarini}, {Rossetto}, {Simon}, {Sparvoli}, {Spillantini}, {Stozhkov},
  {Vacchi}, {Vannuccini}, {Vasilyev}, {Voronov}, {Wu}, {Yurkin}, {Zampa},
  {Zampa}, {Zverev}, \& {PAMELA Collaboration}}]{2010PhRvL.105l1101A}
{Adriani}, O., {Barbarino}, G.~C., {Bazilevskaya}, G.~A., {et~al.} 2010,
  Physical Review Letters, 105, 121101

\bibitem[{{Adriani} {et~al.}(2011{\natexlab{a}}){Adriani}, {Barbarino},
  {Bazilevskaya}, {Bellotti}, {Boezio}, {Bogomolov}, {Bonechi}, {Bongi},
  {Bonvicini}, {Borisov}, {Bottai}, {Bruno}, {Cafagna}, {Campana}, {Carbone},
  {Carlson}, {Casolino}, {Castellini}, {Consiglio}, {De Pascale}, {De Santis},
  {De Simone}, {Di Felice}, {Galper}, {Gillard}, {Grishantseva}, {Jerse},
  {Karelin}, {Koldashov}, {Krutkov}, {Kvashnin}, {Leonov}, {Malakhov},
  {Malvezzi}, {Marcelli}, {Mayorov}, {Menn}, {Mikhailov}, {Mocchiutti},
  {Monaco}, {Mori}, {Nikonov}, {Osteria}, {Palma}, {Papini}, {Pearce},
  {Picozza}, {Pizzolotto}, {Ricci}, {Ricciarini}, {Rossetto}, {Sarkar},
  {Simon}, {Sparvoli}, {Spillantini}, {Stozhkov}, {Vacchi}, {Vannuccini},
  {Vasilyev}, {Voronov}, {Yurkin}, {Wu}, {Zampa}, {Zampa}, \&
  {Zverev}}]{2011Sci...332...69A}
{Adriani}, O., {Barbarino}, G.~C., {Bazilevskaya}, G.~A., {et~al.}
  2011{\natexlab{a}}, Science, 332, 69

\bibitem[{{Adriani} {et~al.}(2009{\natexlab{a}}){Adriani}, {Barbarino},
  {Bazilevskaya}, {Bellotti}, {Boezio}, {Bogomolov}, {Bonechi}, {Bongi},
  {Bonvicini}, {Bottai}, {Bruno}, {Cafagna}, {Campana}, {Carlson}, {Casolino},
  {Castellini}, {de Pascale}, {de Rosa}, {de Simone}, {di Felice}, {Galper},
  {Grishantseva}, {Hofverberg}, {Koldashov}, {Krutkov}, {Kvashnin}, {Leonov},
  {Malvezzi}, {Marcelli}, {Menn}, {Mikhailov}, {Mocchiutti}, {Orsi}, {Osteria},
  {Papini}, {Pearce}, {Picozza}, {Ricci}, {Ricciarini}, {Simon}, {Sparvoli},
  {Spillantini}, {Stozhkov}, {Vacchi}, {Vannuccini}, {Vasilyev}, {Voronov},
  {Yurkin}, {Zampa}, {Zampa}, \& {Zverev}}]{2009Natur.458..607A}
{Adriani}, O., {Barbarino}, G.~C., {Bazilevskaya}, G.~A., {et~al.}
  2009{\natexlab{a}}, \nat, 458, 607

\bibitem[{{Adriani} {et~al.}(2009{\natexlab{b}}){Adriani}, {Barbarino},
  {Bazilevskaya}, {Bellotti}, {Boezio}, {Bogomolov}, {Bonechi}, {Bongi},
  {Bonvicini}, {Bottai}, {Bruno}, {Cafagna}, {Campana}, {Carlson}, {Casolino},
  {Castellini}, {de Pascale}, {de Rosa}, {Fedele}, {Galper}, {Grishantseva},
  {Hofverberg}, {Leonov}, {Koldashov}, {Krutkov}, {Kvashnin}, {Malvezzi},
  {Marcelli}, {Menn}, {Mikhailov}, {Minori}, {Mocchiutti}, {Nagni}, {Orsi},
  {Osteria}, {Papini}, {Pearce}, {Picozza}, {Ricci}, {Ricciarini}, {Simon},
  {Sparvoli}, {Spillantini}, {Stozhkov}, {Taddei}, {Vacchi}, {Vannuccini},
  {Vasilyev}, {Voronov}, {Yurkin}, {Zampa}, {Zampa}, \&
  {Zverev}}]{2009PhRvL.102e1101A}
{Adriani}, O., {Barbarino}, G.~C., {Bazilevskaya}, G.~A., {et~al.}
  2009{\natexlab{b}}, Physical Review Letters, 102, 051101

\bibitem[{{Adriani} {et~al.}(2011{\natexlab{b}}){Adriani}, {Barbarino},
  {Bazilevskaya}, {Bellotti}, {Boezio}, {Bogomolov}, {Bongi}, {Bonvicini},
  {Borisov}, {Bottai}, {Bruno}, {Cafagna}, {Campana}, {Carbone}, {Carlson},
  {Casolino}, {Castellini}, {Consiglio}, {de Pascale}, {de Santis}, {de
  Simone}, {di Felice}, {Galper}, {Gillard}, {Grishantseva}, {Jerse},
  {Karelin}, {Koldashov}, {Krutkov}, {Kvashnin}, {Leonov}, {Malakhov},
  {Malvezzi}, {Marcelli}, {Mayorov}, {Menn}, {Mikhailov}, {Mocchiutti},
  {Monaco}, {Mori}, {Nikonov}, {Osteria}, {Palma}, {Papini}, {Pearce},
  {Picozza}, {Pizzolotto}, {Ricci}, {Ricciarini}, {Rossetto}, {Sarkar},
  {Simon}, {Sparvoli}, {Spillantini}, {Stochaj}, {Stockton}, {Stozhkov},
  {Vacchi}, {Vannuccini}, {Vasilyev}, {Voronov}, {Wu}, {Yurkin}, {Zampa},
  {Zampa}, \& {Zverev}}]{2011PhRvL.106t1101A}
{Adriani}, O., {Barbarino}, G.~C., {Bazilevskaya}, G.~A., {et~al.}
  2011{\natexlab{b}}, Physical Review Letters, 106, 201101

\bibitem[{{Agrinier} {et~al.}(1969){Agrinier}, {Koechlin}, {Parlier}, {Paul},
  {Vasseur}, {Boella}, {Dilworth}, {Scarsi}, {Sironi}, \&
  {Russo}}]{1969NCimL...1...53A}
{Agrinier}, B., {Koechlin}, Y., {Parlier}, B., {et~al.} 1969, Nuovo Cimento
  Lettere, 1, 53

\bibitem[{{Agrinier} {et~al.}(1965){Agrinier}, {Koechlin}, {Parlier},
  {Vasseur}, {Bland}, {Boella}, {Degli Antoni}, {Dilworth}, {Scarsi}, \&
  {Sironi}}]{1965ICRC....1..331A}
{Agrinier}, B., {Koechlin}, Y., {Parlier}, B., {et~al.} 1965, in International
  Cosmic Ray Conference, Vol.~1, International Cosmic Ray Conference, 331

\bibitem[{{Aguilar} {et~al.}(2013){Aguilar}, {Alberti}, {Alpat}, {Alvino},
  {Ambrosi}, {Andeen}, {Anderhub}, {Arruda}, {Azzarello}, {Bachlechner}, \&
  et~al.}]{2013PhRvL.110n1102A}
{Aguilar}, M., {Alberti}, G., {Alpat}, B., {et~al.} 2013, Physical Review
  Letters, 110, 141102

\bibitem[{{Aguilar} {et~al.}(2010){Aguilar}, {Alcaraz}, {Allaby}, {Alpat},
  {Ambrosi}, {Anderhub}, {Ao}, {Arefiev}, {Arruda}, {Azzarello}, {Basile},
  {Barao}, {Barreira}, {Bartoloni}, {Battiston}, {Becker}, {Becker},
  {Bellagamba}, {B{\'e}n{\'e}}, {Berdugo}, {Berges}, {Bertucci}, {Biland},
  {Bindi}, {Boella}, {Boschini}, {Bourquin}, {Bruni}, {Bu{\'e}nerd}, {Burger},
  {Burger}, {Cai}, {Cannarsa}, {Capell}, {Casadei}, {Casaus}, {Castellini},
  {Cernuda}, {Chang}, {Chen}, {Chen}, {Chen}, {Chernoplekov}, {Chiueh}, {Choi},
  {Cindolo}, {Commichau}, {Contin}, {Cortina-Gil}, {Crespo}, {Cristinziani},
  {Dai}, {dela Guia}, {Delgado}, {Di Falco}, {Djambazov}, {D'Antone}, {Dong},
  {Duranti}, {Engelberg}, {Eppling}, {Eronen}, {Extermann}, {Favier},
  {Fiandrini}, {Fisher}, {Fl{\"u}gge}, {Fouque}, {Galaktionov}, {Gervasi},
  {Giovacchini}, {Giusti}, {Grandi}, {Grimm}, {Gu}, {Haino}, {Hangarter},
  {Hasan}, {Hermel}, {Hofer}, {Hungerford}, {Ionica}, {Jongmanns}, {Karlamaa},
  {Karpinski}, {Kenney}, {Kim}, {Kim}, {Kim}, {Kirn}, {Klimentov},
  {Kossakowski}, {Kounine}, {Koutsenko}, {Kraeber}, {Laborie}, {Laitinen},
  {Lamanna}, {Laurenti}, {Lebedev}, {Lechanoine-Leluc}, {Lee}, {Lee}, {Levi},
  {Lin}, {Liu}, {Lu}, {Lu}, {L{\"u}belsmeyer}, {Luckey}, {Lustermann},
  {Ma{\~n}a}, {Margotti}, {Mayet}, {McNeil}, {Menichelli}, {Mihul}, {Mujunen},
  {Oliva}, {Palmonari}, {Park}, {Park}, {Pauluzzi}, {Pauss}, {Pereira},
  {Perrin}, {Pevsner}, {Pilo}, {Pimenta}, {Plyaskin}, {Pojidaev}, {Pohl},
  {Produit}, {Quadrani}, {Rancoita}, {Rapin}, {Ren}, {Ren}, {Ribordy},
  {Richeux}, {Riihonen}, {Ritakari}, {Ro}, {Roeser}, {Sagdeev}, {Santos},
  {Sartorelli}, {Sbarra}, {Schael}, {Schultz von Dratzig}, {Schwering}, {Seo},
  {Shin}, {Shoumilov}, {Shoutko}, {Siedenburg}, {Siedling}, {Son}, {Song},
  {Spada}, {Spinella}, {Steuer}, {Sun}, {Suter}, {Tang}, {Ting}, {Ting},
  {Tomassetti}, {Tornikoski}, {Torsti}, {Tr{\"u}mper}, {Ulbricht}, {Urpo},
  {Valtonen}, {Vandenhirtz}, {Velikhov}, {Verlaat}, {Vetlitsky}, {Vezzu},
  {Vialle}, {Viertel}, {Vit{\'e}}, {Von Gunten}, {Waldmeier Wicki}, {Wallraff},
  {Wang}, {Wiik}, {Williams}, {Wu}, {Xia}, {Xu}, {Xu}, {Yan}, {Yan}, {Yang},
  {Yang}, {Yang}, {Ye}, {Zhang}, {Zhang}, {Zhao}, {Zhou}, {Zhou}, {Zhu}, {Zhu},
  {Zhuang}, {Zichichi}, {Zimmermann}, \& {Zuccon}}]{2010ApJ...724..329A}
{Aguilar}, M., {Alcaraz}, J., {Allaby}, J., {et~al.} 2010, \apj, 724, 329

\bibitem[{{Aguilar} {et~al.}(2011){Aguilar}, {Alcaraz}, {Allaby}, {Alpat},
  {Ambrosi}, {Anderhub}, {Ao}, {Arefiev}, {Arruda}, {Azzarello}, {Basile},
  {Barao}, {Barreira}, {Bartoloni}, {Battiston}, {Becker}, {Becker},
  {Bellagamba}, {Berdugo}, {Berges}, {Bertucci}, {Biland}, {Bindi}, {Boella},
  {Boschini}, {Bourquin}, {Bruni}, {Bu{\'e}nerd}, {Burger}, {Burger}, {Cai},
  {Cannarsa}, {Capell}, {Casadei}, {Casaus}, {Castellini}, {Cernuda}, {Chang},
  {Chen}, {Chen}, {Chen}, {Chernoplekov}, {Chiueh}, {Choi}, {Cindolo},
  {Commichau}, {Contin}, {Cortina-Gil}, {Crespo}, {Cristinziani}, {Dai}, {dela
  Guia}, {Delgado}, {Di Falco}, {Djambazov}, {D'Antone}, {Dong}, {Duranti},
  {Engelberg}, {Eppling}, {Eronen}, {Extermann}, {Favier}, {Fiandrini},
  {Fisher}, {Fl{\"u}gge}, {Fouque}, {Galaktionov}, {Gervasi}, {Giovacchini},
  {Giusti}, {Grandi}, {Grimm}, {Gu}, {Haino}, {Hangarter}, {Hasan}, {Hermel},
  {Hofer}, {Hungerford}, {Ionica}, {Jongmanns}, {Karlamaa}, {Karpinski},
  {Kenney}, {Kim}, {Kim}, {Kim}, {Kirn}, {Klimentov}, {Kossakowski}, {Kounine},
  {Koutsenko}, {Kraeber}, {Laborie}, {Laitinen}, {Lamanna}, {Laurenti},
  {Lebedev}, {Lechanoine-Leluc}, {Lee}, {Lee}, {Levi}, {Lin}, {Liu}, {Lu},
  {Lu}, {L{\"u}belsmeyer}, {Luckey}, {Lustermann}, {Ma{\~n}a}, {Margotti},
  {Mayet}, {McNeil}, {Menichelli}, {Mihul}, {Mujunen}, {Natale}, {Oliva},
  {Palmonari}, {Paniccia}, {Park}, {Park}, {Pauluzzi}, {Pauss}, {Pereira},
  {Perrin}, {Pevsner}, {Pilo}, {Pimenta}, {Plyaskin}, {Pojidaev}, {Pohl},
  {Produit}, {Quadrani}, {Rancoita}, {Rapin}, {Ren}, {Ren}, {Ribordy},
  {Riihonen}, {Ritakari}, {Ro}, {Roeser}, {Sagdeev}, {Santos}, {Sartorelli},
  {Saouter}, {Sbarra}, {Schael}, {Schultz von Dratzig}, {Schwering}, {Seo},
  {Shin}, {Shoumilov}, {Shoutko}, {Siedenburg}, {Siedling}, {Son}, {Song},
  {Spada}, {Spinella}, {Steuer}, {Sun}, {Suter}, {Tang}, {Ting}, {Ting},
  {Tomassetti}, {Tornikoski}, {Torsti}, {Tr{\"u}mper}, {Ulbricht}, {Urpo},
  {Valtonen}, {Vandenhirtz}, {Velikhov}, {Verlaat}, {Vetlitsky}, {Vezzu},
  {Vialle}, {Viertel}, {Vit{\'e}}, {Von Gunten}, {Waldmeier Wicki}, {Wallraff},
  {Wang}, {Wiik}, {Williams}, {Wu}, {Xia}, {Xu}, {Xu}, {Yan}, {Yan}, {Yang},
  {Yang}, {Yang}, {Ye}, {Zhang}, {Zhang}, {Zhao}, {Zhou}, {Zhou}, {Zhu}, {Zhu},
  {Zhuang}, {Zichichi}, {Zimmermann}, \& {Zuccon}}]{2011ApJ...736..105A}
{Aguilar}, M., {Alcaraz}, J., {Allaby}, J., {et~al.} 2011, \apj, 736, 105

\bibitem[{{Aharonian} {et~al.}(2009){Aharonian}, {Akhperjanian}, {Anton},
  {Barres de Almeida}, {Bazer-Bachi}, {Becherini}, {Behera}, {Bernl{\"o}hr},
  {Bochow}, {Boisson}, {Bolmont}, {Borrel}, {Brucker}, {Brun}, {Brun},
  {B{\"u}hler}, {Bulik}, {B{\"u}sching}, {Boutelier}, {Chadwick},
  {Charbonnier}, {Chaves}, {Cheesebrough}, {Chounet}, {Clapson}, {Coignet},
  {Dalton}, {Daniel}, {Davids}, {Degrange}, {Deil}, {Dickinson},
  {Djannati-Ata{\"i}}, {Domainko}, {O'C.~Drury}, {Dubois}, {Dubus}, {Dyks},
  {Dyrda}, {Egberts}, {Emmanoulopoulos}, {Espigat}, {Farnier}, {Feinstein},
  {Fiasson}, {F{\"o}rster}, {Fontaine}, {F{\"u}{\ss}ling}, {Gabici}, {Gallant},
  {G{\'e}rard}, {Gerbig}, {Giebels}, {Glicenstein}, {Gl{\"u}ck}, {Goret},
  {G{\"o}ring}, {Hauser}, {Hauser}, {Heinz}, {Heinzelmann}, {Henri}, {Hermann},
  {Hinton}, {Hoffmann}, {Hofmann}, {Holleran}, {Hoppe}, {Horns},
  {Jacholkowska}, {de Jager}, {Jahn}, {Jung}, {Katarzy{\'n}ski}, {Katz},
  {Kaufmann}, {Kendziorra}, {Kerschhaggl}, {Khangulyan}, {Kh{\'e}lifi},
  {Keogh}, {Klu{\'z}niak}, {Kneiske}, {Komin}, {Kosack}, {Kossakowski},
  {Lamanna}, {Lenain}, {Lohse}, {Marandon}, {Martin}, {Martineau-Huynh},
  {Marcowith}, {Masbou}, {Maurin}, {McComb}, {Medina}, {Moderski}, {Moulin},
  {Naumann-Godo}, {de Naurois}, {Nedbal}, {Nekrassov}, {Nicholas}, {Niemiec},
  {Nolan}, {Ohm}, {Olive}, {de O{\~n}a Wilhelmi}, {Orford}, {Ostrowski},
  {Panter}, {Paz Arribas}, {Pedaletti}, {Pelletier}, {Petrucci}, {Pita},
  {P{\"u}hlhofer}, {Punch}, {Quirrenbach}, {Raubenheimer}, {Raue}, {Rayner},
  {Reimer}, {Renaud}, {Rieger}, {Ripken}, {Rob}, {Rosier-Lees}, {Rowell},
  {Rudak}, {Rulten}, {Ruppel}, {Sahakian}, {Santangelo}, {Schlickeiser},
  {Sch{\"o}ck}, {Schr{\"o}der}, {Schwanke}, {Schwarzburg}, {Schwemmer},
  {Shalchi}, {Sikora}, {Skilton}, {Sol}, {Spangler}, {Stawarz}, {Steenkamp},
  {Stegmann}, {Stinzing}, {Superina}, {Szostek}, {Tam}, {Tavernet}, {Terrier},
  {Tibolla}, {Tluczykont}, {van Eldik}, {Vasileiadis}, {Venter}, {Venter},
  {Vialle}, {Vincent}, {Vivier}, {V{\"o}lk}, {Volpe}, {Wagner}, {Ward},
  {Zdziarski}, \& {Zech}}]{2009A&A...508..561A}
{Aharonian}, F., {Akhperjanian}, A.~G., {Anton}, G., {et~al.} 2009, \aap, 508,
  561

\bibitem[{{Aharonian} {et~al.}(2008){Aharonian}, {Akhperjanian}, {Barres de
  Almeida}, {Bazer-Bachi}, {Becherini}, {Behera}, {Benbow}, {Bernl{\"o}hr},
  {Boisson}, {Bochow}, {Borrel}, {Braun}, {Brion}, {Brucker}, {Brun},
  {B{\"u}hler}, {Bulik}, {B{\"u}sching}, {Boutelier}, {Carrigan}, {Chadwick},
  {Charbonnier}, {Chaves}, {Cheesebrough}, {Chounet}, {Clapson}, {Coignet},
  {Costamante}, {Dalton}, {Degrange}, {Deil}, {Dickinson}, {Djannati-Ata{\"i}},
  {Domainko}, {Drury}, {Dubois}, {Dubus}, {Dyks}, {Dyrda}, {Egberts},
  {Emmanoulopoulos}, {Espigat}, {Farnier}, {Feinstein}, {Fiasson},
  {F{\"o}rster}, {Fontaine}, {F{\"u}{\ss}ling}, {Gabici}, {Gallant},
  {G{\'e}rard}, {Giebels}, {Glicenstein}, {Gl{\"u}ck}, {Goret},
  {Hadjichristidis}, {Hauser}, {Hauser}, {Heinz}, {Heinzelmann}, {Henri},
  {Hermann}, {Hinton}, {Hoffmann}, {Hofmann}, {Holleran}, {Hoppe}, {Horns},
  {Jacholkowska}, {de Jager}, {Jung}, {Katarzy{\'n}ski}, {Kaufmann},
  {Kendziorra}, {Kerschhaggl}, {Khangulyan}, {Kh{\'e}lifi}, {Keogh}, {Komin},
  {Kosack}, {Lamanna}, {Lenain}, {Lohse}, {Marandon}, {Martin},
  {Martineau-Huynh}, {Marcowith}, {Maurin}, {McComb}, {Medina}, {Moderski},
  {Moulin}, {Naumann-Godo}, {de Naurois}, {Nedbal}, {Nekrassov}, {Niemiec},
  {Nolan}, {Ohm}, {Olive}, {de O{\~n}a Wilhelmi}, {Orford}, {Osborne},
  {Ostrowski}, {Panter}, {Pedaletti}, {Pelletier}, {Petrucci}, {Pita},
  {P{\"u}hlhofer}, {Punch}, {Quirrenbach}, {Raubenheimer}, {Raue}, {Rayner},
  {Renaud}, {Rieger}, {Ripken}, {Rob}, {Rosier-Lees}, {Rowell}, {Rudak},
  {Rulten}, {Ruppel}, {Sahakian}, {Santangelo}, {Schlickeiser}, {Sch{\"o}ck},
  {Schr{\"o}der}, {Schwanke}, {Schwarzburg}, {Schwemmer}, {Shalchi}, {Skilton},
  {Sol}, {Spangler}, {Stawarz}, {Steenkamp}, {Stegmann}, {Superina}, {Tam},
  {Tavernet}, {Terrier}, {Tibolla}, {van Eldik}, {Vasileiadis}, {Venter},
  {Vialle}, {Vincent}, {Vivier}, {V{\"o}lk}, {Volpe}, {Wagner}, {Ward},
  {Zdziarski}, \& {Zech}}]{2008PhRvL.101z1104A}
{Aharonian}, F., {Akhperjanian}, A.~G., {Barres de Almeida}, U., {et~al.} 2008,
  Physical Review Letters, 101, 261104

\bibitem[{{Ahlen} {et~al.}(2000){Ahlen}, {Greene}, {Loomba}, {Mitchell},
  {Bower}, {Heinz}, {Mufson}, {Musser}, {Pitts}, {Spiczak}, {Clem}, {Guzik},
  {Lijowski}, {Wefel}, {McKee}, {Nutter}, {Tomasch}, {Beatty}, {Ficenec}, \&
  {Tobias}}]{2000ApJ...534..757A}
{Ahlen}, S.~P., {Greene}, N.~R., {Loomba}, D., {et~al.} 2000, \apj, 534, 757

\bibitem[{{Ahn} {et~al.}(2009){Ahn}, {Allison}, {Bagliesi}, {Barbier},
  {Beatty}, {Bigongiari}, {Brandt}, {Childers}, {Conklin}, {Coutu}, {Du
  Vernois}, {Ganel}, {Han}, {Jeon}, {Kim}, {Lee}, {Maestro}, {Malinine},
  {Marrocchesi}, {Minnick}, {Mognet}, {Nam}, {Nutter}, {Park}, {Park}, {Seo},
  {Sina}, {Walpole}, {Wu}, {Yang}, {Yoon}, {Zei}, \&
  {Zinn}}]{2009ApJ...707..593A}
{Ahn}, H.~S., {Allison}, P., {Bagliesi}, M.~G., {et~al.} 2009, \apj, 707, 593

\bibitem[{{Ahn} {et~al.}(2010){Ahn}, {Allison}, {Bagliesi}, {Barbier},
  {Beatty}, {Bigongiari}, {Brandt}, {Childers}, {Conklin}, {Coutu},
  {DuVernois}, {Ganel}, {Han}, {Jeon}, {Kim}, {Lee}, {Lee}, {Maestro},
  {Malinin}, {Marrocchesi}, {Minnick}, {Mognet}, {Na}, {Nam}, {Nam}, {Nutter},
  {Park}, {Park}, {Seo}, {Sina}, {Walpole}, {Wu}, {Yang}, {Yoon}, {Zei}, \&
  {Zinn}}]{2010ApJ...715.1400A}
{Ahn}, H.~S., {Allison}, P.~S., {Bagliesi}, M.~G., {et~al.} 2010, \apj, 715,
  1400

\bibitem[{{Ahn} {et~al.}(2008){Ahn}, {Allison}, {Bagliesi}, {Beatty},
  {Bigongiari}, {Boyle}, {Brandt}, {Childers}, {Conklin}, {Coutu}, {Duvernois},
  {Ganel}, {Han}, {Hyun}, {Jeon}, {Kim}, {Lee}, {Lee}, {Lutz}, {Maestro},
  {Malinin}, {Marrocchesi}, {Minnick}, {Mognet}, {Nam}, {Nutter}, {Park},
  {Park}, {Seo}, {Sina}, {Swordy}, {Wakely}, {Wu}, {Yang}, {Yoon}, {Zei}, \&
  {Zinn}}]{2008APh....30..133A}
{Ahn}, H.~S., {Allison}, P.~S., {Bagliesi}, M.~G., {et~al.} 2008, Astroparticle
  Physics, 30, 133

\bibitem[{{Alcaraz} {et~al.}(2000{\natexlab{a}}){Alcaraz}, {Alpat}, {Ambrosi},
  {Anderhub}, {Ao}, {Arefiev}, {Azzarello}, {Babucci}, {Baldini}, {Basile},
  {Barancourt}, {Barao}, {Barbier}, {Barreira}, {Battiston}, {Becker},
  {Becker}, {Bellagamba}, {B{\'e}n{\'e}}, {Berdugo}, {Berges}, {Bertucci},
  {Biland}, {Bizzaglia}, {Blasko}, {Boella}, {Boschini}, {Bourquin}, {Brocco},
  {Bruni}, {Buenerd}, {Burger}, {Burger}, {Cai}, {Camps}, {Cannarsa}, {Capell},
  {Casadei}, {Casaus}, {Castellini}, {Cecchi}, {Chang}, {Chen}, {Chen}, {Chen},
  {Chernoplekov}, {Chiueh}, {Chuang}, {Cindolo}, {Commichau}, {Contin},
  {Crespo}, {Cristinziani}, {da Cunha}, {Dai}, {Deus}, {Dinu}, {Djambazov},
  {D'Antone}, {Dong}, {Emonet}, {Engelberg}, {Eppling}, {Eronen}, {Esposito},
  {Extermann}, {Favier}, {Fiandrini}, {Fisher}, {Fluegge}, {Fouque},
  {Galaktionov}, {Gervasi}, {Giusti}, {Grandi}, {Grimm}, {Gu}, {Hangarter},
  {Hasan}, {Hermel}, {Hofer}, {Huang}, {Hungerford}, {Ionica}, {Ionica},
  {Jongmanns}, {Karlamaa}, {Karpinski}, {Kenney}, {Kenny}, {Kim}, {Klimentov},
  {Kossakowski}, {Koutsenko}, {Kraeber}, {Laborie}, {Laitinen}, {Lamanna},
  {Laurenti}, {Lebedev}, {Lee}, {Levi}, {Levtchenko}, {Liu}, {Liu}, {Lopes},
  {Lu}, {Lu}, {L{\"u}belsmeyer}, {Luckey}, {Lustermann}, {Ma{\~n}a},
  {Margotti}, {Mayet}, {McNeil}, {Meillon}, {Menichelli}, {Mihul}, {Mourao},
  {Mujunen}, {Palmonari}, {Papi}, {Park}, {Pauluzzi}, {Pauss}, {Perrin},
  {Pesci}, {Pevsner}, {Pimenta}, {Plyaskin}, {Pojidaev}, {Pohl}, {Postolache},
  {Produit}, {Rancoita}, {Rapin}, {Raupach}, {Ren}, {Ren}, {Ribordy},
  {Richeux}, {Riihonen}, {Ritakari}, {Roeser}, {Roissin}, {Sagdeev},
  {Sartorelli}, {Schultz von Dratzig}, {Schwering}, {Scolieri}, {Seo},
  {Shoutko}, {Shoumilov}, {Siedling}, {Son}, {Song}, {Steuer}, {Sun}, {Suter},
  {Tang}, {Ting}, {Ting}, {Tornikoski}, {Torsti}, {Tr{\"u}mper}, {Ulbricht},
  {Urpo}, {Usoskin}, {Valtonen}, {Vandenhirtz}, {Velcea}, {Velikhov},
  {Verlaat}, {Vetlitsky}, {Vezzu}, {Vialle}, {Viertel}, {Vit{\'e}}, {Von
  Gunten}, {Waldmeier Wicki}, {Wallraff}, {Wang}, {Wang}, {Wang}, {Wiik},
  {Williams}, {Wu}, {Xia}, {Yan}, {Yan}, {Yang}, {Yang}, {Ye}, {Yeh}, {Xu},
  {Zhang}, {Zhang}, {Zhao}, {Zhu}, {Zhu}, {Zhuang}, {Zichichi}, \&
  {Zimmermann}}]{2000PhLB..490...27A}
{Alcaraz}, J., {Alpat}, B., {Ambrosi}, G., {et~al.} 2000{\natexlab{a}}, Physics
  Letters B, 490, 27

\bibitem[{{Alcaraz} {et~al.}(2000{\natexlab{b}}){Alcaraz}, {Alpat}, {Ambrosi},
  {Anderhub}, {Ao}, {Arefiev}, {Azzarello}, {Babucci}, {Baldini}, {Basile},
  {Barancourt}, {Barao}, {Barbier}, {Barreira}, {Battiston}, {Becker},
  {Becker}, {Bellagamba}, {B{\'e}n{\'e}}, {Berdugo}, {Berges}, {Bertucci},
  {Biland}, {Bizzaglia}, {Blasko}, {Boella}, {Boschini}, {Bourquin}, {Brocco},
  {Bruni}, {Buenerd}, {Burger}, {Burger}, {Cai}, {Camps}, {Cannarsa}, {Capell},
  {Casadei}, {Casaus}, {Castellini}, {Cecchi}, {Chang}, {Chen}, {Chen}, {Chen},
  {Chernoplekov}, {Chiueh}, {Chuang}, {Cindolo}, {Commichau}, {Contin},
  {Crespo}, {Cristinziani}, {da Cunha}, {Dai}, {Deus}, {Dinu}, {Djambazov},
  {D'Antone}, {Dong}, {Emonet}, {Engelberg}, {Eppling}, {Eronen}, {Esposito},
  {Extermann}, {Favier}, {Fiandrini}, {Fisher}, {Fluegge}, {Fouque},
  {Galaktionov}, {Gervasi}, {Giusti}, {Grandi}, {Grimm}, {Gu}, {Hangarter},
  {Hasan}, {Hermel}, {Hofer}, {Huang}, {Hungerford}, {Ionica}, {Ionica},
  {Jongmanns}, {Karlamaa}, {Karpinski}, {Kenney}, {Kenny}, {Kim}, {Klimentov},
  {Kossakowski}, {Koutsenko}, {Kraeber}, {Laborie}, {Laitinen}, {Lamanna},
  {Laurenti}, {Lebedev}, {Lee}, {Levi}, {Levtchenko}, {Liu}, {Liu}, {Lopes},
  {Lu}, {Lu}, {L{\"u}belsmeyer}, {Luckey}, {Lustermann}, {Ma{\~n}a},
  {Margotti}, {Mayet}, {McNeil}, {Meillon}, {Menichelli}, {Mihul}, {Mourao},
  {Mujunen}, {Palmonari}, {Papi}, {Park}, {Pauluzzi}, {Pauss}, {Perrin},
  {Pesci}, {Pevsner}, {Pimenta}, {Plyaskin}, {Pojidaev}, {Postolache},
  {Produit}, {Rancoita}, {Rapin}, {Raupach}, {Ren}, {Ren}, {Ribordy},
  {Richeux}, {Riihonen}, {Ritakari}, {Roeser}, {Roissin}, {Sagdeev},
  {Sartorelli}, {Schultz von Dratzig}, {Schwering}, {Scolieri}, {Seo},
  {Shoutko}, {Shoumilov}, {Siedling}, {Son}, {Song}, {Steuer}, {Sun}, {Suter},
  {Tang}, {Ting}, {Ting}, {Tornikoski}, {Torsti}, {Tr{\"u}mper}, {Ulbricht},
  {Urpo}, {Usoskin}, {Valtonen}, {Vandenhirtz}, {Velcea}, {Velikhov},
  {Verlaat}, {Vetlitsky}, {Vezzu}, {Vialle}, {Viertel}, {Vit{\'e}}, {Von
  Gunten}, {Waldmeier Wicki}, {Wallraff}, {Wang}, {Wang}, {Wang}, {Wiik},
  {Williams}, {Wu}, {Xia}, {Yan}, {Yan}, {Yang}, {Yang}, {Ye}, {Yeh}, {Xu},
  {Zhang}, {Zhang}, {Zhao}, {Zhu}, {Zhu}, {Zhuang}, {Zichichi}, \&
  {Zimmermann}}]{2000PhLB..484...10A}
{Alcaraz}, J., {Alpat}, B., {Ambrosi}, G., {et~al.} 2000{\natexlab{b}}, Physics
  Letters B, 484, 10

\bibitem[{{AMS-01 Collaboration} {et~al.}(2007){AMS-01 Collaboration},
  {Aguilar}, {Alcaraz}, {Allaby}, {Alpat}, {Ambrosi}, {Anderhub}, {Ao},
  {Arefiev}, {Azzarello}, {Baldini}, {Basile}, {Barancourt}, {Barao},
  {Barbier}, {Barreira}, {Battiston}, {Becker}, {Becker}, {Bellagamba},
  {B{\'e}n{\'e}}, {Berdugo}, {Berges}, {Bertucci}, {Biland}, {Blasko},
  {Boella}, {Boschini}, {Bourquin}, {Brocco}, {Bruni}, {Bu{\'e}nerd}, {Burger},
  {Burger}, {Cai}, {Camps}, {Cannarsa}, {Capell}, {Cardano}, {Casadei},
  {Casaus}, {Castellini}, {Chang}, {Chen}, {Chen}, {Chen}, {Chernoplekov},
  {Chiueh}, {Cho}, {Choi}, {Choi}, {Cindolo}, {Commichau}, {Contin},
  {Cortina-Gil}, {Cristinziani}, {Dai}, {Delgado}, {Difalco}, {Djambazov},
  {D'Antone}, {Dong}, {Emonet}, {Engelberg}, {Eppling}, {Eronen}, {Esposito},
  {Extermann}, {Favier}, {Fiandrini}, {Fisher}, {Fl{\"u}gge}, {Fouque},
  {Galaktionov}, {Gast}, {Gervasi}, {Giusti}, {Grandi}, {Grimm}, {Gu},
  {Hangarter}, {Hasan}, {Hermel}, {Hofer}, {Hungerford}, {Jongmanns},
  {Karlamaa}, {Karpinski}, {Kenney}, {Kim}, {Kim}, {Kim}, {Kim}, {Klimentov},
  {Kossakowski}, {Kounine}, {Koutsenko}, {Kraeber}, {Laborie}, {Laitinen},
  {Lamanna}, {Lanciotti}, {Laurenti}, {Lebedev}, {Lechanoine-Leluc}, {Lee},
  {Lee}, {Levi}, {Liu}, {Liu}, {Lu}, {Lu}, {L{\"u}belsmeyer}, {Luckey},
  {Lustermann}, {Ma{\~n}a}, {Margotti}, {Mayet}, {McNeil}, {Meillon},
  {Menichelli}, {Mihul}, {Mujunen}, {Oliva}, {Olzem}, {Palmonari}, {Park},
  {Park}, {Pauluzzi}, {Pauss}, {Perrin}, {Pesci}, {Pevsner}, {Pilo}, {Pimenta},
  {Plyaskin}, {Pojidaev}, {Pohl}, {Produit}, {Rancoita}, {Rapin}, {Raupach},
  {Ren}, {Ren}, {Ribordy}, {Richeux}, {Riihonen}, {Ritakari}, {Ro}, {Roeser},
  {Rossin}, {Sagdeev}, {Santos}, {Sartorelli}, {Sbarra}, {Schael}, {Schultz von
  Dratzig}, {Schwering}, {Seo}, {Shin}, {Shoumilov}, {Shoutko}, {Siedenburg},
  {Siedling}, {Son}, {Song}, {Spinella}, {Steuer}, {Sun}, {Suter}, {Tang},
  {Ting}, {Ting}, {Tornikoski}, {Torsti}, {Tr{\"u}mper}, {Ulbricht}, {Urpo},
  {Valtonen}, {Vandenhirtz}, {Velikhov}, {Verlaat}, {Vetlitsky}, {Vezzu},
  {Vialle}, {Viertel}, {Vit{\'e}}, {von Gunten}, {Waldmeier Wicki}, {Wallraff},
  {Wang}, {Wang}, {Wiik}, {Williams}, {Wu}, {Xia}, {Xu}, {Yan}, {Yan}, {Yang},
  {Yang}, {Yang}, {Ye}, {Xu}, {Zhang}, {Zhang}, {Zhao}, {Zhou}, {Zhu}, {Zhu},
  {Zhuang}, {Zichichi}, {Zimmermann}, \& {Zuccon}}]{2007PhLB..646..145A}
{AMS-01 Collaboration}, {Aguilar}, M., {Alcaraz}, J., {et~al.} 2007, Physics
  Letters B, 646, 145

\bibitem[{{AMS Collaboration} {et~al.}(2002){AMS Collaboration}, {Aguilar},
  {Alcaraz}, {Allaby}, {Alpat}, {Ambrosi}, {Anderhub}, {Ao}, {Arefiev},
  {Azzarello}, \& et~al.}]{2002PhR...366..331A}
{AMS Collaboration}, {Aguilar}, M., {Alcaraz}, J., {et~al.} 2002, \physrep,
  366, 331

\bibitem[{{AMS Collaboration} {et~al.}(2000){AMS Collaboration}, {Alcaraz},
  {Alpat}, {Ambrosi}, {Anderhub}, {Ao}, {Arefiev}, {Azzarello}, {Babucci},
  {Baldini}, {Basile}, {Barancourt}, {Barao}, {Barbier}, {Barreira},
  {Battiston}, {Becker}, {Becker}, {Bellagamba}, {B{\'e}n{\'e}}, {Berdugo},
  {Berges}, {Bertucci}, {Biland}, {Bizzaglia}, {Blasko}, {Boella}, {Boschini},
  {Bourquin}, {Brocco}, {Bruni}, {Buenerd}, {Burger}, {Burger}, {Cai}, {Camps},
  {Cannarsa}, {Capell}, {Casadei}, {Casaus}, {Castellini}, {Cecchi}, {Chang},
  {Chen}, {Chen}, {Chen}, {Chernoplekov}, {Chiueh}, {Chuang}, {Cindolo},
  {Commichau}, {Contin}, {Cristinziani}, {da Cunha}, {Dai}, {Deus}, {Dinu},
  {Djambazov}, {D'Antone}, {Dong}, {Emonet}, {Engelberg}, {Eppling}, {Eronen},
  {Esposito}, {Extermann}, {Favier}, {Fiandrini}, {Fisher}, {Fluegge},
  {Fouque}, {Galaktionov}, {Gervasi}, {Giusti}, {Grandi}, {Grimm}, {Gu},
  {Hangarter}, {Hasan}, {Hermel}, {Hofer}, {Huang}, {Hungerford}, {Ionica},
  {Ionica}, {Jongmanns}, {Karlamaa}, {Karpinski}, {Kenney}, {Kenny}, {Kim},
  {Klimentov}, {Kossakowski}, {Koutsenko}, {Kraeber}, {Laborie}, {Laitinen},
  {Lamanna}, {Laurenti}, {Lebedev}, {Lee}, {Levi}, {Levtchenko}, {Liu}, {Liu},
  {Lopes}, {Lu}, {Lu}, {L{\"u}belsmeyer}, {Luckey}, {Lustermann}, {Ma{\~n}a},
  {Margotti}, {Mayet}, {McNeil}, {Meillon}, {Menichelli}, {Mihul}, {Mourao},
  {Mujunen}, {Palmonari}, {Papi}, {Park}, {Pauluzzi}, {Pauss}, {Perrin},
  {Pesci}, {Pevsner}, {Pimenta}, {Plyaskin}, {Pojidaev}, {Pohl}, {Postolache},
  {Produit}, {Rancoita}, {Rapin}, {Raupach}, {Ren}, {Ren}, {Ribordy},
  {Richeux}, {Riihonen}, {Ritakari}, {Roeser}, {Roissin}, {Sagdeev},
  {Sartorelli}, {Schultz von Dratzig}, {Schwering}, {Scolieri}, {Seo},
  {Shoutko}, {Shoumilov}, {Siedling}, {Son}, {Song}, {Steuer}, {Sun}, {Suter},
  {Tang}, {Ting}, {Ting}, {Tornikoski}, {Torsti}, {Tr{\"u}mper}, {Ulbricht},
  {Urpo}, {Usoskin}, {Valtonen}, {Vandenhirtz}, {Velcea}, {Velikhov},
  {Verlaat}, {Vetlitsky}, {Vezzu}, {Vialle}, {Viertel}, {Vit{\'e}}, {Von
  Gunten}, {Waldmeier Wicki}, {Wallraff}, {Wang}, {Wang}, {Wang}, {Wiik},
  {Williams}, {Wu}, {Xia}, {Yan}, {Yan}, {Yang}, {Yang}, {Ye}, {Yeh}, {Xu},
  {Zhang}, {Zhang}, {Zhao}, {Zhu}, {Zhu}, {Zhuang}, {Zichichi}, {Zimmermann},
  \& {Zuccon}}]{2000PhLB..494..193A}
{AMS Collaboration}, {Alcaraz}, J., {Alpat}, B., {et~al.} 2000, Physics Letters
  B, 494, 193

\bibitem[{{Anand} {et~al.}(1968){Anand}, {Daniel}, \&
  {Stephens}}]{1968PhRvL..20..764A}
{Anand}, K.~C., {Daniel}, R.~R., \& {Stephens}, S.~A. 1968, Physical Review
  Letters, 20, 764

\bibitem[{{Anand} {et~al.}(1973){Anand}, {Daniel}, \&
  {Stephens}}]{1973ICRC....1..355A}
{Anand}, K.~C., {Daniel}, R.~R., \& {Stephens}, S.~A. 1973, in International
  Cosmic Ray Conference, Vol.~1, International Cosmic Ray Conference, 355

\bibitem[{{Apparao}(1973)}]{1973ICRC....1..126A}
{Apparao}, K.~M.~V. 1973, in International Cosmic Ray Conference, Vol.~1,
  International Cosmic Ray Conference, 126

\bibitem[{{Aramaki} {et~al.}(2012){Aramaki}, {Chan}, {Hailey}, {Kaplan},
  {Krings}, {Madden}, {Proti{\'c}}, \& {Ross}}]{2012NIMPA.682...90A}
{Aramaki}, T., {Chan}, S.~K., {Hailey}, C.~J., {et~al.} 2012, Nuclear
  Instruments and Methods in Physics Research A, 682, 90

\bibitem[{{Arruda} {et~al.}(2008){Arruda}, {Bar{\~a}o}, \&
  {Pereira}}]{2008arXiv0801.3243A}
{Arruda}, L., {Bar{\~a}o}, F., \& {Pereira}, R. 2008, ArXiv:0801.3243

\bibitem[{{Asakimori} {et~al.}(1998){Asakimori}, {Burnett}, {Cherry}, {Chevli},
  {Christ}, {Dake}, {Derrickson}, {Fountain}, {Fuki}, {Gregory}, {Hayashi},
  {Holynski}, {Iwai}, {Iyono}, {Johnson}, {Kobayashi}, {Lord}, {Miyamura},
  {Moon}, {Nilsen}, {Oda}, {Ogata}, {Olson}, {Parnell}, {Roberts}, {Sengupta},
  {Shiina}, {Strausz}, {Sugitate}, {Takahashi}, {Tominaga}, {Watts}, {Wefel},
  {Wilczynska}, {Wilczynski}, {Wilkes}, {Wolter}, {Yokomi}, \&
  {Zager}}]{1998ApJ...502..278A}
{Asakimori}, K., {Burnett}, T.~H., {Cherry}, M.~L., {et~al.} 1998, \apj, 502,
  278

\bibitem[{{Asaoka} {et~al.}(2002){Asaoka}, {Shikaze}, {Abe}, {Anraku},
  {Fujikawa}, {Fuke}, {Haino}, {Imori}, {Izumi}, {Maeno}, {Makida}, {Matsuda},
  {Matsui}, {Matsukawa}, {Matsumoto}, {Matsunaga}, {Mitchell}, {Mitsui},
  {Moiseev}, {Motoki}, {Nishimura}, {Nozaki}, {Orito}, {Ormes}, {Saeki},
  {Sanuki}, {Sasaki}, {Seo}, {Sonoda}, {Streitmatter}, {Suzuki}, {Tanaka},
  {Tanizaki}, {Ueda}, {Wang}, {Yajima}, {Yamagami}, {Yamamoto}, {Yamamoto},
  {Yamato}, {Yoshida}, \& {Yoshimura}}]{2002PhRvL..88e1101A}
{Asaoka}, Y., {Shikaze}, Y., {Abe}, K., {et~al.} 2002, Physical Review Letters,
  88, 051101

\bibitem[{{Auger} {et~al.}(1939){Auger}, {Ehrenfest}, {Maze}, {Daudin}, \&
  {Fr{\'e}on}}]{1939RvMP...11..288A}
{Auger}, P., {Ehrenfest}, P., {Maze}, R., {Daudin}, J., \& {Fr{\'e}on}, R.~A.
  1939, Reviews of Modern Physics, 11, 288

\bibitem[{{Ave} {et~al.}(2008){Ave}, {Boyle}, {Gahbauer}, {H{\"o}ppner},
  {H{\"o}randel}, {Ichimura}, {M{\"u}ller}, \&
  {Romero-Wolf}}]{2008ApJ...678..262A}
{Ave}, M., {Boyle}, P.~J., {Gahbauer}, F., {et~al.} 2008, \apj, 678, 262

\bibitem[{{Barwick} {et~al.}(1997){Barwick}, {Beatty}, {Bhattacharyya},
  {Bower}, {Chaput}, {Coutu}, {de Nolfo}, {Knapp}, {Lowder}, {McKee},
  {Mueller}, {Musser}, {Nutter}, {Schneider}, {Swordy}, {Tarle}, {Tomasch},
  {Torbet}, \& {HEAT Collaboration}}]{1997ApJ...482L.191B}
{Barwick}, S.~W., {Beatty}, J.~J., {Bhattacharyya}, A., {et~al.} 1997, \apjl,
  482, L191

\bibitem[{{Barwick} {et~al.}(1995){Barwick}, {Beatty}, {Bower}, {Chaput},
  {Coutu}, {de Nolfo}, {Ficenec}, {Knapp}, {Lowder}, {McKee}, {M{\"u}ller},
  {Musser}, {Nutter}, {Schneider}, {Swordy}, {Tang}, {Tarl{\'e}}, {Tomasch}, \&
  {Torbet}}]{1995PhRvL..75..390B}
{Barwick}, S.~W., {Beatty}, J.~J., {Bower}, C.~R., {et~al.} 1995, Physical
  Review Letters, 75, 390

\bibitem[{{Barwick} {et~al.}(1998){Barwick}, {Beatty}, {Bower}, {Chaput},
  {Coutu}, {de Nolfo}, {Duvernois}, {Ellithorpe}, {Ficenec}, {Knapp}, {Lowder},
  {McKee}, {Mueller}, {Musser}, {Nutter}, {Schneider}, {Swordy}, {Tarle},
  {Tomasch}, \& {Torbet}}]{1998ApJ...498..779B}
{Barwick}, S.~W., {Beatty}, J.~J., {Bower}, C.~R., {et~al.} 1998, \apj, 498,
  779

\bibitem[{{Basini}(1999)}]{1999ICRC....3...77B}
{Basini}, G. 1999, in International Cosmic Ray Conference, Vol.~3,
  International Cosmic Ray Conference, 77

\bibitem[{{Beach} {et~al.}(2001){Beach}, {Beatty}, {Bhattacharyya}, {Bower},
  {Coutu}, {Duvernois}, {Labrador}, {McKee}, {Minnick}, {M{\"u}ller}, {Musser},
  {Nutter}, {Schubnell}, {Swordy}, {Tarl{\'e}}, \&
  {Tomasch}}]{2001PhRvL..87A1101B}
{Beach}, A.~S., {Beatty}, J.~J., {Bhattacharyya}, A., {et~al.} 2001, Physical
  Review Letters, 87, A261101

\bibitem[{{Beatty} {et~al.}(2004){Beatty}, {Bhattacharyya}, {Bower}, {Coutu},
  {Duvernois}, {McKee}, {Minnick}, {M{\"u}ller}, {Musser}, {Nutter},
  {Labrador}, {Schubnell}, {Swordy}, {Tarl{\'e}}, \&
  {Tomasch}}]{2004PhRvL..93x1102B}
{Beatty}, J.~J., {Bhattacharyya}, A., {Bower}, C., {et~al.} 2004, Physical
  Review Letters, 93, 241102

\bibitem[{{Beatty} {et~al.}(1993){Beatty}, {Ficenec}, {Tobias}, {Mitchell},
  {McKee}, {Nutter}, {Tarle}, {Tomasch}, {Clem}, {Guzik}, {Lijowski}, {Wefel},
  {Bower}, {Heinz}, {Mufson}, {Musser}, {Pitts}, {Spiczak}, {Ahlen}, \&
  {Zhou}}]{1993ApJ...413..268B}
{Beatty}, J.~J., {Ficenec}, D.~J., {Tobias}, S., {et~al.} 1993, \apj, 413, 268

\bibitem[{{Beatty} {et~al.}(1985){Beatty}, {Garcia-Munoz}, \&
  {Simpson}}]{1985ApJ...294..455B}
{Beatty}, J.~J., {Garcia-Munoz}, M., \& {Simpson}, J.~A. 1985, \apj, 294, 455

\bibitem[{{Beedle} \& {Webber}(1968)}]{1968CaJPS..46.1014B}
{Beedle}, R.~E. \& {Webber}, W.~R. 1968, Canadian Journal of Physics
  Supplement, 46, 1014

\bibitem[{{Bellotti} {et~al.}(1999){Bellotti}, {Cafagna}, {Circella}, {de
  Marzo}, {Golden}, {Stochaj}, {de Pascale}, {Morselli}, {Picozza}, {Stephens},
  {Hof}, {Menn}, {Simon}, {Mitchell}, {Ormes}, {Streitmatter}, {Finetti},
  {Grimani}, {Papini}, {Piccardi}, {Spillantini}, {Basini}, \&
  {Ricci}}]{1999PhRvD..60e2002B}
{Bellotti}, R., {Cafagna}, F., {Circella}, M., {et~al.} 1999, \prd, 60, 052002

\bibitem[{{BESS Collaboration} {et~al.}(2008){BESS Collaboration}, {Abe},
  {Fuke}, {Haino}, {Hams}, {Itazaki}, {Kim}, {Kumazawa}, {Lee}, {Makida},
  {Matsuda}, {Matsumoto}, {Mitchell}, {Moiseev}, {Myers}, {Nishimura},
  {Nozaki}, {Orito}, {Ormes}, {Sasaki}, {Seo}, {Shikaze}, {Streitmatter},
  {Suzuki}, {Takasugi}, {Takeuchi}, {Tanaka}, {Yamagami}, {Yamamoto},
  {Yoshida}, \& {Yoshimura}}]{2008PhLB..670..103B}
{BESS Collaboration}, {Abe}, K., {Fuke}, H., {et~al.} 2008, Physics Letters B,
  670, 103

\bibitem[{{Beuermann} {et~al.}(1969){Beuermann}, {Rice}, {Stone}, \&
  {Vogt}}]{1969PhRvL..22..412B}
{Beuermann}, K.~P., {Rice}, C.~J., {Stone}, E.~C., \& {Vogt}, R.~E. 1969,
  Physical Review Letters, 22, 412

\bibitem[{{Bjarle} {et~al.}(1979){Bjarle}, {Herrstr{\"o}m}, {Jonsson}, \&
  {Kristiansson}}]{1979ZPhyA.291..383B}
{Bjarle}, C., {Herrstr{\"o}m}, N.-Y., {Jonsson}, G., \& {Kristiansson}, K.
  1979, Zeitschrift fur Physik A Hadrons and Nuclei, 291, 383

\bibitem[{{Blanford} {et~al.}(1969){Blanford}, {Friedlander}, {Klarmann},
  {Walker}, {Wefel}, {Wells}, {Fleischer}, {Nichols}, \&
  {Price}}]{1969PhRvL..23..338B}
{Blanford}, G.~E., {Friedlander}, M.~W., {Klarmann}, J., {et~al.} 1969,
  Physical Review Letters, 23, 338

\bibitem[{{Bleeker} {et~al.}(1970){Bleeker}, {Burger}, {Deerenberg}, {van de
  Hulst}, {Scheepmaker}, {Swanenburg}, \& {Tanaka}}]{1970ICRC....1..209B}
{Bleeker}, J.~A.~M., {Burger}, J.~J., {Deerenberg}, A.~J.~M., {et~al.} 1970, in
  International Cosmic Ray Conference, Vol.~1, International Cosmic Ray
  Conference, 209

\bibitem[{{Bleeker} {et~al.}(1965){Bleeker}, {Burger}, {Scheepmaker},
  {Swanenburg}, \& {Tanaka}}]{1965ICRC....1..327B}
{Bleeker}, J.~A.~M., {Burger}, J.~J., {Scheepmaker}, A., {Swanenburg}, B.~N.,
  \& {Tanaka}, Y. 1965, in International Cosmic Ray Conference, Vol.~1,
  International Cosmic Ray Conference, 327

\bibitem[{{Boezio} {et~al.}(2001{\natexlab{a}}){Boezio}, {Barbiellini},
  {Bonvicini}, {Schiavon}, {Vacchi}, {Zampa}, {Bergstr{\"o}m}, {Carlson},
  {Francke}, {Grinstein}, {Weber}, {Suffert}, {Hof}, {Kremer}, {Menn}, {Simon},
  {Stephens}, {Ambriola}, {Bellotti}, {Cafagna}, {Ciacio}, {Circella}, {de
  Marzo}, {Finetti}, {Papini}, {Piccardi}, {Spillantini}, {Bartalucci},
  {Ricci}, {Grimani}, {Casolino}, {de Pascale}, {Morselli}, {Picozza},
  {Sparvoli}, {Mitchell}, {Ormes}, {Streitmatter}, {Bravar}, \&
  {Stochaj}}]{2001AdSpR..27..669B}
{Boezio}, M., {Barbiellini}, G., {Bonvicini}, V., {et~al.} 2001{\natexlab{a}},
  Advances in Space Research, 27, 669

\bibitem[{{Boezio} {et~al.}(2001{\natexlab{b}}){Boezio}, {Bonvicini},
  {Schiavon}, {Vacchi}, {Zampa}, {Bergstr{\"o}m}, {Carlson}, {Francke},
  {Grinstein}, {Suffert}, {Hof}, {Kremer}, {Menn}, {Simon}, {Stephens},
  {Ambriola}, {Bellotti}, {Cafagna}, {Ciacio}, {Circella}, {De Marzo},
  {Finetti}, {Papini}, {Piccardi}, {Spillantini}, {Vannuccini}, {Bartalucci},
  {Ricci}, {Casolino}, {De Pascale}, {Morselli}, {Picozza}, {Sparvoli},
  {Mitchell}, {Ormes}, {Streitmatter}, {Bravar}, \&
  {Stochaj}}]{2001ApJ...561..787B}
{Boezio}, M., {Bonvicini}, V., {Schiavon}, P., {et~al.} 2001{\natexlab{b}},
  \apj, 561, 787

\bibitem[{{Boezio} {et~al.}(2003){Boezio}, {Bonvicini}, {Schiavon}, {Vacchi},
  {Zampa}, {Bergstr{\"o}m}, {Carlson}, {Francke}, {Hansen}, {Mocchiutti},
  {Suffert}, {Hof}, {Kremer}, {Menn}, {Simon}, {Ambriola}, {Bellotti},
  {Cafagna}, {Ciacio}, {Circella}, {de Marzo}, {Finetti}, {Papini}, {Piccardi},
  {Spillantini}, {Vannuccini}, {Bartalucci}, {Ricci}, {Casolino}, {de Pascale},
  {Morselli}, {Picozza}, {Sparvoli}, {Mitchell}, {Ormes}, {Stephens},
  {Streitmatter}, {Bravar}, \& {Stochaj}}]{2003APh....19..583B}
{Boezio}, M., {Bonvicini}, V., {Schiavon}, P., {et~al.} 2003, Astroparticle
  Physics, 19, 583

\bibitem[{{Boezio} {et~al.}(1997){Boezio}, {Carlson}, {Francke}, {Weber},
  {Suffert}, {Hof}, {Menn}, {Simon}, {Stephens}, {Bellotti}, {Cafagna},
  {Castellano}, {Circella}, {de Cataldo}, {de Marzo}, {Giglietto}, {Spinelli},
  {Bocciolini}, {Papini}, {Perego}, {Piccardi}, {Spillantini}, {Basini},
  {Ricci}, {Codino}, {Finetti}, {Grimani}, {Candusso}, {Casolino}, {de
  Pascale}, {Morselli}, {Picozza}, {Sparvoli}, {Barbiellini}, {Bravar},
  {Schiavon}, {Vacchi}, {Zampa}, {Mitchell}, {Ormes}, {Streitmatter}, {Golden},
  \& {Stochaj}}]{1997ApJ...487..415B}
{Boezio}, M., {Carlson}, P., {Francke}, T., {et~al.} 1997, \apj, 487, 415

\bibitem[{{Boezio} {et~al.}(1999){Boezio}, {Carlson}, {Francke}, {Weber},
  {Suffert}, {Hof}, {Menn}, {Simon}, {Stephens}, {Bellotti}, {Cafagna},
  {Castellano}, {Circella}, {de Marzo}, {Finetti}, {Papini}, {Piccardi},
  {Spillantini}, {Ricci}, {Casolino}, {de Pascale}, {Morselli}, {Picozza},
  {Sparvoli}, {Barbiellini}, {Bravar}, {Schiavon}, {Vacchi}, {Zampa},
  {Mitchell}, {Ormes}, {Streitmatter}, {Golden}, \&
  {Stochaj}}]{1999ApJ...518..457B}
{Boezio}, M., {Carlson}, P., {Francke}, T., {et~al.} 1999, \apj, 518, 457

\bibitem[{{Boezio} {et~al.}(2000){Boezio}, {Carlson}, {Francke}, {Weber},
  {Suffert}, {Hof}, {Menn}, {Simon}, {Stephens}, {Bellotti}, {Cafagna},
  {Castellano}, {Circella}, {De Marzo}, {Finetti}, {Papini}, {Piccardi},
  {Spillantini}, {Ricci}, {Casolino}, {De Pascale}, {Morselli}, {Picozza},
  {Sparvoli}, {Barbiellini}, {Bravar}, {Schiavon}, {Vacchi}, {Zampa},
  {Grimani}, {Mitchell}, {Ormes}, {Streitmatter}, {Golden}, \&
  {Stochaj}}]{2000ApJ...532..653B}
{Boezio}, M., {Carlson}, P., {Francke}, T., {et~al.} 2000, \apj, 532, 653

\bibitem[{{Bogomolov} {et~al.}(1979){Bogomolov}, {Lubyanaya}, {Romanov},
  {Stepanov}, \& {Shulakova}}]{1979ICRC....1..330B}
{Bogomolov}, E.~A., {Lubyanaya}, N.~D., {Romanov}, V.~A., {Stepanov}, S.~V., \&
  {Shulakova}, M.~S. 1979, in International Cosmic Ray Conference, Vol.~1,
  International Cosmic Ray Conference, 330

\bibitem[{{Bogomolov} {et~al.}(1995){Bogomolov}, {Vasilyev}, {Yu}, {Krut'kov},
  {Stepanov}, \& {Shulakova}}]{1995ICRC....2..598B}
{Bogomolov}, E.~A., {Vasilyev}, G.~I., {Yu}, S., {et~al.} 1995, in
  International Cosmic Ray Conference, Vol.~2, International Cosmic Ray
  Conference, 598

\bibitem[{{Buckley} {et~al.}(1994){Buckley}, {Dwyer}, {Mueller}, {Swordy}, \&
  {Tang}}]{1994ApJ...429..736B}
{Buckley}, J., {Dwyer}, J., {Mueller}, D., {Swordy}, S., \& {Tang}, K.~K. 1994,
  \apj, 429, 736

\bibitem[{{Buffington} {et~al.}(1978){Buffington}, {Orth}, \&
  {Mast}}]{1978ApJ...226..355B}
{Buffington}, A., {Orth}, C.~D., \& {Mast}, T.~S. 1978, \apj, 226, 355

\bibitem[{{Buffington} {et~al.}(1974){Buffington}, {Orth}, \&
  {Smoot}}]{1974PhRvL..33...34B}
{Buffington}, A., {Orth}, C.~D., \& {Smoot}, G.~F. 1974, Physical Review
  Letters, 33, 34

\bibitem[{{Buffington} {et~al.}(1975){Buffington}, {Orth}, \&
  {Smoot}}]{1975ApJ...199..669B}
{Buffington}, A., {Orth}, C.~D., \& {Smoot}, G.~F. 1975, \apj, 199, 669

\bibitem[{{Burger} \& {Swanenburg}(1974)}]{1974JGR....79.1533B}
{Burger}, J.~J. \& {Swanenburg}, B.~N. 1974, \jgr, 79, 1533

\bibitem[{{Caballero-Lopez} \& {Moraal}(2004)}]{2004JGRA..109.1101C}
{Caballero-Lopez}, R.~A. \& {Moraal}, H. 2004, Journal of Geophysical Research
  (Space Physics), 109, 1101

\bibitem[{{Caballero-Lopez} {et~al.}(2010){Caballero-Lopez}, {Moraal}, \&
  {McDonald}}]{2010ApJ...725..121C}
{Caballero-Lopez}, R.~A., {Moraal}, H., \& {McDonald}, F.~B. 2010, \apj, 725,
  121

\bibitem[{{Caldwell} {et~al.}(1975){Caldwell}, {Evenson}, {Jordan}, \&
  {Meyer}}]{1975ICRC....3.1000C}
{Caldwell}, J., {Evenson}, P., {Jordan}, S., \& {Meyer}, P. 1975, in
  International Cosmic Ray Conference, Vol.~3, International Cosmic Ray
  Conference, 1000

\bibitem[{{Caldwell} {et~al.}(1977){Caldwell}, {Evenson}, {Jordan}, \&
  {Meyer}}]{1977ICRC...11..203C}
{Caldwell}, J.~H., {Evenson}, P., {Jordan}, S., \& {Meyer}, P. 1977, in
  International Cosmic Ray Conference, Vol.~11, International Cosmic Ray
  Conference, 203

\bibitem[{{Casolino} {et~al.}(2011){Casolino}, {de Santis}, {de Simone},
  {Formato}, {Nikonov}, {Picozza}, \& {PAMELA
  Collaboration}}]{2011ASTRA...7..465C}
{Casolino}, M., {de Santis}, C., {de Simone}, N., {et~al.} 2011, Astrophysics
  and Space Sciences Transactions, 7, 465

\bibitem[{{Chang} {et~al.}(2008){Chang}, {Adams}, {Ahn}, {Bashindzhagyan},
  {Christl}, {Ganel}, {Guzik}, {Isbert}, {Kim}, {Kuznetsov}, {Panasyuk},
  {Panov}, {Schmidt}, {Seo}, {Sokolskaya}, {Watts}, {Wefel}, {Wu}, \&
  {Zatsepin}}]{2008Natur.456..362C}
{Chang}, J., {Adams}, J.~H., {Ahn}, H.~S., {et~al.} 2008, \nat, 456, 362

\bibitem[{{Chardonnet} {et~al.}(1997){Chardonnet}, {Orloff}, \&
  {Salati}}]{1997PhLB..409..313C}
{Chardonnet}, P., {Orloff}, J., \& {Salati}, P. 1997, Physics Letters B, 409,
  313

\bibitem[{{Choutko} \& {Giovacchini}(2008)}]{2008ICRC....4..765C}
{Choutko}, V. \& {Giovacchini}, F. 2008, in International Cosmic Ray
  Conference, Vol.~4, International Cosmic Ray Conference, 765--768

\bibitem[{{Clem} \& {Evenson}(2004)}]{2004JGRA..109.7107C}
{Clem}, J. \& {Evenson}, P. 2004, Journal of Geophysical Research (Space
  Physics), 109, 7107

\bibitem[{{Clem} \& {Evenson}(2009)}]{2009JGRA..11410108C}
{Clem}, J. \& {Evenson}, P. 2009, Journal of Geophysical Research (Space
  Physics), 114, 10108

\bibitem[{{Clem} {et~al.}(1996){Clem}, {Clements}, {Esposito}, {Evenson},
  {Huber}, {L'Heureux}, {Meyer}, \& {Constantin}}]{1996ApJ...464..507C}
{Clem}, J.~M., {Clements}, D.~P., {Esposito}, J., {et~al.} 1996, \apj, 464, 507

\bibitem[{{Clem} {et~al.}(2000){Clem}, {Evenson}, {Huber}, {Pyle}, {Lopate}, \&
  {Simpson}}]{2000JGR...10523099C}
{Clem}, J.~M., {Evenson}, P., {Huber}, D., {et~al.} 2000, \jgr, 105, 23099

\bibitem[{{Clem} \& {Evenson}(2002)}]{2002ApJ...568..216C}
{Clem}, J.~M. \& {Evenson}, P.~A. 2002, \apj, 568, 216

\bibitem[{{Cline} {et~al.}(1964){Cline}, {Ludwig}, \&
  {McDonald}}]{1964PhRvL..13..786C}
{Cline}, T.~L., {Ludwig}, G.~H., \& {McDonald}, F.~B. 1964, Physical Review
  Letters, 13, 786

\bibitem[{{Connell}(1999)}]{1999ICRC....3...33C}
{Connell}, J. 1999, in International Cosmic Ray Conference, Vol.~3,
  International Cosmic Ray Conference, 33

\bibitem[{{Connell}(1998)}]{1998ApJ...501L..59C}
{Connell}, J.~J. 1998, \apjl, 501, L59

\bibitem[{{Connell} {et~al.}(1998){Connell}, {Duvernois}, \&
  {Simpson}}]{1998ApJ...509L..97C}
{Connell}, J.~J., {Duvernois}, M.~A., \& {Simpson}, J.~A. 1998, \apjl, 509, L97

\bibitem[{{Connell} \& {Simpson}(1997)}]{1997ApJ...475L..61C}
{Connell}, J.~J. \& {Simpson}, J.~A. 1997, \apjl, 475, L61

\bibitem[{{Daniel} \& {Stephens}(1965{\natexlab{a}})}]{1965PhRvL..15..769D}
{Daniel}, R.~R. \& {Stephens}, S.~A. 1965{\natexlab{a}}, Physical Review
  Letters, 15, 769

\bibitem[{{Daniel} \& {Stephens}(1965{\natexlab{b}})}]{1965ICRC....1..335D}
{Daniel}, R.~R. \& {Stephens}, S.~A. 1965{\natexlab{b}}, in International
  Cosmic Ray Conference, Vol.~1, International Cosmic Ray Conference, 335

\bibitem[{{Danjo} {et~al.}(1968){Danjo}, {Hayakawa}, {Makino}, \&
  {Tanaka}}]{1968CaJPh..46..530D}
{Danjo}, A., {Hayakawa}, S., {Makino}, F., \& {Tanaka}, Y. 1968, Canadian
  Journal of Physics, 46, 530

\bibitem[{{Daugherty} {et~al.}(1975){Daugherty}, {Hartman}, \&
  {Schmidt}}]{1975ApJ...198..493D}
{Daugherty}, J.~K., {Hartman}, R.~C., \& {Schmidt}, P.~J. 1975, \apj, 198, 493

\bibitem[{{de Nolfo} {et~al.}(2000){de Nolfo}, {Barbier}, {Christian}, {Davis},
  {Golden}, {Hof}, {Krombel}, {Labrador}, {Menn}, {Mewaldt}, {Mitchell},
  {Ormes}, {Rasmussen}, {Reimer}, {Schindler}, {Simon}, {Stochaj},
  {Streitmatter}, \& {Webber}}]{2000AIPC..528..425D}
{de Nolfo}, G.~A., {Barbier}, L.~M., {Christian}, E.~R., {et~al.} 2000, in
  American Institute of Physics Conference Series, Vol. 528, Acceleration and
  Transport of Energetic Particles Observed in the Heliosphere, ed. R.~A.
  {Mewaldt}, J.~R. {Jokipii}, M.~A. {Lee}, E.~{M{\"o}bius}, \& T.~H.
  {Zurbuchen}, 425--428

\bibitem[{{de Nolfo} {et~al.}(2006){de Nolfo}, {Moskalenko}, {Binns},
  {Christian}, {Cummings}, {Davis}, {George}, {Hink}, {Israel}, {Leske},
  {Lijowski}, {Mewaldt}, {Stone}, {Strong}, {von Rosenvinge}, {Wiedenbeck}, \&
  {Yanasak}}]{2006AdSpR..38.1558D}
{de Nolfo}, G.~A., {Moskalenko}, I.~V., {Binns}, W.~R., {et~al.} 2006, Advances
  in Space Research, 38, 1558

\bibitem[{{de Shong} {et~al.}(1964){de Shong}, {Hildebrand}, \&
  {Meyer}}]{1964PhRvL..12....3D}
{de Shong}, J.~A., {Hildebrand}, R.~H., \& {Meyer}, P. 1964, Physical Review
  Letters, 12, 3

\bibitem[{{Derbina} {et~al.}(2005){Derbina}, {Galkin}, {Hareyama}, {Hirakawa},
  {Horiuchi}, {Ichimura}, {Inoue}, {Kamioka}, {Kobayashi}, {Kopenkin},
  {Kuramata}, {Managadze}, {Matsutani}, {Misnikova}, {Mukhamedshin},
  {Nagasawa}, {Nakano}, {Namiki}, {Nakazawa}, {Nanjo}, {Nazarov}, {Ohata},
  {Ohtomo}, {Osedlo}, {Oshuev}, {Publichenko}, {Rakobolskaya}, {Roganova},
  {Saito}, {Sazhina}, {Semba}, {Shibata}, {Shuto}, {Sugimoto}, {Suzuki},
  {Sveshnikova}, {Taran}, {Yajima}, {Yamagami}, {Yashin}, {Zamchalova},
  {Zatsepin}, {Zayarnaya}, \& {Runjob Collaboration}}]{2005ApJ...628L..41D}
{Derbina}, V.~A., {Galkin}, V.~I., {Hareyama}, M., {et~al.} 2005, \apjl, 628,
  L41

\bibitem[{{Diehl} {et~al.}(2003){Diehl}, {Ellithorpe}, {M{\"u}ller}, \&
  {Swordy}}]{2003APh....18..487D}
{Diehl}, E., {Ellithorpe}, D., {M{\"u}ller}, D., \& {Swordy}, S.~P. 2003,
  Astroparticle Physics, 18, 487

\bibitem[{{Duvernois}(1997)}]{1997ApJ...481..241D}
{Duvernois}, M.~A. 1997, \apj, 481, 241

\bibitem[{{DuVernois} {et~al.}(2001){DuVernois}, {Barwick}, {Beatty},
  {Bhattacharyya}, {Bower}, {Chaput}, {Coutu}, {de Nolfo}, {Lowder}, {McKee},
  {M{\"u}ller}, {Musser}, {Nutter}, {Schneider}, {Swordy}, {Tarl{\'e}},
  {Tomasch}, \& {Torbet}}]{2001ApJ...559..296D}
{DuVernois}, M.~A., {Barwick}, S.~W., {Beatty}, J.~J., {et~al.} 2001, \apj,
  559, 296

\bibitem[{{Duvernois} {et~al.}(1996{\natexlab{a}}){Duvernois}, {Garcia-Munoz},
  {Pyle}, {Simpson}, \& {Thayer}}]{1996ApJ...466..457D}
{Duvernois}, M.~A., {Garcia-Munoz}, M., {Pyle}, K.~R., {Simpson}, J.~A., \&
  {Thayer}, M.~R. 1996{\natexlab{a}}, \apj, 466, 457

\bibitem[{{Duvernois} {et~al.}(1996{\natexlab{b}}){Duvernois}, {Simpson}, \&
  {Thayer}}]{1996A&A...316..555D}
{Duvernois}, M.~A., {Simpson}, J.~A., \& {Thayer}, M.~R. 1996{\natexlab{b}},
  \aap, 316, 555

\bibitem[{{Duvernois} \& {Thayer}(1996)}]{1996ApJ...465..982D}
{Duvernois}, M.~A. \& {Thayer}, M.~R. 1996, \apj, 465, 982

\bibitem[{{Dwyer}(1978)}]{1978ApJ...224..691D}
{Dwyer}, R. 1978, \apj, 224, 691

\bibitem[{{Dwyer} \& {Meyer}(1987)}]{1987ApJ...322..981D}
{Dwyer}, R. \& {Meyer}, P. 1987, \apj, 322, 981

\bibitem[{{Earl}(1961)}]{1961PhRvL...6..125E}
{Earl}, J.~A. 1961, Physical Review Letters, 6, 125

\bibitem[{{Earl} {et~al.}(1972){Earl}, {Neely}, \&
  {Rygg}}]{1972JGR....77.1087E}
{Earl}, J.~A., {Neely}, D.~E., \& {Rygg}, T.~A. 1972, \jgr, 77, 1087

\bibitem[{{Engelmann} {et~al.}(1990){Engelmann}, {Ferrando}, {Soutoul},
  {Goret}, \& {Juliusson}}]{1990A&A...233...96E}
{Engelmann}, J.~J., {Ferrando}, P., {Soutoul}, A., {Goret}, P., \& {Juliusson},
  E. 1990, \aap, 233, 96

\bibitem[{{Esposito} {et~al.}(1992){Esposito}, {Christian}, {Balasubrahmanyan},
  {Barbier}, {Ormes}, {Streitmatter}, {Acharya}, {Luzietti}, {Hesse}, \&
  {Heinbach}}]{1992APh.....1...33E}
{Esposito}, J.~A., {Christian}, E.~R., {Balasubrahmanyan}, V.~K., {et~al.}
  1992, Astroparticle Physics, 1, 33

\bibitem[{{Evenson} {et~al.}(1995){Evenson}, {Huber}, {Patterson}, {Esposito},
  {Clements}, \& {Clem}}]{1995JGR...100.7873E}
{Evenson}, P., {Huber}, D., {Patterson}, E.~T., {et~al.} 1995, \jgr, 100, 7873

\bibitem[{{Evenson} {et~al.}(1981){Evenson}, {Krawczyk}, {Moses}, \&
  {Meyer}}]{1981ICRC...10...77E}
{Evenson}, P., {Krawczyk}, L., {Moses}, D., \& {Meyer}, P. 1981, in
  International Cosmic Ray Conference, Vol.~10, International Cosmic Ray
  Conference, 77--80

\bibitem[{{Evenson} \& {Meyer}(1984)}]{1984JGR....89.2647E}
{Evenson}, P. \& {Meyer}, P. 1984, \jgr, 89, 2647

\bibitem[{{Evenson} {et~al.}(1979){Evenson}, {Meyer}, \&
  {Nandkumar}}]{1979ICRC....1..462E}
{Evenson}, P., {Meyer}, P., \& {Nandkumar}, R. 1979, in International Cosmic
  Ray Conference, Vol.~1, International Cosmic Ray Conference, 462

\bibitem[{{Evoli} {et~al.}(2008){Evoli}, {Gaggero}, {Grasso}, \&
  {Maccione}}]{2008JCAP...10..018E}
{Evoli}, C., {Gaggero}, D., {Grasso}, D., \& {Maccione}, L. 2008, \jcap, 10, 18

\bibitem[{{Fan} {et~al.}(1968){Fan}, {Gloeckler}, {Simpson}, \&
  {Verma}}]{1968ApJ...151..737F}
{Fan}, C.~Y., {Gloeckler}, G., {Simpson}, J.~A., \& {Verma}, S.~D. 1968, \apj,
  151, 737

\bibitem[{{Fanselow}(1968)}]{1968ApJ...152..783F}
{Fanselow}, J.~L. 1968, \apj, 152, 783

\bibitem[{{Fanselow} {et~al.}(1969){Fanselow}, {Hartman}, {Hildebrad}, \&
  {Meyer}}]{1969ApJ...158..771F}
{Fanselow}, J.~L., {Hartman}, R.~C., {Hildebrad}, R.~H., \& {Meyer}, P. 1969,
  \apj, 158, 771

\bibitem[{{Fanselow} {et~al.}(1971){Fanselow}, {Hartman}, {Meyer}, \&
  {Schmidt}}]{1971Ap&SS..14..301F}
{Fanselow}, J.~L., {Hartman}, R.~C., {Meyer}, P., \& {Schmidt}, P.~J. 1971,
  \apss, 14, 301

\bibitem[{{Ferrando} {et~al.}(1988){Ferrando}, {Engelmann}, {Goret},
  {Koch-Miramond}, \& {Petrou}}]{1988A&A...193...69F}
{Ferrando}, P., {Engelmann}, J.~J., {Goret}, P., {Koch-Miramond}, L., \&
  {Petrou}, N. 1988, \aap, 193, 69

\bibitem[{{Ferrando} {et~al.}(1991){Ferrando}, {Lal}, {McDonald}, \&
  {Webber}}]{1991A&A...247..163F}
{Ferrando}, P., {Lal}, N., {McDonald}, F.~B., \& {Webber}, W.~R. 1991, \aap,
  247, 163

\bibitem[{{Fisher} {et~al.}(1976){Fisher}, {Hagen}, {Maehl}, {Ormes}, \&
  {Arens}}]{1976ApJ...205..938F}
{Fisher}, A.~J., {Hagen}, F.~A., {Maehl}, R.~C., {Ormes}, J.~F., \& {Arens},
  J.~F. 1976, \apj, 205, 938

\bibitem[{{Fisk}(1971)}]{1971JGR....76..221F}
{Fisk}, L.~A. 1971, \jgr, 76, 221

\bibitem[{{Fisk} \& {Axford}(1969)}]{1969JGR....74.4973F}
{Fisk}, L.~A. \& {Axford}, W.~I. 1969, \jgr, 74, 4973

\bibitem[{{Fowler} {et~al.}(1967){Fowler}, {Adams}, {Cowen}, \&
  {Kidd}}]{1967RSPSA.301...39F}
{Fowler}, P.~H., {Adams}, R.~A., {Cowen}, V.~G., \& {Kidd}, J.~M. 1967, Royal
  Society of London Proceedings Series A, 301, 39

\bibitem[{{Fowler} {et~al.}(1970){Fowler}, {Clapham}, {Cowen}, {Kidd}, \&
  {Moses}}]{1970RSPSA.318....1F}
{Fowler}, P.~H., {Clapham}, V.~M., {Cowen}, V.~G., {Kidd}, J.~M., \& {Moses},
  R.~T. 1970, Royal Society of London Proceedings Series A, 318, 1

\bibitem[{{Freier} {et~al.}(1977){Freier}, {Gilman}, \&
  {Waddington}}]{1977ApJ...213..588F}
{Freier}, P., {Gilman}, C., \& {Waddington}, C.~J. 1977, \apj, 213, 588

\bibitem[{{Freier} {et~al.}(1948{\natexlab{a}}){Freier}, {Lofgren}, {Ney}, \&
  {Oppenheimer}}]{1948PhRv...74.1818F}
{Freier}, P., {Lofgren}, E.~J., {Ney}, E.~P., \& {Oppenheimer}, F.
  1948{\natexlab{a}}, Physical Review, 74, 1818

\bibitem[{{Freier} {et~al.}(1948{\natexlab{b}}){Freier}, {Lofgren}, {Ney},
  {Oppenheimer}, {Bradt}, \& {Peters}}]{1948PhRv...74..213F}
{Freier}, P., {Lofgren}, E.~J., {Ney}, E.~P., {et~al.} 1948{\natexlab{b}},
  Physical Review, 74, 213

\bibitem[{{Freier} {et~al.}(1980){Freier}, {Young}, \&
  {Waddington}}]{1980ApJ...240L..53F}
{Freier}, P.~S., {Young}, J.~S., \& {Waddington}, C.~J. 1980, \apjl, 240, L53

\bibitem[{{Fuke} {et~al.}(2005){Fuke}, {Maeno}, {Abe}, {Haino}, {Makida},
  {Matsuda}, {Matsumoto}, {Mitchell}, {Moiseev}, {Nishimura}, {Nozaki},
  {Orito}, {Ormes}, {Sasaki}, {Seo}, {Shikaze}, {Streitmatter}, {Suzuki},
  {Tanaka}, {Tanizaki}, {Yamagami}, {Yamamoto}, {Yamamoto}, {Yamato},
  {Yoshida}, \& {Yoshimura}}]{2005PhRvL..95h1101F}
{Fuke}, H., {Maeno}, T., {Abe}, K., {et~al.} 2005, Physical Review Letters, 95,
  081101

\bibitem[{{Fulks}(1975)}]{1975JGR....80.1701F}
{Fulks}, G.~J. 1975, \jgr, 80, 1701

\bibitem[{{Garcia-Munoz} {et~al.}(1975){Garcia-Munoz}, {Mason}, \&
  {Simpson}}]{1975ApJ...202..265G}
{Garcia-Munoz}, M., {Mason}, G.~M., \& {Simpson}, J.~A. 1975, \apj, 202, 265

\bibitem[{{Garcia-Munoz} {et~al.}(1977){Garcia-Munoz}, {Mason}, \&
  {Simpson}}]{1977ApJ...217..859G}
{Garcia-Munoz}, M., {Mason}, G.~M., \& {Simpson}, J.~A. 1977, \apj, 217, 859

\bibitem[{{Garcia-Munoz} {et~al.}(1986){Garcia-Munoz}, {Meyer}, {Pyle},
  {Simpson}, \& {Evenson}}]{1986JGR....91.2858G}
{Garcia-Munoz}, M., {Meyer}, P., {Pyle}, K.~R., {Simpson}, J.~A., \& {Evenson},
  P. 1986, \jgr, 91, 2858

\bibitem[{{Garcia-Munoz} {et~al.}(1987){Garcia-Munoz}, {Simpson}, {Guzik},
  {Wefel}, \& {Margolis}}]{1987ApJS...64..269G}
{Garcia-Munoz}, M., {Simpson}, J.~A., {Guzik}, T.~G., {Wefel}, J.~P., \&
  {Margolis}, S.~H. 1987, \apjs, 64, 269

\bibitem[{{Garcia-Munoz} {et~al.}(1979){Garcia-Munoz}, {Simpson}, \&
  {Wefel}}]{1979ApJ...232L..95G}
{Garcia-Munoz}, M., {Simpson}, J.~A., \& {Wefel}, J.~P. 1979, \apjl, 232, L95

\bibitem[{{Garcia-Munoz} {et~al.}(1981){Garcia-Munoz}, {Simpson}, \&
  {Wefel}}]{1981ICRC....2...72G}
{Garcia-Munoz}, M., {Simpson}, J.~A., \& {Wefel}, J.~P. 1981, in International
  Cosmic Ray Conference, Vol.~2, International Cosmic Ray Conference, 72--75

\bibitem[{{George} {et~al.}(2009){George}, {Lave}, {Wiedenbeck}, {Binns},
  {Cummings}, {Davis}, {de Nolfo}, {Hink}, {Israel}, {Leske}, {Mewaldt},
  {Scott}, {Stone}, {von Rosenvinge}, \& {Yanasak}}]{2009ApJ...698.1666G}
{George}, J.~S., {Lave}, K.~A., {Wiedenbeck}, M.~E., {et~al.} 2009, \apj, 698,
  1666

\bibitem[{{Gibner} {et~al.}(1992){Gibner}, {Mewaldt}, {Schindler}, {Stone}, \&
  {Webber}}]{1992ApJ...391L..89G}
{Gibner}, P.~S., {Mewaldt}, R.~A., {Schindler}, S.~M., {Stone}, E.~C., \&
  {Webber}, W.~R. 1992, \apjl, 391, L89

\bibitem[{{Gleeson} \& {Axford}(1967)}]{1967ApJ...149L.115G}
{Gleeson}, L.~J. \& {Axford}, W.~I. 1967, \apjl, 149, L115

\bibitem[{{Gleeson} \& {Axford}(1968)}]{1968ApJ...154.1011G}
{Gleeson}, L.~J. \& {Axford}, W.~I. 1968, \apj, 154, 1011

\bibitem[{{Golden} {et~al.}(1994){Golden}, {Grimani}, {Kimbell}, {Stephens},
  {Stochaj}, {Webber}, {Basini}, {Bongiorno}, {Massimo Brancaccio}, {Ricci},
  {Ormes}, {Streitmatter}, {Papini}, {Spillantini}, {Brunetti}, {Codino},
  {Menichelli}, {Salvatori}, {de Pascale}, {Morselli}, \&
  {Picozza}}]{1994ApJ...436..769G}
{Golden}, R.~L., {Grimani}, C., {Kimbell}, B.~L., {et~al.} 1994, \apj, 436, 769

\bibitem[{{Golden} {et~al.}(1979){Golden}, {Horan}, {Mauger}, {Badhwar},
  {Lacy}, {Stephens}, {Daniel}, \& {Zipse}}]{1979PhRvL..43.1196G}
{Golden}, R.~L., {Horan}, S., {Mauger}, B.~G., {et~al.} 1979, Physical Review
  Letters, 43, 1196

\bibitem[{{Golden} {et~al.}(1984){Golden}, {Mauger}, {Badhwar}, {Daniel},
  {Lacy}, {Stephens}, \& {Zipse}}]{1984ApJ...287..622G}
{Golden}, R.~L., {Mauger}, B.~G., {Badhwar}, G.~D., {et~al.} 1984, \apj, 287,
  622

\bibitem[{{Golden} {et~al.}(1987){Golden}, {Mauger}, {Horan}, {Stephens},
  {Daniel}, {Badhwar}, {Lacy}, \& {Zipse}}]{1987A&A...188..145G}
{Golden}, R.~L., {Mauger}, B.~G., {Horan}, S., {et~al.} 1987, \aap, 188, 145

\bibitem[{{Golden} {et~al.}(1996){Golden}, {Stochaj}, {Stephens}, {Aversa},
  {Barbiellini}, {Boezio}, {Bravar}, {Colavita}, {Fratnik}, {Schiavon},
  {Vacchi}, {Zampa}, {Mitchell}, {Ormes}, {Steitmatter}, {Bellotti}, {Cafagna},
  {Castellano}, {Circella}, {de Cataldo}, {de Marzo}, {Giglietto},
  {Marangelli}, {Raino}, {Spinelli}, {Bocciolini}, {Finetti}, {Papini},
  {Perego}, {Piccardi}, {Spillantini}, {Basini}, {Massimo Brancaccio}, {Ricci},
  {Brunetti}, {Codino}, {Grimani}, {Menichelli}, {Bidoli}, {Candusso},
  {Casolino}, {de Pascale}, {Morselli}, {Picozza}, {Sparvoli}, {Hof}, {Menn},
  \& {Simon}}]{1996ApJ...457L.103G}
{Golden}, R.~L., {Stochaj}, S.~J., {Stephens}, S.~A., {et~al.} 1996, \apjl,
  457, L103

\bibitem[{{Grimani} {et~al.}(2002){Grimani}, {Stephens}, {Cafagna}, {Basini},
  {Bellotti}, {Brunetti}, {Circella}, {Codino}, {De Marzo}, {De Pascale},
  {Finetti}, {Golden}, {Hof}, {Menn}, {Mitchell}, {Morselli}, {Ormes},
  {Papini}, {Pfeifer}, {Piccardi}, {Picozza}, {Ricci}, {Simon}, {Spillantini},
  {Stochaj}, \& {Streitmatter}}]{2002A&A...392..287G}
{Grimani}, C., {Stephens}, S.~A., {Cafagna}, F.~S., {et~al.} 2002, \aap, 392,
  287

\bibitem[{{Guzik}(1981)}]{1981ApJ...244..695G}
{Guzik}, T.~G. 1981, \apj, 244, 695

\bibitem[{{Hagen} {et~al.}(1977){Hagen}, {Fisher}, \&
  {Ormes}}]{1977ApJ...212..262H}
{Hagen}, F.~A., {Fisher}, A.~J., \& {Ormes}, J.~F. 1977, \apj, 212, 262

\bibitem[{{Haino} {et~al.}(2005){Haino}, {Abe}, {Fuke}, {Maeno}, {Makida},
  {Matsumoto}, {Mitchell}, {Moiseev}, {Nishimura}, {Nozaki}, {Orito}, {Ormes},
  {Sanuki}, {Sasaki}, {Seo}, {Shikaze}, {Streitmatter}, {Suzuki}, {Tanaka},
  {Yamagami}, {Yamamoto}, {Yoshida}, \& {Yoshimura}}]{2005ICRC....3...13H}
{Haino}, S., {Abe}, K., {Fuke}, H., {et~al.} 2005, in International Cosmic Ray
  Conference, Vol.~3, International Cosmic Ray Conference, 13

\bibitem[{{Haino} {et~al.}(2004){Haino}, {Sanuki}, {Abe}, {Anraku}, {Asaoka},
  {Fuke}, {Imori}, {Itasaki}, {Maeno}, {Makida}, {Matsuda}, {Matsui},
  {Matsumoto}, {Mitchell}, {Moiseev}, {Nishimura}, {Nozaki}, {Orito}, {Ormes},
  {Sasaki}, {Seo}, {Shikaze}, {Streitmatter}, {Suzuki}, {Takasugi}, {Tanaka},
  {Tanizaki}, {Yamagami}, {Yamamoto}, {Yamamoto}, {Yamato}, {Yoshida}, \&
  {Yoshimura}}]{2004PhLB..594...35H}
{Haino}, S., {Sanuki}, T., {Abe}, K., {et~al.} 2004, Physics Letters B, 594, 35

\bibitem[{{Hams} {et~al.}(2004){Hams}, {Barbier}, {Bremerich}, {Christian}, {de
  Nolfo}, {Geier}, {G{\"o}bel}, {Gupta}, {Hof}, {Menn}, {Mewaldt}, {Mitchell},
  {Schindler}, {Simon}, \& {Streitmatter}}]{2004ApJ...611..892H}
{Hams}, T., {Barbier}, L.~M., {Bremerich}, M., {et~al.} 2004, \apj, 611, 892

\bibitem[{{Hartman}(1967)}]{1967ApJ...150..371H}
{Hartman}, R.~C. 1967, \apj, 150, 371

\bibitem[{{Hartman} {et~al.}(1965){Hartman}, {Meyer}, \&
  {Hildebrand}}]{1965JGR....70.2713H}
{Hartman}, R.~C., {Meyer}, P., \& {Hildebrand}, R.~H. 1965, \jgr, 70, 2713

\bibitem[{{Hartman} \& {Pellerin}(1976)}]{1976ApJ...204..927H}
{Hartman}, R.~C. \& {Pellerin}, C.~J. 1976, \apj, 204, 927

\bibitem[{{Hartmann} {et~al.}(1977){Hartmann}, {Mueller}, \&
  {Prince}}]{1977PhRvL..38.1368H}
{Hartmann}, G., {Mueller}, D., \& {Prince}, T. 1977, Physical Review Letters,
  38, 1368

\bibitem[{{Hatano} {et~al.}(1995){Hatano}, {Fukada}, {Saito}, {Oda}, \&
  {Yanagita}}]{1995PhRvD..52.6219H}
{Hatano}, Y., {Fukada}, Y., {Saito}, T., {Oda}, H., \& {Yanagita}, T. 1995,
  \prd, 52, 6219

\bibitem[{{Hesse} {et~al.}(1996){Hesse}, {Acharya}, {Heinbach}, {Heinrich},
  {Henkel}, {Luzietti}, {Simon}, {Christian}, {Esposito}, {Balasubrahmanyan},
  {Barbier}, {Ormes}, \& {Streitmatter}}]{1996A&A...314..785H}
{Hesse}, A., {Acharya}, B.~S., {Heinbach}, U., {et~al.} 1996, \aap, 314, 785

\bibitem[{{Hof} {et~al.}(1996){Hof}, {Menn}, {Pfeifer}, {Simon}, {Golden},
  {Stochaj}, {Stephens}, {Basini}, {Ricci}, {Brancaccio}, {Papini}, {Piccardi},
  {Spillantini}, {de Pascale}, {Morselli}, {Picozza}, {Brunetti}, {Codino},
  {Grimani}, {Menichelli}, {Mitchell}, {Ormes}, \&
  {Streitmatter}}]{1996ApJ...467L..33H}
{Hof}, M., {Menn}, W., {Pfeifer}, C., {et~al.} 1996, \apjl, 467, L33

\bibitem[{{Hovestadt} {et~al.}(1971){Hovestadt}, {Meyer}, \&
  {Schmidt}}]{1971ApL.....9..165H}
{Hovestadt}, D., {Meyer}, P., \& {Schmidt}, P.~J. 1971, \aplett, 9, 165

\bibitem[{{Hsieh} {et~al.}(1971){Hsieh}, {Mason}, \&
  {Simpson}}]{1971ApJ...166..221H}
{Hsieh}, K.~C., {Mason}, G.~M., \& {Simpson}, J.~A. 1971, \apj, 166, 221

\bibitem[{{Ichimura} {et~al.}(1993){Ichimura}, {Kogawa}, {Kuramata}, {Mito},
  {Murabayashi}, {Nanjo}, {Nakamura}, {Ohba}, {Ohuchi}, {Ozawa}, {Yamada},
  {Matsutani}, {Watanabe}, {Kamioka}, {Kirii}, {Kitazawa}, {Kobayashi},
  {Mihashi}, {Shibata}, {Shibuta}, {Sugimoto}, \&
  {Nakazawa}}]{1993PhRvD..48.1949I}
{Ichimura}, M., {Kogawa}, M., {Kuramata}, S., {et~al.} 1993, \prd, 48, 1949

\bibitem[{{Ishii} {et~al.}(1973){Ishii}, {Kobayashi}, {Shigihara}, {Yokoi},
  {Matsuo}, {Nishimura}, {Taira}, \& {Niu}}]{1973ICRC....5.3073I}
{Ishii}, C., {Kobayashi}, T., {Shigihara}, N., {et~al.} 1973, in International
  Cosmic Ray Conference, Vol.~5, International Cosmic Ray Conference, 3073

\bibitem[{{Israel} \& {Vogt}(1968)}]{1968PhRvL..20.1053I}
{Israel}, M.~H. \& {Vogt}, R.~E. 1968, Physical Review Letters, 20, 1053

\bibitem[{{Ivanenko} {et~al.}(1993){Ivanenko}, {Shestoperov}, {Chikova},
  {Fateeva}, {Khein}, {Podoroznyi}, {Rapoport}, {Samsonov}, {Sobinyakov},
  {Turundaevskyi}, \& {Yashin}}]{1993ICRC....2...17I}
{Ivanenko}, I.~P., {Shestoperov}, V.~Y., {Chikova}, L.~O., {et~al.} 1993, in
  International Cosmic Ray Conference, Vol.~2, International Cosmic Ray
  Conference, 17

\bibitem[{{Jokipii} \& {Kopriva}(1979)}]{1979ApJ...234..384J}
{Jokipii}, J.~R. \& {Kopriva}, D.~A. 1979, \apj, 234, 384

\bibitem[{{Jones} {et~al.}(2001){Jones}, {Lukasiak}, {Ptuskin}, \&
  {Webber}}]{2001ApJ...547..264J}
{Jones}, F.~C., {Lukasiak}, A., {Ptuskin}, V., \& {Webber}, W. 2001, \apj, 547,
  264

\bibitem[{{Jordan}(1985)}]{1985ApJ...291..207J}
{Jordan}, S.~P. 1985, \apj, 291, 207

\bibitem[{{Juliusson}(1974)}]{1974ApJ...191..331J}
{Juliusson}, E. 1974, \apj, 191, 331

\bibitem[{{Kamioka} {et~al.}(1997){Kamioka}, {Hareyama}, {Ichimura},
  {Ishihara}, {Kobayashi}, {Komatsu}, {Kuramata}, {Maruguchi}, {Matsutani},
  {Mihashi}, {Mito}, {Nakamura}, {Nanjo}, {Ouchi}, {Ozawa}, {Shibata},
  {Sugimoto}, \& {Watanabe}}]{1997APh.....6..155K}
{Kamioka}, E., {Hareyama}, M., {Ichimura}, M., {et~al.} 1997, Astroparticle
  Physics, 6, 155

\bibitem[{{Kobayashi}(1999)}]{1999ICRC....3...61K}
{Kobayashi}, T. 1999, in International Cosmic Ray Conference, Vol.~3,
  International Cosmic Ray Conference, 61

\bibitem[{{Kobayashi} {et~al.}(2012){Kobayashi}, {Komori}, {Yoshida},
  {Yanagisawa}, {Nishimura}, {Yamagami}, {Saito}, {Tateyama}, {Yuda}, \&
  {Wilkes}}]{2012ApJ...760..146K}
{Kobayashi}, T., {Komori}, Y., {Yoshida}, K., {et~al.} 2012, \apj, 760, 146

\bibitem[{{Kraushaar} {et~al.}(1972){Kraushaar}, {Clark}, {Garmire}, {Borken},
  {Higbie}, {Leong}, \& {Thorsos}}]{1972ApJ...177..341K}
{Kraushaar}, W.~L., {Clark}, G.~W., {Garmire}, G.~P., {et~al.} 1972, \apj, 177,
  341

\bibitem[{{Kroeger}(1986)}]{1986ApJ...303..816K}
{Kroeger}, R. 1986, \apj, 303, 816

\bibitem[{{Krombel} \& {Wiedenbeck}(1988)}]{1988ApJ...328..940K}
{Krombel}, K.~E. \& {Wiedenbeck}, M.~E. 1988, \apj, 328, 940

\bibitem[{{Lavalle} \& {Salati}(2012)}]{2012CRPhy..13..740L}
{Lavalle}, J. \& {Salati}, P. 2012, Comptes Rendus Physique, 13, 740

\bibitem[{{Leech} \& {Ogallagher}(1978)}]{1978ApJ...221.1110L}
{Leech}, H.~W. \& {Ogallagher}, J.~J. 1978, \apj, 221, 1110

\bibitem[{{Leske}(1993)}]{1993ApJ...405..567L}
{Leske}, R.~A. 1993, \apj, 405, 567

\bibitem[{{Lezniak} \& {Webber}(1978)}]{1978ApJ...223..676L}
{Lezniak}, J.~A. \& {Webber}, W.~R. 1978, \apj, 223, 676

\bibitem[{{L'Heureux}(1967)}]{1967ApJ...148..399L}
{L'Heureux}, J. 1967, \apj, 148, 399

\bibitem[{{L'Heureux} {et~al.}(1972){L'Heureux}, {Fan}, \&
  {Meyer}}]{1972ApJ...171..363L}
{L'Heureux}, J., {Fan}, C.~Y., \& {Meyer}, P. 1972, \apj, 171, 363

\bibitem[{{L'Heureux} \& {Meyer}(1968)}]{1968CaJPS..46..892L}
{L'Heureux}, J. \& {Meyer}, P. 1968, Canadian Journal of Physics Supplement,
  46, 892

\bibitem[{{Lukasiak}(1999)}]{1999ICRC....3...41L}
{Lukasiak}, A. 1999, in International Cosmic Ray Conference, Vol.~3,
  International Cosmic Ray Conference, 41

\bibitem[{{Lukasiak} {et~al.}(1994{\natexlab{a}}){Lukasiak}, {Ferrando},
  {McDonald}, \& {Webber}}]{1994ApJ...426..366L}
{Lukasiak}, A., {Ferrando}, P., {McDonald}, F.~B., \& {Webber}, W.~R.
  1994{\natexlab{a}}, \apj, 426, 366

\bibitem[{{Lukasiak} {et~al.}(1994{\natexlab{b}}){Lukasiak}, {Ferrando},
  {McDonald}, \& {Webber}}]{1994ApJ...423..426L}
{Lukasiak}, A., {Ferrando}, P., {McDonald}, F.~B., \& {Webber}, W.~R.
  1994{\natexlab{b}}, \apj, 423, 426

\bibitem[{{Lukasiak} {et~al.}(1994{\natexlab{c}}){Lukasiak}, {McDonald}, \&
  {Webber}}]{1994ApJ...430L..69L}
{Lukasiak}, A., {McDonald}, F.~B., \& {Webber}, W.~R. 1994{\natexlab{c}},
  \apjl, 430, L69

\bibitem[{{Lukasiak} {et~al.}(1997{\natexlab{a}}){Lukasiak}, {McDonald}, \&
  {Webber}}]{1997ICRC....3..389L}
{Lukasiak}, A., {McDonald}, F.~B., \& {Webber}, W.~R. 1997{\natexlab{a}}, in
  International Cosmic Ray Conference, Vol.~3, International Cosmic Ray
  Conference, 389

\bibitem[{{Lukasiak} {et~al.}(1997{\natexlab{b}}){Lukasiak}, {McDonald}, \&
  {Webber}}]{1997ApJ...488..454L}
{Lukasiak}, A., {McDonald}, F.~B., \& {Webber}, W.~R. 1997{\natexlab{b}}, \apj,
  488, 454

\bibitem[{{Maeno} {et~al.}(2001){Maeno}, {Orito}, {Matsunaga}, {Abe}, {Anraku},
  {Asaoka}, {Fujikawa}, {Imori}, {Makida}, {Matsui}, {Matsumoto}, {Mitchell},
  {Mitsui}, {Moiseev}, {Motoki}, {Nishimura}, {Nozaki}, {Ormes}, {Saeki},
  {Sanuki}, {Sasaki}, {Seo}, {Shikaze}, {Sonoda}, {Streitmatter}, {Suzuki},
  {Tanaka}, {Ueda}, {Wang}, {Yajima}, {Yamagami}, {Yamamoto}, {Yoshida}, \&
  {Yoshimura}}]{2001APh....16..121M}
{Maeno}, T., {Orito}, S., {Matsunaga}, H., {et~al.} 2001, Astroparticle
  Physics, 16, 121

\bibitem[{{Matsunaga} {et~al.}(1998){Matsunaga}, {Orito}, {Matsumoto},
  {Yoshimura}, {Moiseev}, {Anraku}, {Golden}, {Imori}, {Makida}, {Mitchell},
  {Motoki}, {Nishimura}, {Nozaki}, {Ormes}, {Saeki}, {Sanuki}, {Streitmatter},
  {Suzuki}, {Tanaka}, {Ueda}, {Yajima}, {Yamagami}, {Yamamoto}, \&
  {Yoshida}}]{1998PhRvL..81.4052M}
{Matsunaga}, H., {Orito}, S., {Matsumoto}, H., {et~al.} 1998, Physical Review
  Letters, 81, 4052

\bibitem[{{Maurin} {et~al.}(2014){Maurin}, {Cheminet}, {Derome}, {Ghelfi}, \&
  {Hubert}}]{2014arXiv1403.1612M}
{Maurin}, D., {Cheminet}, A., {Derome}, L., {Ghelfi}, A., \& {Hubert}, G. 2014,
  ArXiv e-prints

\bibitem[{{Maurin} {et~al.}(2001){Maurin}, {Donato}, {Taillet}, \&
  {Salati}}]{2001ApJ...555..585M}
{Maurin}, D., {Donato}, F., {Taillet}, R., \& {Salati}, P. 2001, \apj, 555, 585

\bibitem[{{Meegan} \& {Earl}(1975)}]{1975ApJ...197..219M}
{Meegan}, C.~A. \& {Earl}, J.~A. 1975, \apj, 197, 219

\bibitem[{{Menn} {et~al.}(2000){Menn}, {Hof}, {Reimer}, {Simon}, {Davis},
  {Labrador}, {Mewaldt}, {Schindler}, {Barbier}, {Christian}, {Krombel},
  {Krizmanic}, {Mitchell}, {Ormes}, {Streitmatter}, {Golden}, {Stochaj},
  {Webber}, \& {Rasmussen}}]{2000ApJ...533..281M}
{Menn}, W., {Hof}, M., {Reimer}, O., {et~al.} 2000, \apj, 533, 281

\bibitem[{{Mewaldt}(1986)}]{1986ApJ...311..979M}
{Mewaldt}, R.~A. 1986, \apj, 311, 979

\bibitem[{{Mewaldt} {et~al.}(1980{\natexlab{a}}){Mewaldt}, {Spalding}, {Stone},
  \& {Vogt}}]{1980ApJ...235L..95M}
{Mewaldt}, R.~A., {Spalding}, J.~D., {Stone}, E.~C., \& {Vogt}, R.~E.
  1980{\natexlab{a}}, \apjl, 235, L95

\bibitem[{{Mewaldt} {et~al.}(1980{\natexlab{b}}){Mewaldt}, {Spalding}, {Stone},
  \& {Vogt}}]{1980ApJ...236L.121M}
{Mewaldt}, R.~A., {Spalding}, J.~D., {Stone}, E.~C., \& {Vogt}, R.~E.
  1980{\natexlab{b}}, \apjl, 236, L121

\bibitem[{{Mewaldt} {et~al.}(1981){Mewaldt}, {Spalding}, {Stone}, \&
  {Vogt}}]{1981ApJ...251L..27M}
{Mewaldt}, R.~A., {Spalding}, J.~D., {Stone}, E.~C., \& {Vogt}, R.~E. 1981,
  \apjl, 251, L27

\bibitem[{{Meyer} \& {Vogt}(1961)}]{1961PhRvL...6..193M}
{Meyer}, P. \& {Vogt}, R. 1961, Physical Review Letters, 6, 193

\bibitem[{{Minagawa}(1981)}]{1981ApJ...248..847M}
{Minagawa}, G. 1981, \apj, 248, 847

\bibitem[{{Mitchell} {et~al.}(1996){Mitchell}, {Barbier}, {Christian},
  {Krizmanic}, {Krombel}, {Ormes}, {Streitmatter}, {Labrador}, {Davis},
  {Mewaldt}, {Schindler}, {Golden}, {Stochaj}, {Webber}, {Menn}, {Hof},
  {Reimer}, {Simon}, \& {Rasmussen}}]{1996PhRvL..76.3057M}
{Mitchell}, J.~W., {Barbier}, L.~M., {Christian}, E.~R., {et~al.} 1996,
  Physical Review Letters, 76, 3057

\bibitem[{{Mocchiutti} \& {Wizard/Caprice
  Collaboration}(2003)}]{2003ICRC....4.1809M}
{Mocchiutti}, E. \& {Wizard/Caprice Collaboration}. 2003, in International
  Cosmic Ray Conference, Vol.~4, International Cosmic Ray Conference, 1809

\bibitem[{{Moiseev} {et~al.}(1997){Moiseev}, {Yoshimura}, {Ueda}, {Anraku},
  {Golden}, {Imori}, {Inaba}, {Kimball}, {Kimura}, {Makida}, {Matsumoto},
  {Matsunaga}, {Mitchell}, {Motoki}, {Nishimura}, {Nozaki}, {Orito}, {Ormes},
  {Saeki}, {Seo}, {Stochaj}, {Streitmatter}, {Suzuki}, {Tanaka}, {Yajima},
  {Yamagami}, {Yamamoto}, {Yoshida}, \& {BESS
  Collaboration}}]{1997ApJ...474..479M}
{Moiseev}, A., {Yoshimura}, K., {Ueda}, I., {et~al.} 1997, \apj, 474, 479

\bibitem[{{Mueller} {et~al.}(1991){Mueller}, {Swordy}, {Meyer}, {L'Heureux}, \&
  {Grunsfeld}}]{1991ApJ...374..356M}
{Mueller}, D., {Swordy}, S.~P., {Meyer}, P., {L'Heureux}, J., \& {Grunsfeld},
  J.~M. 1991, \apj, 374, 356

\bibitem[{{Mueller} \& {Tang}(1987)}]{1987ApJ...312..183M}
{Mueller}, D. \& {Tang}, K.-K. 1987, \apj, 312, 183

\bibitem[{{Muller} \& {Meyer}(1973)}]{1973ApJ...186..841M}
{Muller}, D. \& {Meyer}, P. 1973, \apj, 186, 841

\bibitem[{{Myers} {et~al.}(2003){Myers}, {Seo}, {Abe}, {Anraku}, {Imori},
  {Maeno}, {Makida}, {Matsumoto}, {Mitchell}, {Moiseev}, {Nishimura}, {Nozaki},
  {Ormes}, {Orito}, {Sanuki}, {Sasaki}, {Shizake}, {Streitmatter}, {Suzuki},
  {Tanaka}, {Yamagami}, {Yamamoto}, {Yoshida}, \&
  {Yoshimura}}]{2003ICRC....4.1805M}
{Myers}, Z.~D., {Seo}, E.~S., {Abe}, K., {et~al.} 2003, in International Cosmic
  Ray Conference, Vol.~4, International Cosmic Ray Conference, 1805

\bibitem[{{Nishimura} {et~al.}(1980){Nishimura}, {Fujii}, {Taira}, {Aizu},
  {Hiraiwa}, {Kobayashi}, {Niu}, {Ohta}, {Golden}, \&
  {Koss}}]{1980ApJ...238..394N}
{Nishimura}, J., {Fujii}, M., {Taira}, T., {et~al.} 1980, \apj, 238, 394

\bibitem[{{Nishimura} {et~al.}(1985){Nishimura}, {Fujii}, {Yoshida}, {Taira},
  {Aizu}, {Nomura}, {Kobayashi}, {Kazuno}, {Nishio}, {Golden}, {Koss}, {Lord},
  \& {Wilkes}}]{1985ICRC....9..539N}
{Nishimura}, J., {Fujii}, M., {Yoshida}, A., {et~al.} 1985, in International
  Cosmic Ray Conference, Vol.~9, International Cosmic Ray Conference, ed. F.~C.
  {Jones}, 539

\bibitem[{{Nishimura} {et~al.}(2001){Nishimura}, {Kobayashi}, {Komori},
  {Shirai}, {Tateyama}, \& {Yoshida}}]{2001AdSpR..26.1827N}
{Nishimura}, J., {Kobayashi}, T., {Komori}, Y., {et~al.} 2001, Advances in
  Space Research, 26, 1827

\bibitem[{{Obermeier} {et~al.}(2011){Obermeier}, {Ave}, {Boyle}, {H{\"o}ppner},
  {H{\"o}randel}, \& {M{\"u}ller}}]{2011ApJ...742...14O}
{Obermeier}, A., {Ave}, M., {Boyle}, P., {et~al.} 2011, \apj, 742, 14

\bibitem[{{Obermeier} {et~al.}(2012){Obermeier}, {Boyle}, {H{\"o}randel}, \&
  {M{\"u}ller}}]{2012ApJ...752...69O}
{Obermeier}, A., {Boyle}, P., {H{\"o}randel}, J., \& {M{\"u}ller}, D. 2012,
  \apj, 752, 69

\bibitem[{{Orito} {et~al.}(2000){Orito}, {Maeno}, {Matsunaga}, {Abe}, {Anraku},
  {Asaoka}, {Fujikawa}, {Imori}, {Ishino}, {Makida}, {Matsui}, {Matsumoto},
  {Mitchell}, {Mitsui}, {Moiseev}, {Motoki}, {Nishimura}, {Nozaki}, {Ormes},
  {Saeki}, {Sanuki}, {Sasaki}, {Seo}, {Shikaze}, {Sonoda}, {Streitmatter},
  {Suzuki}, {Tanaka}, {Ueda}, {Yajima}, {Yamagami}, {Yamamoto}, {Yoshida}, \&
  {Yoshimura}}]{2000PhRvL..84.1078O}
{Orito}, S., {Maeno}, T., {Matsunaga}, H., {et~al.} 2000, Physical Review
  Letters, 84, 1078

\bibitem[{{Orth} {et~al.}(1978){Orth}, {Buffington}, {Smoot}, \&
  {Mast}}]{1978ApJ...226.1147O}
{Orth}, C.~D., {Buffington}, A., {Smoot}, G.~F., \& {Mast}, T.~S. 1978, \apj,
  226, 1147

\bibitem[{{O'Sullivan} {et~al.}(1971){O'Sullivan}, {Price}, {Shirk}, {Fowler},
  {Kidd}, {Kobetich}, \& {Thorne}}]{1971PhRvL..26..463O}
{O'Sullivan}, D., {Price}, P.~B., {Shirk}, E.~K., {et~al.} 1971, Physical
  Review Letters, 26, 463

\bibitem[{{Panov} {et~al.}(2009){Panov}, {Adams}, {Ahn}, {Bashinzhagyan},
  {Watts}, {Wefel}, {Wu}, {Ganel}, {Guzik}, {Zatsepin}, {Isbert}, {Kim},
  {Christl}, {Kouznetsov}, {Panasyuk}, {Seo}, {Sokolskaya}, {Chang}, {Schmidt},
  \& {Fazely}}]{2009BRASP..73..564P}
{Panov}, A.~D., {Adams}, J.~H., {Ahn}, H.~S., {et~al.} 2009, Bulletin of the
  Russian Academy of Science, Phys., 73, 564

\bibitem[{{Panov} {et~al.}(2008){Panov}, {Sokolskaya}, {Adams}, \& {et
  al.}}]{2008ICRC....2....3P}
{Panov}, A.~D., {Sokolskaya}, N.~V., {Adams}, Jr., J.~H., \& {et al.} 2008, in
  International Cosmic Ray Conference, Vol.~2, International Cosmic Ray
  Conference, 3--6

\bibitem[{{Papini} {et~al.}(2004){Papini}, {Piccardi}, {Spillantini},
  {Vannuccini}, {Ambriola}, {Bellotti}, {Cafagna}, {Ciacio}, {Circella}, {De
  Marzo}, {Bartalucci}, {Ricci}, {Bergstr{\"o}m}, {Carlson}, {Francke},
  {Hansen}, {Mocchiutti}, {Boezio}, {Bonvicini}, {Schiavon}, {Vacchi}, {Zampa},
  {Bravar}, {Stochaj}, {Casolino}, {De Pascale}, {Morselli}, {Picozza},
  {Sparvoli}, {Hof}, {Kremer}, {Menn}, {Simon}, {Mitchell}, {Ormes},
  {Stephens}, {Streitmatter}, \& {Suffert}}]{2004ApJ...615..259P}
{Papini}, P., {Piccardi}, S., {Spillantini}, P., {et~al.} 2004, \apj, 615, 259

\bibitem[{{Parker}(1958{\natexlab{a}})}]{1958PhRv..110.1445P}
{Parker}, E.~N. 1958{\natexlab{a}}, Physical Review, 110, 1445

\bibitem[{{Parker}(1958{\natexlab{b}})}]{1958PhRv..109.1874P}
{Parker}, E.~N. 1958{\natexlab{b}}, Physical Review, 109, 1874

\bibitem[{{Perko}(1987)}]{1987A&A...184..119P}
{Perko}, J.~S. 1987, \aap, 184, 119

\bibitem[{{Porter} {et~al.}(2011){Porter}, {Johnson}, \&
  {Graham}}]{2011ARA&A..49..155P}
{Porter}, T.~A., {Johnson}, R.~P., \& {Graham}, P.~W. 2011, \araa, 49, 155

\bibitem[{{Potgieter}(2013)}]{2013LRSP...10....3P}
{Potgieter}, M. 2013, Living Reviews in Solar Physics, 10, 3

\bibitem[{{Potgieter} \& {Moraal}(1985)}]{1985ApJ...294..425P}
{Potgieter}, M.~S. \& {Moraal}, H. 1985, \apj, 294, 425

\bibitem[{{Price} {et~al.}(1971){Price}, {Fowler}, {Kidd}, {Kobetich},
  {Fleischer}, \& {Nichols}}]{1971PhRvD...3..815P}
{Price}, P.~B., {Fowler}, P.~H., {Kidd}, J.~M., {et~al.} 1971, \prd, 3, 815

\bibitem[{{Prince}(1979)}]{1979ApJ...227..676P}
{Prince}, T.~A. 1979, \apj, 227, 676

\bibitem[{{Rastoin} {et~al.}(1996){Rastoin}, {Ferrando}, {Raviart}, {Ducros},
  {Petrucci}, {Paizis}, {Kunow}, {Mueller-Mellin}, {Sierks}, \&
  {Wibberenz}}]{1996A&A...307..981R}
{Rastoin}, C., {Ferrando}, P., {Raviart}, A., {et~al.} 1996, \aap, 307, 981

\bibitem[{{Reimer} {et~al.}(1998){Reimer}, {Menn}, {Hof}, {Simon}, {Davis},
  {Labrador}, {Mewaldt}, {Schindler}, {Barbier}, {Christian}, {Krombel},
  {Mitchell}, {Ormes}, {Streitmatter}, {Golden}, {Stochaj}, {Webber}, \&
  {Rasmussen}}]{1998ApJ...496..490R}
{Reimer}, O., {Menn}, W., {Hof}, M., {et~al.} 1998, \apj, 496, 490

\bibitem[{{Sanuki} {et~al.}(2000){Sanuki}, {Motoki}, {Matsumoto}, {Seo},
  {Wang}, {Abe}, {Anraku}, {Asaoka}, {Fujikawa}, {Imori}, {Maeno}, {Makida},
  {Matsui}, {Matsunaga}, {Mitchell}, {Mitsui}, {Moiseev}, {Nishimura},
  {Nozaki}, {Orito}, {Ormes}, {Saeki}, {Sasaki}, {Shikaze}, {Sonoda},
  {Streitmatter}, {Suzuki}, {Tanaka}, {Ueda}, {Yajima}, {Yamagami}, {Yamamoto},
  {Yoshida}, \& {Yoshimura}}]{2000ApJ...545.1135S}
{Sanuki}, T., {Motoki}, M., {Matsumoto}, H., {et~al.} 2000, \apj, 545, 1135

\bibitem[{{Scheepmaker} \& {Tanaka}(1971)}]{1971A&A....11...53S}
{Scheepmaker}, A. \& {Tanaka}, Y. 1971, \aap, 11, 53

\bibitem[{{Schmidt}(1972)}]{1972JGR....77.3295S}
{Schmidt}, P.~J. 1972, \jgr, 77, 3295

\bibitem[{{Seo} \& {McDonald}(1995)}]{1995ApJ...451L..33S}
{Seo}, E.~S. \& {McDonald}, F.~B. 1995, \apjl, 451, L33

\bibitem[{{Seo} {et~al.}(1994){Seo}, {McDonald}, {Lal}, \&
  {Webber}}]{1994ApJ...432..656S}
{Seo}, E.~S., {McDonald}, F.~B., {Lal}, N., \& {Webber}, W.~R. 1994, \apj, 432,
  656

\bibitem[{{Shikaze} {et~al.}(2007){Shikaze}, {Haino}, {Abe}, {Fuke}, {Hams},
  {Kim}, {Makida}, {Matsuda}, {Mitchell}, {Moiseev}, {Nishimura}, {Nozaki},
  {Orito}, {Ormes}, {Sanuki}, {Sasaki}, {Seo}, {Streitmatter}, {Suzuki},
  {Tanaka}, {Yamagami}, {Yamamoto}, {Yoshida}, \&
  {Yoshimura}}]{2007APh....28..154S}
{Shikaze}, Y., {Haino}, S., {Abe}, K., {et~al.} 2007, Astroparticle Physics,
  28, 154

\bibitem[{{Silverberg} {et~al.}(1973){Silverberg}, {Ormes}, \&
  {Balasubrahmanyan}}]{1973JGR....78.7165S}
{Silverberg}, R.~F., {Ormes}, J.~F., \& {Balasubrahmanyan}, V.~K. 1973, \jgr,
  78, 7165

\bibitem[{{Simon} {et~al.}(1980){Simon}, {Spiegelhauer}, {Schmidt}, {Siohan},
  {Ormes}, {Balasubrahmanyan}, \& {Arens}}]{1980ApJ...239..712S}
{Simon}, M., {Spiegelhauer}, H., {Schmidt}, W.~K.~H., {et~al.} 1980, \apj, 239,
  712

\bibitem[{{Simpson} \& {Connell}(1998)}]{1998ApJ...497L..85S}
{Simpson}, J.~A. \& {Connell}, J.~J. 1998, \apjl, 497, L85

\bibitem[{{Strong} \& {Moskalenko}(1998)}]{1998ApJ...509..212S}
{Strong}, A.~W. \& {Moskalenko}, I.~V. 1998, \apj, 509, 212

\bibitem[{{Strong} \& {Moskalenko}(2009)}]{2009arXiv0907.0565S}
{Strong}, A.~W. \& {Moskalenko}, I.~V. 2009, ArXiv e-prints

\bibitem[{{Strong} {et~al.}(2007){Strong}, {Moskalenko}, \&
  {Ptuskin}}]{2007ARNPS..57..285S}
{Strong}, A.~W., {Moskalenko}, I.~V., \& {Ptuskin}, V.~S. 2007, Annual Review
  of Nuclear and Particle Science, 57, 285

\bibitem[{{Swordy} {et~al.}(1990){Swordy}, {Mueller}, {Meyer}, {L'Heureux}, \&
  {Grunsfeld}}]{1990ApJ...349..625S}
{Swordy}, S.~P., {Mueller}, D., {Meyer}, P., {L'Heureux}, J., \& {Grunsfeld},
  J.~M. 1990, \apj, 349, 625

\bibitem[{{Tang}(1984)}]{1984ApJ...278..881T}
{Tang}, K.-K. 1984, \apj, 278, 881

\bibitem[{{Teegarden} {et~al.}(1975){Teegarden}, {von Rosenvinge}, {McDonald},
  {Trainor}, \& {Webber}}]{1975ApJ...202..815T}
{Teegarden}, B.~J., {von Rosenvinge}, T.~T., {McDonald}, F.~B., {Trainor},
  J.~H., \& {Webber}, W.~R. 1975, \apj, 202, 815

\bibitem[{{Thayer}(1997)}]{1997ApJ...482..792T}
{Thayer}, M.~R. 1997, \apj, 482, 792

\bibitem[{{Torii} {et~al.}(2001){Torii}, {Tamura}, {Tateyama}, {Yoshida},
  {Nishimura}, {Yamagami}, {Murakami}, {Kobayashi}, {Komori}, {Kasahara}, \&
  {Yuda}}]{2001ApJ...559..973T}
{Torii}, S., {Tamura}, T., {Tateyama}, N., {et~al.} 2001, \apj, 559, 973

\bibitem[{{Usoskin} {et~al.}(2002){Usoskin}, {Alanko}, {Mursula}, \&
  {Kovaltsov}}]{2002SoPh..207..389U}
{Usoskin}, I.~G., {Alanko}, K., {Mursula}, K., \& {Kovaltsov}, G.~A. 2002,
  \solphys, 207, 389

\bibitem[{{Usoskin} {et~al.}(2005){Usoskin}, {Alanko-Huotari}, {Kovaltsov}, \&
  {Mursula}}]{2005JGRA..11012108U}
{Usoskin}, I.~G., {Alanko-Huotari}, K., {Kovaltsov}, G.~A., \& {Mursula}, K.
  2005, Journal of Geophysical Research (Space Physics), 110, 12108

\bibitem[{{Usoskin} {et~al.}(2011){Usoskin}, {Bazilevskaya}, \&
  {Kovaltsov}}]{2011JGRA..11602104U}
{Usoskin}, I.~G., {Bazilevskaya}, G.~A., \& {Kovaltsov}, G.~A. 2011, Journal of
  Geophysical Research (Space Physics), 116, 2104

\bibitem[{{Wang} {et~al.}(2002){Wang}, {Seo}, {Anraku}, {Fujikawa}, {Imori},
  {Maeno}, {Matsui}, {Matsunaga}, {Motoki}, {Orito}, {Saeki}, {Sanuki}, {Ueda},
  {Yoshimura}, {Makida}, {Suzuki}, {Tanaka}, {Yamamoto}, {Yoshida}, {Mitsui},
  {Matsumoto}, {Nozaki}, {Sasaki}, {Mitchell}, {Moiseev}, {Ormes},
  {Streitmatter}, {Nishimura}, {Yajima}, \& {Yamagami}}]{2002ApJ...564..244W}
{Wang}, J.~Z., {Seo}, E.~S., {Anraku}, K., {et~al.} 2002, \apj, 564, 244

\bibitem[{{Webber}(1982)}]{1982ApJ...252..386W}
{Webber}, W.~R. 1982, \apj, 252, 386

\bibitem[{{Webber} \& {Chotkowski}(1967)}]{1967JGR....72.2783W}
{Webber}, W.~R. \& {Chotkowski}, C. 1967, \jgr, 72, 2783

\bibitem[{{Webber} {et~al.}(1987){Webber}, {Golden}, \&
  {Mewaldt}}]{1987ApJ...312..178W}
{Webber}, W.~R., {Golden}, R.~L., \& {Mewaldt}, R.~A. 1987, \apj, 312, 178

\bibitem[{{Webber} {et~al.}(1991){Webber}, {Golden}, {Stochaj}, {Ormes}, \&
  {Strittmatter}}]{1991ApJ...380..230W}
{Webber}, W.~R., {Golden}, R.~L., {Stochaj}, S.~J., {Ormes}, J.~F., \&
  {Strittmatter}, R.~E. 1991, \apj, 380, 230

\bibitem[{{Webber} \& {Higbie}(2009)}]{2009JGRA..11402103W}
{Webber}, W.~R. \& {Higbie}, P.~R. 2009, Journal of Geophysical Research (Space
  Physics), 114, 2103

\bibitem[{{Webber} \& {Kish}(1979)}]{1979ICRC....1..389W}
{Webber}, W.~R. \& {Kish}, J. 1979, in International Cosmic Ray Conference,
  Vol.~1, International Cosmic Ray Conference, 389

\bibitem[{{Webber} {et~al.}(1973{\natexlab{a}}){Webber}, {Kish}, \&
  {Rockstroh}}]{1973ICRC....2..760W}
{Webber}, W.~R., {Kish}, J., \& {Rockstroh}, J.~M. 1973{\natexlab{a}}, in
  International Cosmic Ray Conference, Vol.~2, International Cosmic Ray
  Conference, 760

\bibitem[{{Webber} {et~al.}(1979){Webber}, {Kish}, \&
  {Simpson}}]{1979ICRC....1..430W}
{Webber}, W.~R., {Kish}, J., \& {Simpson}, G. 1979, in International Cosmic Ray
  Conference, Vol.~1, International Cosmic Ray Conference, 430

\bibitem[{{Webber} {et~al.}(1985{\natexlab{a}}){Webber}, {Kish}, \&
  {Schrier}}]{1985ICRC....2...16W}
{Webber}, W.~R., {Kish}, J.~C., \& {Schrier}, D.~A. 1985{\natexlab{a}}, in
  International Cosmic Ray Conference, Vol.~2, International Cosmic Ray
  Conference, ed. F.~C. {Jones}, 16--19

\bibitem[{{Webber} {et~al.}(1985{\natexlab{b}}){Webber}, {Kish}, \&
  {Schrier}}]{1985ICRC....2...88W}
{Webber}, W.~R., {Kish}, J.~C., \& {Schrier}, D.~A. 1985{\natexlab{b}}, in
  International Cosmic Ray Conference, Vol.~2, International Cosmic Ray
  Conference, ed. F.~C. {Jones}, 88--91

\bibitem[{{Webber} {et~al.}(1973{\natexlab{b}}){Webber}, {Lezniak}, \&
  {Damle}}]{1973JGR....78.1487W}
{Webber}, W.~R., {Lezniak}, J.~A., \& {Damle}, S.~V. 1973{\natexlab{b}}, \jgr,
  78, 1487

\bibitem[{{Webber} {et~al.}(1973{\natexlab{c}}){Webber}, {Lezniak}, {Kish}, \&
  {Damle}}]{1973Ap&SS..24...17W}
{Webber}, W.~R., {Lezniak}, J.~A., {Kish}, J., \& {Damle}, S.~V.
  1973{\natexlab{c}}, \apss, 24, 17

\bibitem[{{Webber} {et~al.}(1997){Webber}, {Lukasiak}, \&
  {McDonald}}]{1997ApJ...476..766W}
{Webber}, W.~R., {Lukasiak}, A., \& {McDonald}, F.~B. 1997, \apj, 476, 766

\bibitem[{{Webber} {et~al.}(1996){Webber}, {Lukasiak}, {McDonald}, \&
  {Ferrando}}]{1996ApJ...457..435W}
{Webber}, W.~R., {Lukasiak}, A., {McDonald}, F.~B., \& {Ferrando}, P. 1996,
  \apj, 457, 435

\bibitem[{{Webber} \& {McDonald}(1994)}]{1994ApJ...435..464W}
{Webber}, W.~R. \& {McDonald}, F.~B. 1994, \apj, 435, 464

\bibitem[{{Webber} \& {Rockstroh}(1997)}]{1997AdSpR..19..817W}
{Webber}, W.~R. \& {Rockstroh}, J.~M. 1997, Advances in Space Research, 19, 817

\bibitem[{{Webber} \& {Schofield}(1975)}]{1975ICRC....1..312W}
{Webber}, W.~R. \& {Schofield}, N.~J. 1975, in International Cosmic Ray
  Conference, Vol.~1, International Cosmic Ray Conference, 312--317

\bibitem[{{Webber} \& {Yushak}(1983)}]{1983ApJ...275..391W}
{Webber}, W.~R. \& {Yushak}, S.~M. 1983, \apj, 275, 391

\bibitem[{{Westphal} {et~al.}(1996){Westphal}, {Afanasyev}, {Price}, {Solarz},
  {Akimov}, {Rodin}, \& {Shvets}}]{1996ApJ...468..679W}
{Westphal}, A.~J., {Afanasyev}, V.~G., {Price}, P.~B., {et~al.} 1996, \apj,
  468, 679

\bibitem[{{Wiedenbeck}(1983)}]{1983ICRC....9..147W}
{Wiedenbeck}, M.~E. 1983, in International Cosmic Ray Conference, Vol.~9,
  International Cosmic Ray Conference, 147--150

\bibitem[{{Wiedenbeck} {et~al.}(1999){Wiedenbeck}, {Binns}, {Christian},
  {Cummings}, {Dougherty}, {Hink}, {Klarmann}, {Leske}, {Lijowski}, {Mewaldt},
  {Stone}, {Thayer}, {von Rosenvinge}, \& {Yanasak}}]{1999ApJ...523L..61W}
{Wiedenbeck}, M.~E., {Binns}, W.~R., {Christian}, E.~R., {et~al.} 1999, \apjl,
  523, L61

\bibitem[{{Wiedenbeck} \& {Greiner}(1980)}]{1980ApJ...239L.139W}
{Wiedenbeck}, M.~E. \& {Greiner}, D.~E. 1980, \apjl, 239, L139

\bibitem[{{Wiedenbeck} \& {Greiner}(1981{\natexlab{a}})}]{1981ApJ...247L.119W}
{Wiedenbeck}, M.~E. \& {Greiner}, D.~E. 1981{\natexlab{a}}, \apjl, 247, L119

\bibitem[{{Wiedenbeck} \& {Greiner}(1981{\natexlab{b}})}]{1981PhRvL..46..682W}
{Wiedenbeck}, M.~E. \& {Greiner}, D.~E. 1981{\natexlab{b}}, Physical Review
  Letters, 46, 682

\bibitem[{{Xiong} {et~al.}(2003){Xiong}, {Chen}, {Yang}, {Yang}, {Chen},
  {Chen}, {L{\"u}}, {Zhuang}, \& {Tang}}]{2003JHEP...11..048X}
{Xiong}, Z., {Chen}, H., {Yang}, C., {et~al.} 2003, Journal of High Energy
  Physics, 11, 48

\bibitem[{{Yanasak} {et~al.}(2001){Yanasak}, {Wiedenbeck}, {Mewaldt}, {Davis},
  {Cummings}, {George}, {Leske}, {Stone}, {Christian}, {von Rosenvinge},
  {Binns}, {Hink}, \& {Israel}}]{2001ApJ...563..768Y}
{Yanasak}, N.~E., {Wiedenbeck}, M.~E., {Mewaldt}, R.~A., {et~al.} 2001, \apj,
  563, 768

\bibitem[{{Yoon} {et~al.}(2011){Yoon}, {Ahn}, {Allison}, {Bagliesi}, {Beatty},
  {Bigongiari}, {Boyle}, {Childers}, {Conklin}, {Coutu}, {DuVernois}, {Ganel},
  {Han}, {Jeon}, {Kim}, {Lee}, {Lutz}, {Maestro}, {Malinine}, {Marrocchesi},
  {Minnick}, {Mognet}, {Nam}, {Nutter}, {Park}, {Park}, {Seo}, {Sina},
  {Swordy}, {Wakely}, {Wu}, {Yang}, {Zei}, \& {Zinn}}]{2011ApJ...728..122Y}
{Yoon}, Y.~S., {Ahn}, H.~S., {Allison}, P.~S., {et~al.} 2011, \apj, 728, 122

\bibitem[{{Yoshida} {et~al.}(2008){Yoshida}, {Torii}, {Yamagami}, {Tamura},
  {Kitamura}, {Chang}, {Iijima}, {Kadokura}, {Kasahara}, {Katayose},
  {Kobayashi}, {Komori}, {Matsuzaka}, {Mizutani}, {Murakami}, {Namiki},
  {Nishimura}, {Ohta}, {Saito}, {Shibata}, {Tateyama}, {Yamagishi}, \&
  {Yuda}}]{2008AdSpR..42.1670Y}
{Yoshida}, K., {Torii}, S., {Yamagami}, T., {et~al.} 2008, Advances in Space
  Research, 42, 1670

\bibitem[{{Young} {et~al.}(1981){Young}, {Freier}, {Waddington}, {Brewster}, \&
  {Fickle}}]{1981ApJ...246.1014Y}
{Young}, J.~S., {Freier}, P.~S., {Waddington}, C.~J., {Brewster}, N.~R., \&
  {Fickle}, R.~K. 1981, \apj, 246, 1014

\bibitem[{{Zatsepin} {et~al.}(1993){Zatsepin}, {Zamchalova}, {Varkovitskaya},
  {Sokolskaya}, {Sazhina}, \& {Lazareva}}]{1993ICRC....2...13Z}
{Zatsepin}, V.~I., {Zamchalova}, E.~A., {Varkovitskaya}, A.~Y., {et~al.} 1993,
  in International Cosmic Ray Conference, Vol.~2, International Cosmic Ray
  Conference, 13

\end{thebibliography}
\end{document}